\documentclass[12pt,english,aps,superscriptaddress]{revtex4}
\usepackage[T1]{fontenc}
\usepackage[latin9]{inputenc}
\usepackage{babel}
\usepackage{float}
\usepackage{amsmath}
\usepackage{amssymb}
\usepackage{graphicx}
\usepackage{esint}
\usepackage[unicode=true,
 bookmarks=true,bookmarksnumbered=true,bookmarksopen=true,bookmarksopenlevel=1,
 breaklinks=false,pdfborder={0 0 1},backref=false,colorlinks=false]
 {hyperref}
\usepackage{breakurl}

\makeatletter

\providecommand{\tabularnewline}{\\}

\@ifundefined{textcolor}{}
{%
 \definecolor{BLACK}{gray}{0}
 \definecolor{WHITE}{gray}{1}
 \definecolor{RED}{rgb}{1,0,0}
 \definecolor{GREEN}{rgb}{0,1,0}
 \definecolor{BLUE}{rgb}{0,0,1}
 \definecolor{CYAN}{cmyk}{1,0,0,0}
 \definecolor{MAGENTA}{cmyk}{0,1,0,0}
 \definecolor{YELLOW}{cmyk}{0,0,1,0}
}

\makeatother

\begin{document}

\title{A microscopic description for polarization in particle scatterings}

\author{Jun-jie Zhang}

\affiliation{Department of Modern Physics, University of Science and Technology
of China, Hefei, Anhui 230026, China}

\author{Ren-hong Fang}

\affiliation{Key Laboratory of Quark and Lepton Physics (MOE) and Institute of
Particle Physics, Central China Normal University, Wuhan, Hubei 430079,
China}

\author{Qun Wang}

\affiliation{Department of Modern Physics, University of Science and Technology
of China, Hefei, Anhui 230026, China}

\author{Xin-Nian Wang}

\affiliation{Key Laboratory of Quark and Lepton Physics (MOE) and Institute of
Particle Physics, Central China Normal University, Wuhan, Hubei 430079,
China}

\affiliation{Nuclear Science Division, MS 70R0319, Lawrence Berkeley National
Laboratory, Berkeley, California 94720}
\begin{abstract}
We propose a microscopic description for the polarization from the
first principle through the spin-orbit coupling in particle collisions.
The model is different from previous ones based on local equilibrium
assumptions for the spin degree of freedom. It is based on scatterings
of particles as wave packets, an effective method to deal with particle
scatterings at specified impact parameters. The polarization is then
the consequence of particle collisions in a non-equilibrium state
of spins. The spin-vorticity coupling naturally emerges from the spin-orbit
one encoded in polarized scattering amplitudes of collisional integrals
when one assumes local equilibrium in momentum but not in spin. 
\end{abstract}
\maketitle

\section{Introduction}

A very large orbital angular momentum (OAM) can be created in peripheral
heavy ion collisions \cite{Liang:2004ph,Liang:2004xn,Voloshin:2004ha,Betz2007,Becattini:2007sr,Gao2008,Wang:2017jpl}.
Such a huge OAM can be transferred to the hot and dense matter produced
in collisions and make particles with spins polarized along the direction
of OAM \cite{Liang:2004ph,Gao2008,Huang:2011ru,Wang:2017jpl}. Recently
the STAR collaboration has measured the global polarization of $\Lambda$
and $\bar{\Lambda}$ for the first time in Au+Au collisions at $\sqrt{s_{NN}}=7.7-200$
GeV \cite{STAR:2017ckg,Adam:2018ivw,Niida:2018hfw}. The global polarization
is the net polarization of local ones in an event which is aligned
in the direction of the event plane. The results show that the magnitude
of the global $\Lambda$ and $\bar{\Lambda}$ polarization is of the
order a few percent and decreases with collisional energies. The difference
between the global polarization of $\Lambda$ and $\bar{\Lambda}$
may possibly indicates the effect from the strong magnetic field formed
in high energy heavy ion collisions. 

Several theoretical models have been developed to study the global
polarization. If the spin degree of freedom is thermalized, one can
construct the statistic-hydro model by including the spin-vorticity
coupling $S_{\mu\nu}\omega^{\mu\nu}$ into the thermal distribution
function \cite{Becattini:2013fla,Becattini:2015nva,Becattini:2016gvu,Florkowski:2017ruc,Florkowski:2017dyn,Florkowski:2018ahw}.
Here $S_{\mu\nu}$ is the spin tensor, $\omega^{\mu\nu}=-(1/2)(\partial^{\mu}\beta^{\nu}-\partial^{\nu}\beta^{\mu})$
is the thermal vorticity, the macroscopic analog of the local OAM,
and $\beta^{\mu}\equiv\beta u^{\mu}$ is the thermal velocity with
$\beta=1/T$ being the inverse of the temperature and $u^{\mu}$ being
the fluid velocity. It turns out that the average spin or polarization
is proportional to the thermal vorticity if the spin-vorticity coupling
is weak. 

Similar to the statistic-hydro model, another approach to the global
polarization assuming local equilibrium is the the Wigner function
(WF) formalism. The WF formalism for spin-1/2 fermions \cite{Heinz:1983nx,Elze:1986qd,Vasak:1987um,Zhuang:1995pd,Florkowski:1995ei,Blaizot:2001nr,Wang:2001dm}
has recently been revived to study the chiral magnetic effect (CME)
\cite{Vilenkin:1980fu,Kharzeev:2007jp,Fukushima:2008xe,Kharzeev:2015znc}
(for reviews, see, e.g., Ref. \cite{Kharzeev:2013jha,Kharzeev:2015znc,Huang:2015oca})
and chiral vortical effect (CVE) \cite{Vilenkin:1978hb,Erdmenger:2008rm,Banerjee:2008th,Son:2009tf,Gao:2012ix,Hou:2012xg}
for massless fermions \cite{Gao:2012ix,Chen:2012ca,Gao:2015zka,Hidaka:2016yjf,Gao:2017gfq,Gao:2018wmr,Huang:2018wdl,Gao:2018jsi}.
The Wigner functions for spin-1/2 fermions are $4\times4$ matrices.
The axial vector component gives the spin phase space distribution
of fermions near thermal equilibrium \cite{Fang:2016vpj,Weickgenannt:2019dks,Gao:2019znl,Hattori:2019ahi}.
It can be shown that when the thermal vorticity is small, the spin
polarization of fermions from the WF is proportional to the thermal
vorticity vector. So the WF can also be applied to the study of the
global polarization of hyperons. 

In order to describe the STAR data on the global $\Lambda/\bar{\Lambda}$
polarization, the hydrodynamic or transport models have been used
to calculate the vorticity fields in heavy ion collisions \cite{Baznat:2013zx,Csernai:2013bqa,Csernai:2014ywa,Teryaev:2015gxa,Jiang:2016woz,Deng:2016gyh,Ivanov:2017dff}.
Then the polarization of $\Lambda/\bar{\Lambda}$ can be obtained
from vorticity fields at the freezeout when the $\Lambda/\bar{\Lambda}$
hyperons are decoupled from the rest of the hot and dense matter \cite{Karpenko:2016jyx,Xie:2017upb,Li:2017slc,Sun:2017xhx}. 

All these models are based on the assumption that the spin degree
of freedom has reached local equilbrium. But this assumption is not
justified. The recent disagreement between some theoretical models
and data on the longitudinal polarization indicates that the spins
might not be in local equilibrium \cite{Niida:2018hfw,Becattini:2017gcx,Xia:2018tes}.
Although one model of the chiral kinetic theory can explain the sign
of the data \cite{Sun:2018bjl}, it cannot reproduce the magnitude
of the data. If the spins are not in local equilbrium, how is the
polarization generated in particle collisions? This is also related
to the role of the spin-orbit coupling which is regarded as the microscopic
mechanism for the global polarization. In one particle scattering
such as a 2-to-2 scattering at fixed impact parameter the effect of
spin-orbit coupling in the polarized cross section is obvious \cite{Liang:2004ph,Gao2008},
but how does the spin-vorticity coupling naturally emerge from the
spin-orbit one? It is far from easy and obvious as it involves the
treatment of particle scatterings at different space-ime points in
a system of particles in randomly distributed momentum. To the best
of our knowledge, this problem has not been seriously investigated
due to such a difficulty. In this paper we will construct a microscopic
model for the global polarization based on the spin-orbit coupling.
We will show that the spin-vorticity coupling naturally emerges from
scatterings of particles at different space-time points incorporating
polarized scattering amplitudes with the spin-orbit coupling. This
provides a microscopic mechanism for the global polarization from
the first principle through particle collisions in non-equilibrium. 

The paper is organized as follows. In Section II we will introduce
scatterings of two wave packets for spin-0 particles. The wave packet
method is necessary to describe particle scatterings at different
space-time points. In Section III we will study collisions of spin-0
particles as wave packets which take place at different space-time
in a multi-particle system. In Section IV we will derive the polarization
rate for spin-1/2 particles from particle collisions. As an example,
we will apply in Section V the formalism to derive the quark polarization
rate in a quark-gluon plasma in local equilibrium in momentum. In
Section VI we will discuss the numerical method to calculate the quark
polarization rate, a challenging task to deal with collision integrals
in very high dimensions. We will present the numerical results in
Section VII. Finally we will give a summary of the work and an outlook
for future studies. 

Throughout the paper we use natural units $\hbar=c=k_{B}=1$. The
convention for the metric tensor is $g^{\mu\nu}=\mathrm{diag}(+1,-1,-1,-1)$.
We also use the notation $a^{\mu}b_{\mu}\equiv a\cdot b$ for the
scalar product of two four-vectors $a^{\mu}$, $b^{\mu}$ and $\mathbf{a}\cdot\mathbf{b}$
for the corresponding scalar product of two spatial vectors $\mathbf{a}$,
$\mathbf{b}$. The direction of a three-vector $\mathbf{a}$ is denoted
as $\hat{\mathbf{a}}$. Sometimes we denote the components of a three-vector
by indices $(1,2,3)$ or $(x,y,z)$.

\section{Scatterings of wave packets for spin-0 particles}

In this section we will consider the scattering process $A+B\rightarrow1+2\cdots+n$,
where the incident particles $A$ and $B$ in the remote past are
localized in some region and can be described by wave packets. The
details of this section can be found in the textbook by Peskin and
Schroeder \cite{Peskin:1995ev}. The purpose of this section is to
give an idea of how the wave packets displaced by an impact parameter
are treated in the scattering process, and to provide the basis for
the discussion in the next section. We work in the frame in which
the central momenta of two wave packets are collinear or in the same
direction which we denote as the longitudinal direction. We assume
that the wave packet $B$ is displaced by an impact parameter vector
$\mathbf{b}$ in the transverse direction, so the \textit{in} state
can be written as 
\begin{equation}
|\phi_{A}\phi_{B}\rangle_{\text{in}}=\int\frac{d^{3}k_{A}}{(2\pi)^{3}}\frac{d^{3}k_{B}}{(2\pi)^{3}}\frac{\phi_{A}(\mathbf{k}_{A})\phi_{B}(\mathbf{k}_{B})e^{-i\mathbf{k}_{B}\cdot\mathbf{b}}}{\sqrt{4E_{A}E_{B}}}\left|\mathbf{k}_{A}\mathbf{k}_{B}\right\rangle _{\text{in}}.
\end{equation}
Here we see that the incident particles are treated as two wave packets
$\left|\phi_{A}\right\rangle $ and $\left|\phi_{B}\right\rangle $
defined in Appendix \ref{sec:one-part-wave-pack}. The definition
of the single particle states $\left|\mathbf{k}_{A}\right\rangle $
and $\left|\mathbf{k}_{B}\right\rangle $ can also be found in Appendix
\ref{sec:one-part-wave-pack}. As we have mentioned that the amplitudes
$\phi_{i}(\mathbf{k}_{i})$ center at $\mathbf{p}_{i}=(0,0,p_{iz})$
for $i=A,B$. We assume that the \textit{out }state is a pure momentum
state $|\mathbf{p}_{1}\mathbf{p}_{2}\cdots\mathbf{p}_{n}\rangle_{\text{out}}$
in the far future. This is physically reasonable as long as the detectors
of final-state particles mainly measure momentum or they do not resolve
positions at the level of de Broglie wavelengths. Taking into account
the normalization factors for the in-state and out-state, the scattering
probability is given by 
\begin{eqnarray}
\mathcal{P}(AB\rightarrow12\cdots n) & = & \sum_{\mathbf{p}_{1}}\sum_{\mathbf{p}_{2}}\cdots\sum_{\mathbf{p}_{n}}\frac{|_{\text{out}}\langle\mathbf{p}_{1}\mathbf{p}_{2}\cdots\mathbf{p}_{n}|\phi_{A}\phi_{B}\rangle_{\text{in}}|^{2}}{\prod_{f=1}^{n}\langle\mathbf{p}_{f}|\mathbf{p}_{f}\rangle\langle\phi_{A}|\phi_{A}\rangle\langle\phi_{B}|\phi_{B}\rangle}\nonumber \\
 & = & \left(\prod_{f=1}^{n}\int\frac{\Omega d^{3}p_{f}}{(2\pi)^{3}}\right)\frac{|_{\text{out}}\langle\mathbf{p}_{1}\mathbf{p}_{2}\cdots\mathbf{p}_{n}|\phi_{A}\phi_{B}\rangle_{\text{in}}|^{2}}{\prod_{f=1}^{n}(2E_{f}\Omega)}\nonumber \\
 & = & \left(\prod_{f=1}^{n}\int\frac{d^{3}p_{f}}{(2\pi)^{3}2E_{f}}\right)|{}_{\text{out}}\langle\mathbf{p}_{1}\mathbf{p}_{2}\cdots\mathbf{p}_{n}|\phi_{A}\phi_{B}\rangle_{\text{in}}|^{2},\label{eq:prob-1}
\end{eqnarray}
where the normalization of single particle states and wave packets
is given in Appendix \ref{sec:one-part-wave-pack}. Since $\mathcal{P}(AB\rightarrow12\cdots n)$
depends on the impact parameter $\mathbf{b}$, we can rewrite it as
$\mathcal{P}(\mathbf{b})$. This probability gives the differential
cross section at the impact parameter $\mathbf{b}$, 
\begin{eqnarray}
\frac{d\sigma}{d^{2}b} & = & \mathcal{P}(\mathbf{b}).
\end{eqnarray}
The total cross section is then an integral over the impact parameter 

\begin{eqnarray}
\sigma & = & \int d^{2}b\mathcal{P}(\mathbf{b})\nonumber \\
 & = & \left(\prod_{f=1}^{n}\int\frac{d^{3}p_{f}}{(2\pi)^{3}2E_{f}}\right)\prod_{i=A,B}\int\frac{d^{3}k_{i}}{(2\pi)^{3}}\frac{\phi_{i}(\mathbf{k}_{i})}{\sqrt{2E_{i}}}\int\frac{d^{3}k_{i}^{\prime}}{(2\pi)^{3}}\frac{\phi_{i}^{*}(\mathbf{k}_{i}^{\prime})}{\sqrt{2E_{i}^{\prime}}}\nonumber \\
 &  & \times\int d^{2}be^{i(\mathbf{k}_{B}^{\prime}-\mathbf{k}_{B})\cdot\mathbf{b}}\left(_{\text{out}}\langle\{\mathbf{p}_{f}\}|\{\mathbf{k}_{i}\}\rangle_{\text{in}}\right)\left(_{\text{out}}\langle\{\mathbf{p}_{f}\}|\{\mathbf{k}_{i}^{\prime}\}\rangle_{\text{in}}\right)^{*}\nonumber \\
 & = & \left(\prod_{f=1}^{n}\int\frac{d^{3}p_{f}}{(2\pi)^{3}2E_{f}}\right)\left(\prod_{i=A,B}\int\frac{d^{3}k_{i}}{(2\pi)^{3}}\frac{\phi_{i}(\mathbf{k}_{i})}{\sqrt{2E_{ki}}}\int\frac{d^{3}k_{i}^{\prime}}{(2\pi)^{3}}\frac{\phi_{i}^{*}(\mathbf{k}_{i}^{\prime})}{\sqrt{2E_{ki}^{\prime}}}\right)(2\pi)^{2}\delta^{(2)}\left(\mathbf{k}_{B,\perp}^{\prime}-\mathbf{k}_{B,\perp}\right)\nonumber \\
 &  & \times(2\pi)^{4}\delta^{(4)}(k_{A}^{\prime}+k_{B}^{\prime}-\sum_{f=1}^{n}p_{f})(2\pi)^{4}\delta^{(4)}(k_{A}+k_{B}-\sum_{f=1}^{n}p_{f})\nonumber \\
 &  & \times\mathcal{M}\left(\{k_{A},k_{B}\}\rightarrow\{p_{1},p_{2},\cdots,p_{n}\}\right)\mathcal{M}^{*}\left(\{k_{A}^{\prime},k_{B}^{\prime}\}\rightarrow\{p_{1},p_{2},\cdots,p_{n}\}\right),
\end{eqnarray}
where $E_{ki}=\sqrt{|\mathbf{k}_{i}|^{2}+m_{i}^{2}}$, $E_{ki}^{\prime}=\sqrt{|\mathbf{k}_{i}^{\prime}|^{2}+m_{i}^{2}}$
with $i=A,B$, $\mathbf{k}_{B,\perp}$ denotes the transverse part
of the momentum, $\mathcal{M}$ denotes the invariant amplitude of
the scattering process. We can integrate out six delta functions involving
$\mathbf{k}_{A}^{\prime}$ and $\mathbf{k}_{B}^{\prime}$, i.e. $\delta^{(2)}\left(\mathbf{k}_{B,\perp}^{\prime}-\mathbf{k}_{B,\perp}\right)$
and $\delta^{(4)}\left(k_{A}^{\prime}+k_{B}^{\prime}-\sum_{f=1}^{n}p_{f}\right)$.
By integrating over $\mathbf{k}_{B,\bot}^{\prime}$ to remove $\delta^{(2)}\left(\mathbf{k}_{B,\perp}^{\prime}-\mathbf{k}_{B,\perp}\right)$,
we can replace $\mathbf{k}_{B,\bot}^{\prime}$ by $\mathbf{k}_{B,\bot}$
in the integrand. By integrating over $\mathbf{k}_{A,\perp}^{\prime}$
to remove $\delta^{(2)}\left(\mathbf{k}_{A,\perp}^{\prime}+\mathbf{k}_{B,\perp}^{\prime}-\sum_{f=1}^{n}\mathbf{p}_{f,\perp}\right)$,
we can replace $\mathbf{k}_{A,\bot}^{\prime}$ by $-\mathbf{k}_{B,\perp}+\sum_{f=1}^{n}\mathbf{k}_{f,\perp}$
in the integrand. Then we can integrate over $k_{B,z}^{\prime}$ to
remove $\delta(k_{A,z}^{\prime}+k_{B,z}^{\prime}-p_{1,z}-p_{2,z})$,
in which $k_{B,z}^{\prime}$ is replaced by $\sum_{f=1}^{n}p_{f,z}-k_{A,z}^{\prime}$.
The last variable that can be integrated over is $k_{A,z}^{\prime}$
in the delta function for the energy conservation $\delta(E_{A}^{\prime}+E_{B}^{\prime}-E_{p1}-E_{p2})$.
We can solve $k_{A,z}^{\prime}$ as the root of the equation $E_{A}^{\prime}+E_{B}^{\prime}=E_{p1}+E_{p2}$.
Note that $E_{A}^{\prime}$ and $E_{B}^{\prime}$ are given by 
\begin{eqnarray}
E_{A}^{\prime} & = & \sqrt{(-\mathbf{k}_{B,\perp}+\sum_{f=1}^{n}\mathbf{k}_{f,\perp})^{2}+k_{A,z}^{\prime2}+m_{A}^{2}},\nonumber \\
E_{B}^{\prime} & = & \sqrt{\mathbf{k}_{B,\perp}^{2}+(\sum_{f=1}^{n}p_{f,z}-k_{A,z}^{\prime})^{2}+m_{B}^{2}}.
\end{eqnarray}
The delta function can be rewritten as 
\begin{equation}
\delta\left(E_{A}^{\prime}+E_{B}^{\prime}-\sum_{f=1}^{n}E_{f}\right)=\sum_{j}\left|\frac{k_{A,z,j}^{\prime}}{E_{A}^{\prime}}-\frac{k_{B,z,j}^{\prime}}{E_{B}^{\prime}}\right|^{-1}\delta(k_{A,z}^{\prime}-k_{A,z,j}^{\prime}),
\end{equation}
where $k_{A,z,j}^{\prime}$ are the roots of the equation $E_{A}^{\prime}+E_{B}^{\prime}=E_{p1}+E_{p2}$. 

If we assume that the incident wave packets are narrow in momentum
and centered at momenta $\mathbf{p}_{A}$ and $\mathbf{p}_{B}$, i.e.
$\phi_{i}(\mathbf{k}_{i})$ are close to delta functions $\delta(\mathbf{k}_{i}-\mathbf{p}_{i})$,
we can approximate $(E_{kA}^{\prime},\mathbf{k}_{A}^{\prime})\approx(E_{kA},\mathbf{k}_{A})\approx(E_{A},\mathbf{p}_{A})$
and $(E_{B}^{\prime},\mathbf{k}_{B}^{\prime})\approx(E_{kB},\mathbf{k}_{B})\approx(E_{B},\mathbf{p}_{B})$.
We can also approximate $v_{i}=p_{i,z}/E_{i}\approx k_{i,z}^{\prime}/E_{i}^{\prime}$
with $i=A,B$. Then we obtain
\begin{eqnarray}
\sigma & \approx & \left(\prod_{f=1}^{n}\int\frac{d^{3}p_{f}}{(2\pi)^{3}2E_{f}}\right)\int\frac{d^{3}k_{A}}{(2\pi)^{3}}\frac{|\phi_{A}(\mathbf{k}_{A})|^{2}}{2E_{A}}\int\frac{d^{3}k_{B}}{(2\pi)^{3}}\frac{|\phi_{B}(\mathbf{k}_{B})|^{2}}{2E_{B}}\left|v_{A}-v_{B}\right|^{-1}\nonumber \\
 &  & \times(2\pi)^{4}\delta(p_{A}+p_{B}-\sum_{f=1}^{n}p_{f})\left|\mathcal{M}(\{p_{i}\}\rightarrow\{p_{f}\})\right|^{2}\nonumber \\
 & = & \frac{1}{4E_{A}E_{B}|v_{A}-v_{B}|}\left(\prod_{f=1}^{n}\int\frac{d^{3}p_{f}}{(2\pi)^{3}2E_{f}}\right)\nonumber \\
 &  & \times(2\pi)^{4}\delta(p_{A}+p_{B}-\sum_{f=1}^{n}p_{f})\left|\mathcal{M}(\{p_{i}\}\rightarrow\{p_{f}\})\right|^{2}.
\end{eqnarray}
Here we have used the normalization condition for the wave amplitude
(\ref{eq:wavepack}). We note that the above formula is derived in
the frame in which incident particles are collinear in momemtum. We
can boost the frame to the center-of-mass frame of the incident particles
and the cross section is invariant. 

If the number densities of $A$ and $B$ in coordinate space are $n_{A}$
and $n_{B}$ respectively, the collision rate, i.e. the number of
scatterings per unit time and unit volume is given by 
\begin{eqnarray}
R & = & n_{A}n_{B}|v_{A}-v_{B}|\sigma\nonumber \\
 & = & \frac{n_{A}n_{B}}{4E_{A}E_{B}}4E_{A}E_{B}|v_{A}-v_{B}|\sigma,\label{eq:rate-sigma}
\end{eqnarray}
where we have rewritten the rate in a Lorentz invariant way by making
use of the fact that $4E_{A}E_{B}|v_{A}-v_{B}|$, $n_{A}/E_{A}$ and
$n_{B}/E_{B}$ are Lorentz invariant along the collision axis.

\section{Collision rate for spin-0 particles in a multi-particle system}

In this section we will derive the collision rate in a system of spin-0
particles of multi-species. We will generalize the result of the previous
section by treating the incident particles as wave packets. The emphasis
is put on the collision of two particles at two different space-time
points. 

\begin{figure}[H]
\caption{\label{fig:coll-heat-bath}A collision or scattering in the Lab frame
(left) and center-of-mass frame (right). }

\begin{center}\includegraphics[scale=0.5]{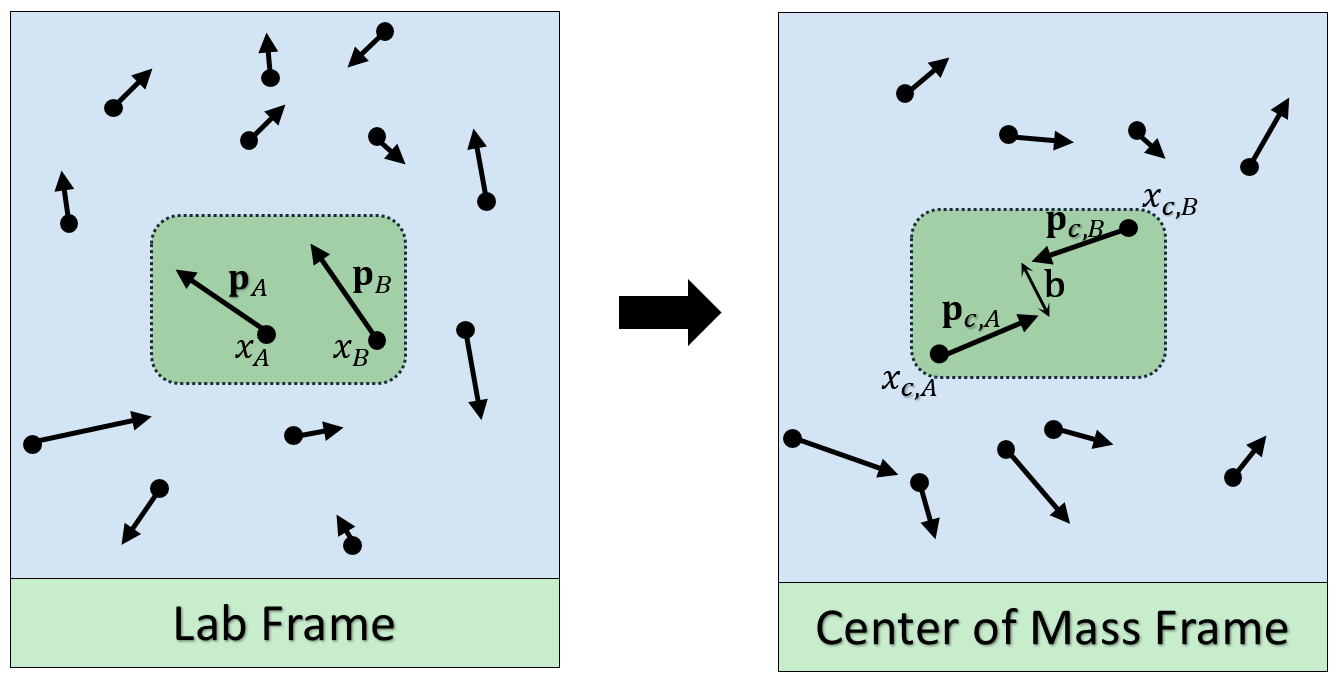}\end{center}
\end{figure}

We will frequently use two frames in this and the next section: the
lab frame and the center-of-mass system (CMS) of one specific collision.
In the lab frame, the movement of one species of particles follows
their phase space distribution $f(x,p)$. There are many collisions
taking place in the system. Figure \ref{fig:coll-heat-bath} shows
one collision of two incident particles at $x_{A}=(t_{A},\mathbf{x}_{A})$
and $x_{B}=(t_{B},\mathbf{x}_{B})$ in the lab frame and CMS. We see
that $\mathbf{p}_{A}$ and $\mathbf{p}_{B}$ are not aligned in the
same direction in the lab frame. When boosted to the CMS of this collision
with the boost velocity determined by $\mathbf{v}_{\mathrm{bst}}=(\mathbf{p}_{A}+\mathbf{p}_{B})/(E_{A}+E_{B})$,
we have $\mathbf{p}_{c,A}+\mathbf{p}_{c,B}=0$ as shown in the right
panel of Fig. \ref{fig:coll-heat-bath}, see Appendix \ref{sec:lorentz}
for more details of such a Lorentz transformation. Hereafter we denote
the quantities in the CMS by the index 'c'. There is an inherent problem
in the collision of incident particles located at different space-time
points: the collision time is not well defined. If we assume that
the collision takes place at the same time in the lab frame, i.e.
$t_{A}=t_{B}$, after being boosted to the CMS, the time will be mis-matched,
i.e. $t_{c,A}\neq t_{c,B}$, since $\mathbf{x}_{A}$ and $\mathbf{x}_{B}$
are different. The reverse statement is also true: if $t_{c,A}=t_{c,B}$
then $t_{A}\neq t_{B}$ due to $\mathbf{x}_{c,A}\neq\mathbf{x}_{c,B}$.
Such an ambiguity in the collision time cannot be avoided but can
be constrained by the requirement that the difference $\Delta t_{c}=t_{c,A}-t_{c,B}$
should not be large, otherwise the incident particles are irrelevant
or the collision is un-causal in the CMS. In the calculation of this
paper, we will put a simple constraint $\Delta t_{c}=0$. In the right
panel of Fig. \ref{fig:coll-heat-bath}, we also see that the impact
parameter $\mathbf{b}$ is given by the distance of $\mathbf{x}_{c,A}$
and $\mathbf{x}_{c,B}$ in the transverse direction which is perpendicular
to $\mathbf{p}_{c,A}$ or $\mathbf{p}_{c,B}$. In the longitudinal
direction or the direction of $\mathbf{p}_{c,A}$ or $\mathbf{p}_{c,B}$,
two space points are also different in general, i.e. $\hat{\mathbf{p}}_{c,A}\cdot\mathbf{x}_{c,A}\neq\hat{\mathbf{p}}_{c,A}\cdot\mathbf{x}_{c,B}$.
In the calculation we also require that the distance between two space
points in the longitudinal direction, $\Delta x_{c,L}=\hat{\mathbf{p}}_{c,A}\cdot(\mathbf{x}_{c,A}-\mathbf{x}_{c,B})$,
should not be large, otherwise the incident particles as wave packets
lose coherence and cannot interact in the CMS. In the calculation,
we will also put a simple constraint $\Delta x_{c,L}=0$. The CMS
constraint $\Delta t_{c}=0$ and $\Delta x_{c,L}=0$ is equivalent
to the condition $\Delta t=\mathbf{v}_{\mathrm{bst}}\cdot\Delta\mathbf{x}$
and $(\mathbf{v}_{A}-\mathbf{v}_{B})\cdot\Delta\mathbf{x}=0$ in the
lab frame, see Appendix \ref{sec:lorentz} for the derivation. 

Since we will work in the CMS of incident particles in each collision,
for notational simplicity, we will suppress the index 'c' (standing
for the CMS) of all quantities in the rest part of this section. So
all quantities are implied in the CMS if not explicitly stated here. 

We know that the momentum integral of the distribution function gives
the number density in the coordinate space. Similar to Eq. (\ref{eq:rate-sigma}),
the collision rate in corresponding momentum and space-time intervals
can be written as 
\begin{equation}
R_{AB\rightarrow12}=\frac{d^{3}p_{A}}{(2\pi)^{3}}\frac{d^{3}p_{B}}{(2\pi)^{3}}f_{A}(x_{A},p_{A})f_{B}(x_{B},p_{B})|v_{A}-v_{B}|\Delta\sigma,\label{eq:diff-rate}
\end{equation}
where $v_{A}=|\mathbf{p}_{A}|/E_{A}$ and $v_{B}=-|\mathbf{p}_{B}|/E_{B}$
are the longitudinal velocities with $\mathbf{p}_{A}=-\mathbf{p}_{B}$
in the CMS, $f_{A}$ and $f_{B}$ are the phase space distributions
for the incident particle $A$ and $B$ respectively, and $\Delta\sigma$
denotes the infinitesimal element of the cross section given by 
\begin{eqnarray}
\Delta\sigma & = & \frac{1}{C_{AB}}d^{4}x_{A}d^{4}x_{B}\delta(\Delta t)\delta(\Delta x_{L})\nonumber \\
 &  & \times\frac{d^{3}p_{1}}{(2\pi)^{3}2E_{1}}\frac{d^{3}p_{2}}{(2\pi)^{3}2E_{2}}\frac{1}{(2E_{A})(2E_{B})}K.\label{eq:delta-sigma}
\end{eqnarray}
Here we have assumed that the scattering takes place at the same time
and the same longitudinal position in the CMS, so we put two delta
functions to implement these constraints. The constant $C_{AB}$ is
to make $\Delta\sigma$ have the right dimension of the cross section
and will be defined later. In Eq. (\ref{eq:delta-sigma}) $K$ is
given by 
\begin{eqnarray}
K & = & (2E_{A})(2E_{B})|{}_{\text{out}}\langle p_{1}p_{2}|\phi_{A}(x_{A},p_{A})\phi_{B}(x_{B},p_{B})\rangle_{\text{in}}|^{2}\nonumber \\
 & = & \frac{4E_{A}E_{B}}{(2\pi)^{12}}G_{1}G_{2}\int d^{3}k_{A}d^{3}k_{B}d^{3}k_{A}^{\prime}d^{3}k_{B}^{\prime}\nonumber \\
 &  & \times\frac{\phi_{A}(\mathbf{k}_{A}-\mathbf{p}_{A})\phi_{B}(\mathbf{k}_{B}-\mathbf{p}_{B})\phi_{A}^{*}(\mathbf{k}_{A}^{\prime}-\mathbf{p}_{A})\phi_{B}^{*}(\mathbf{k}_{B}^{\prime}-\mathbf{p}_{B})}{\sqrt{16E_{A,k}E_{B,k}E_{A,k^{\prime}}E_{B,k^{\prime}}}}\nonumber \\
 &  & \times\exp\left(-i\mathbf{k}_{A}\cdot\mathbf{x}_{A}-i\mathbf{k}_{B}\cdot\mathbf{x}_{B}+i\mathbf{k}_{A}^{\prime}\cdot\mathbf{x}_{A}+i\mathbf{k}_{B}^{\prime}\cdot\mathbf{x}_{B}\right)\nonumber \\
 &  & \times(2\pi)^{4}\delta^{(4)}(k_{A}^{\prime}+k_{B}^{\prime}-p_{1}-p_{2})(2\pi)^{4}\delta^{(4)}(k_{A}+k_{B}-p_{1}-p_{2})\nonumber \\
 &  & \times\mathcal{M}\left(\{k_{A},k_{B}\}\rightarrow\{p_{1},p_{2}\}\right)\mathcal{M}^{*}\left(\{k_{A}^{\prime},k_{B}^{\prime}\}\rightarrow\{p_{1},p_{2}\}\right),\label{eq:k-factor}
\end{eqnarray}
where $\phi_{i}(\mathbf{k}_{i}-\mathbf{p}_{i})$ and $\phi_{i}(\mathbf{k}_{i}^{\prime}-\mathbf{p}_{i})$
for $i=A,B$ denote the incident wave packet amplitudes centered at
$\mathbf{p}_{i}$, $E_{i,k}=\sqrt{|\mathbf{k}_{i}|^{2}+m_{i}^{2}}$,
$E_{i,k^{\prime}}=\sqrt{|\mathbf{k}_{i}^{\prime}|^{2}+m_{i}^{2}}$
and $E_{i}=\sqrt{|\mathbf{p}_{i}|^{2}+m_{i}^{2}}$ are energies for
$i=A,B$. In Eq. (\ref{eq:k-factor}) $G_{i}$ $(i=1,2)$ denote distribution
factors depending on particle types in the final state, we have $G_{i}=1$
for the Boltzmann particles and $G_{i}=1\pm f_{i}(p_{i})$ for bosons
(upper sign) and fermions (lower sign). Note that $f_{i}(p_{i})$
can be in any other form in non-equilibrium cases. In (\ref{eq:k-factor})
we have taken the following form for $|\phi_{i}(x_{i},p_{i})\rangle_{\text{in}}$
with $i=A,B$, 
\begin{eqnarray}
|\phi_{i}(x_{i},p_{i})\rangle_{\text{in}} & = & \int\frac{d^{3}k_{i}}{(2\pi)^{3}}\frac{1}{\sqrt{2E_{i,k}}}\phi_{i}(\mathbf{k}_{i}-\mathbf{p}_{i})e^{-i\mathbf{k}_{i}\cdot\mathbf{x}_{i}}|\mathbf{k}_{i}\rangle_{\text{in}}.
\end{eqnarray}
Here we take the Gaussian form for the wave packet amplitude $\phi_{i}(\mathbf{k}_{i}-\mathbf{p}_{i})$
as in (\ref{eq:width-wavepacket}), 
\begin{equation}
\phi_{i}(\mathbf{k}_{i}-\mathbf{p}_{i})=\frac{(8\pi)^{3/4}}{\alpha_{i}^{3/2}}\exp\left[-\frac{(\mathbf{k}_{i}-\mathbf{p}_{i})^{2}}{\alpha_{i}^{2}}\right],\label{eq:wave-packet-gs}
\end{equation}
where $\alpha_{i}$ denote the width parameters of the wave packet
$A$ or $B$. For simplicity we will set equal width for two incident
particles (even for different species), $\alpha_{A}=\alpha_{B}=\alpha$. 

We can also make the approximation of narrow wave packets, so we have
$|\mathbf{k}_{i}|\approx|\mathbf{k}_{i}^{\prime}|\approx|\mathbf{p}_{i}|$
for $i=A,B$ and then $\sqrt{E_{A,k}E_{A,k}^{\prime}}\approx E_{A}$
and $\sqrt{E_{B,k}E_{B,k}^{\prime}}\approx E_{B}$, and the energy
factors in (\ref{eq:k-factor}) drop out. By taking the integral over
$x_{A}$ and $x_{B}$ and then the integral over on-shell momenta
$p_{A}$, $p_{B}$, $p_{1}$ and $p_{2}$, we obtain the scattering
or collision rate per unit volume, 
\begin{eqnarray}
R_{AB\rightarrow12} & = & \int\frac{d^{3}p_{A}}{(2\pi)^{3}2E_{A}}\frac{d^{3}p_{B}}{(2\pi)^{3}2E_{B}}\frac{d^{3}p_{1}}{(2\pi)^{3}2E_{1}}\frac{d^{3}p_{2}}{(2\pi)^{3}2E_{2}}\nonumber \\
 &  & \times\frac{1}{C_{AB}}\int d^{4}x_{A}d^{4}x_{B}\delta(\Delta t)\delta(\Delta x_{L})\nonumber \\
 &  & \times f_{A}(x_{A},p_{A})f_{B}(x_{B},p_{B})G_{1}G_{2}|v_{A}-v_{B}|K.\label{eq:rate}
\end{eqnarray}
Now we use new variables to replace $x_{A}$ and $x_{B}$, 
\begin{eqnarray}
X & = & \frac{1}{2}(x_{A}+x_{B}),\nonumber \\
y & = & x_{A}-x_{B}.
\end{eqnarray}
We can rewrite the integral over $x_{A}$ and $x_{B}$ in Eq. (\ref{eq:rate})
as 
\begin{eqnarray}
I & = & \int d^{4}x_{A}d^{4}x_{B}\delta(\Delta t)\delta(\Delta x_{L})f_{A}(x_{A},p_{A})f_{B}(x_{B},p_{B})\nonumber \\
 &  & \times\exp\left(-i\mathbf{k}_{A}\cdot\mathbf{x}_{A}-i\mathbf{k}_{B}\cdot\mathbf{x}_{B}+i\mathbf{k}_{A}^{\prime}\cdot\mathbf{x}_{A}+i\mathbf{k}_{B}^{\prime}\cdot\mathbf{x}_{B}\right)\nonumber \\
 & \approx & \int d^{4}Xd^{2}\mathbf{b}f_{A}\left(X+\frac{y_{T}}{2},p_{A}\right)f_{B}\left(X-\frac{y_{T}}{2},p_{B}\right)\nonumber \\
 &  & \times\exp\left[i(\mathbf{k}_{A}^{\prime}-\mathbf{k}_{A})\cdot\mathbf{b}\right],\label{eq:xa-xb-int}
\end{eqnarray}
where we have used $\mathbf{k}_{A}+\mathbf{k}_{B}-\mathbf{k}_{A}^{\prime}-\mathbf{k}_{B}^{\prime}=0$
and $-\mathbf{k}_{A}+\mathbf{k}_{B}+\mathbf{k}_{A}^{\prime}-\mathbf{k}_{B}^{\prime}=2(\mathbf{k}_{A}^{\prime}-\mathbf{k}_{A})$
implied by two delta functions in Eq. (\ref{eq:k-factor}). In Eq.
(\ref{eq:xa-xb-int}) we have integrated over $y^{0}=\Delta t=t_{A}-t_{B}$
and $y_{L}=\Delta x_{L}=\hat{\mathbf{p}}_{A}\cdot(\mathbf{x}_{A}-\mathbf{x}_{B})$
to remove two detla functions, then we are left with the integral
over the transverse part $y_{T}^{\mu}=(0,\mathbf{b})$ with $\mathbf{b}$
being in the transverse direction. Because we work in the CMS in which
all kinematic variables depend on the incident momenta in the lab
frame, the impact parameter $\mathbf{b}$ in the CMS depends on $(x_{A},x_{B})$
as well as $(\mathbf{p}_{A},\mathbf{p}_{B})$ in the lab frame through
a boost velocity. 

Now we define the constant $C_{AB}$ in (\ref{eq:diff-rate},\ref{eq:rate})
as $C_{AB}\equiv\int d^{4}X=t_{X}\Omega_{\mathrm{int}}$ so that the
final results have the right dimension. Here $t_{X}$ and $\Omega_{\mathrm{int}}$
are the local time and space volume for the interaction respectively.
Note that $C_{AB}^{-1}\int d^{4}X\left(\cdots\right)$ plays the role
of the average over $X$ or $\left\langle \left(\cdots\right)\right\rangle _{X}$.
If we take the limit $t_{X}\Omega_{\mathrm{int}}\rightarrow0$, we
obtain the local rate per unit volume from Eq. (\ref{eq:rate}), 
\begin{eqnarray}
\frac{d^{4}N_{AB\rightarrow12}}{dX^{4}} & = & \frac{1}{(2\pi)^{4}}\int\frac{d^{3}p_{A}}{(2\pi)^{3}2E_{A}}\frac{d^{3}p_{B}}{(2\pi)^{3}2E_{B}}\frac{d^{3}p_{1}}{(2\pi)^{3}2E_{1}}\frac{d^{3}p_{2}}{(2\pi)^{3}2E_{2}}\nonumber \\
 &  & \times|v_{A}-v_{B}|G_{1}G_{2}\int d^{3}k_{A}d^{3}k_{B}d^{3}k_{A}^{\prime}d^{3}k_{B}^{\prime}\nonumber \\
 &  & \times\phi_{A}(\mathbf{k}_{A}-\mathbf{p}_{A})\phi_{B}(\mathbf{k}_{B}-\mathbf{p}_{B})\phi_{A}^{*}(\mathbf{k}_{A}^{\prime}-\mathbf{p}_{A})\phi_{B}^{*}(\mathbf{k}_{B}^{\prime}-\mathbf{p}_{B})\nonumber \\
 &  & \times\delta^{(4)}(k_{A}^{\prime}+k_{B}^{\prime}-p_{1}-p_{2})\delta^{(4)}(k_{A}+k_{B}-p_{1}-p_{2})\nonumber \\
 &  & \times\mathcal{M}\left(\{k_{A},k_{B}\}\rightarrow\{p_{1},p_{2}\}\right)\mathcal{M}^{*}\left(\{k_{A}^{\prime},k_{B}^{\prime}\}\rightarrow\{p_{1},p_{2}\}\right)\nonumber \\
 &  & \times\int d^{2}\mathbf{b}f_{A}\left(X+\frac{y_{T}}{2},p_{A}\right)f_{B}\left(X-\frac{y_{T}}{2},p_{B}\right)\exp\left[i(\mathbf{k}_{A}^{\prime}-\mathbf{k}_{A})\cdot\mathbf{b}\right],\label{eq:rate-ab12}
\end{eqnarray}
where $N_{AB\rightarrow12}$ is the number of scatterings. We emphasize
again that all quantities in Eq. (\ref{eq:rate-ab12}) are defined
in the CMS of two incident particles (we have suppressed the index
'c').

\section{Polarization rate for spin-1/2 particles from collisions}

In this section we will generalize the previous section for spin-0
particles to spin-1/2 ones. Our purpose is to derive the polarization
rate from collisions in a system of particles of multi-species. We
assume that particle distributions in phase space are independent
of spin states, so the spin dependence comes only from scatterings
of particles carrying the spin degree of freedom. 

As a simple example to illustrate the idea of the polarization arising
from collisions, we consider a fluid with the three-vector fluid velocity
in the z direction $v_{z}$ that depends on $x$, which we denote
as $v_{z}(x)$. We assume $dv_{z}(x)/dx>0$. In the comoving frame
of any fluid cell in the range $\left[x-\Delta x/2,x+\Delta x/2\right]$
where $\Delta x$ is a small distance, the fluid velocity at $x\pm\Delta x/2$
is $\pm(dv_{z}(x)/dx)\Delta x$, forming a rotation or local orbital
angular momentum (OAM) pointing to the $-y$ direction. Due to the
spin-orbit coupling, the scattering of two unpolarized particles with
velocity $\pm(dv_{z}(x)/dx)\Delta x$ and impact parameter $\Delta x$
will polarize the particles in the final state along the direction
of the local OAM. It has been proved that the polarization cross section
is proportional to $\mathbf{s}\cdot\mathbf{n}_{c}$, where $\mathbf{s}$
is the spin quantization (polarization) direction and $\mathbf{n}_{c}=\hat{\mathbf{b}}_{c}\times\hat{\mathbf{p}}_{c}$
is the direction of the reaction plane (the local OAM) in the CMS
of the scattering, where $\hat{\mathbf{b}}_{c}$ and $\hat{\mathbf{p}}_{c}$
are the direction of the impact parameter and the incident momentum
respectively. This is what happens in one scattering. In a thermal
system with collective motion, there are many scatterings whose reaction
planes point to almost random directions, but in average the direction
of the reaction plane points to that of the local rotation or vorticity.
To calculate the polarization in a thermal system with collective
motion, we have to take a convolution of distribution functions and
polarized scattering amplitudes similar to (\ref{eq:rate-ab12}). 

In this section we will distinguish quantities in the CMS and lab
frame, i.e. we will resume the subscript 'c' for all CMS quantities,
while quantities in the lab frame do not have the subscript 'c'. 

Now we consider a scattering process $A+B\rightarrow1+2$ where the
incident and outgoing particles are in the spin state labeled by $s_{A}$,
$s_{B}$, $s_{1}$ and $s_{2}$ ($s_{i}=\pm1/2$, $i=A,B,1,2$) respectively.
The quantization direction of the spin state is chosen to be along
the direction of the reaction plane in the CMS of the scattering.
The polarization rate per unit volume for particle 2 in the final
state is given by 
\begin{eqnarray}
\frac{d^{4}\mathbf{P}_{AB\rightarrow12}(X)}{dX^{4}} & = & \frac{1}{(2\pi)^{4}}\int\frac{d^{3}p_{c,A}}{(2\pi)^{3}2E_{c,A}}\frac{d^{3}p_{c,B}}{(2\pi)^{3}2E_{c,B}}\frac{d^{3}p_{c,1}}{(2\pi)^{3}2E_{c,1}}\frac{d^{3}p_{c,2}}{(2\pi)^{3}2E_{c,2}}\nonumber \\
 &  & \times|v_{c,A}-v_{c,B}|G_{1}G_{2}\int d^{3}k_{c,A}d^{3}k_{c,B}d^{3}k_{c,A}^{\prime}d^{3}k_{c,B}^{\prime}\nonumber \\
 &  & \times\phi_{A}(\mathbf{k}_{c,A}-\mathbf{p}_{c,A})\phi_{B}(\mathbf{k}_{c,B}-\mathbf{p}_{c,B})\phi_{A}^{*}(\mathbf{k}_{c,A}^{\prime}-\mathbf{p}_{c,A})\phi_{B}^{*}(\mathbf{k}_{c,B}^{\prime}-\mathbf{p}_{c,B})\nonumber \\
 &  & \times\delta^{(4)}(k_{c,A}^{\prime}+k_{c,B}^{\prime}-p_{c,1}-p_{c,2})\delta^{(4)}(k_{c,A}+k_{c,B}-p_{c,1}-p_{c,2})\nonumber \\
 &  & \times\int d^{2}\mathbf{b}_{c}f_{A}\left(X_{c}+\frac{y_{c,T}}{2},p_{c,A}\right)f_{B}\left(X_{c}-\frac{y_{c,T}}{2},p_{c,B}\right)\exp\left[i(\mathbf{k}_{c,A}^{\prime}-\mathbf{k}_{c,A})\cdot\mathbf{b}_{c}\right]\nonumber \\
 &  & \times\sum_{s_{A},s_{B},s_{1},s_{2}}2s_{2}\mathbf{n}_{c}\mathcal{M}\left(\{s_{A},k_{c,A};s_{B},k_{c,B}\}\rightarrow\{s_{1},p_{c,1};s_{2},p_{c,2}\}\right)\nonumber \\
 &  & \times\mathcal{M}^{*}\left(\{s_{A},k_{c,A}^{\prime};s_{B},k_{c,B}^{\prime}\}\rightarrow\{s_{1},p_{c,1};s_{2},p_{c,2}\}\right),\label{eq:rate-pol}
\end{eqnarray}
where $\mathbf{P}_{AB\rightarrow12}$ denotes the polarization vector
and $\mathbf{n}_{c}=\hat{\mathbf{b}}_{c}\times\hat{\mathbf{p}}_{c,A}$
is the direction of the reaction plane in the CMS of the scattering
which is also the quantization direction of the spin. In the second
to the last line of Eq. (\ref{eq:rate-pol}), the summation of $2s_{2}\mathcal{M}(\cdots,s_{2})\mathcal{M}^{*}(\cdots,s_{2})$
over $s_{2}=\pm1/2$ gives the polarized amplitude squared for particle
2 in the final state, and the factor 2 arises from the normalization
convention for the polarization that makes it in the range $[-1,1]$
instead of $[-1/2,1/2]$. Equation (\ref{eq:rate-pol}) is one of
our main results.

\section{Quark/antiquark polarization rate in a quark-gluon plasma of local
equilibrium in momentum}

\label{sec:quark-rate}In this section we will calculate the quark/antiquark
polarization rate from all 2-to-2 parton (quark or gluon) collisions
in a quark-gluon plasma (QGP) of local equilibrium in momentum but
not in spin. We assume that the QGP is a multi-component fluid with
the same fluid velocity $u(x)$ as a function of space-time for all
partons. The partons in a fluid cell follow a thermal distribution
in momentum in its comoving frame with the local temperature $T(x)$.
We assume that the phase space distribution $f(x,p)$ depends on $x^{\mu}=(t,\mathbf{x})$
through the fluid velocity $u^{\mu}(x)$ in the form $f(x,p)=f[\beta(x)p\cdot u(x)]$
where $p^{\mu}=(E_{p},\mathbf{p})$ is an on-shell four-momentum of
the parton and $\beta(x)\equiv1/T(x)$. 

We consider the scattering, $A+B\rightarrow1+2$, where $A$ and $B$
denote two incident partons in the wave packet form localized at $x_{A}$
and $x_{B}$ respectively, and '1' and '2' denote two outgoing partons
in momentum states. In order to calculate the polarization rate from
the collision of two wave packets displaced by an impact parameter
by Eq. (\ref{eq:rate-pol}), we must work in the CMS of the incident
partons for each collision. Note that many collisions take place in
the system at different space-time, the CMS of each collision depends
on the momenta of incident partons which vary from collision to collision.
In one collision, the phase space distributions for incident partons
(denoted as $i=A,B$) can be written in the form 
\begin{eqnarray}
f_{i}(x_{c},p_{c}) & = & f_{i}[\beta(x_{c})p_{c}\cdot u_{c}(x_{c})]\nonumber \\
 & = & f_{i}[\beta(x)p\cdot u(x)]\nonumber \\
 & = & f_{i}(x,p),
\end{eqnarray}
where $x,p$ are the space-time and momentum in the lab frame respectively,
while $x_{c},p_{c}$ are their corresponding values in the CMS of
$A$ and $B$ in this collision which depend on $p_{A}$ and $p_{B}$
in the heat bath (lab frame) through the boost velocity, and $u_{c}^{\mu}(x_{c})$
denotes the fluid velocity in the CMS as a function of the space-time
in the CMS.

\subsection{Polarization rate}

We now apply Eq. (\ref{eq:rate-pol}) to 2-to-2 parton scatterings.
For simplicity we assume that the phase space distributions of incident
partons follow the Boltzmann distribution, i.e. $f(x,p)=\exp[-\beta(x)p\cdot u(x)]$,
so we have $G_{1}G_{2}=1$ in (\ref{eq:rate-pol}). Also we assume
that $y_{c,T}$ is small compared with $X_{c}$ so that we can make
an expansion in $y_{c,T}$ for the distributions, the details are
given in Appendix \ref{sec:Derivation-of-Expansion}. The relevant
contribution in the linear or first order in $y_{c,T}$ involves the
term $y_{c,T}^{\mu}[\partial(\beta u_{c,\rho})/\partial X_{c}^{\mu}]p_{c,A}^{\rho}$
which can be rewritten as 
\begin{eqnarray}
y_{c,T}^{\mu}p_{c,A}^{\rho}\frac{\partial(\beta u_{\rho})}{\partial X_{c}^{\mu}} & = & -\frac{1}{2}L_{(c)}^{\mu\rho}\omega_{\mu\rho}^{(c)}+\frac{1}{4}y_{c,T}^{\{\mu}p_{c,A}^{\rho\}}\left[\frac{\partial(\beta u_{c,\rho})}{\partial X_{c}^{\mu}}+\frac{\partial(\beta u_{c,\mu})}{\partial X_{c}^{\rho}}\right],\label{eq:oam-vorticity-s}
\end{eqnarray}
where $L_{(c)}^{\mu\rho}\equiv y_{c,T}^{[\mu}p_{c,A}^{\rho]}$ is
the OAM tensor, $\omega_{\mu\rho}^{(c)}\equiv-(1/2)[\partial_{\mu}^{X_{c}}(\beta u_{c,\rho})-\partial_{\rho}^{X_{c}}(\beta u_{c,\mu})]$
is the thermal vorticity tensor, and $y_{c,T}^{\{\mu}p_{c,A}^{\rho\}}\equiv y_{c,T}^{\mu}p_{c,A}^{\rho}+y_{c,T}^{\rho}p_{c,A}^{\mu}$,
all in the CMS. The derivation of Eq. (\ref{eq:oam-vorticity-s})
is given in Eq. (\ref{eq:oam-vorticity}). Note that the OAM-vorticity
coupling $L_{(c)}^{\mu\rho}\omega_{\mu\rho}^{(c)}$ shows up in the
$y_{c,T}$ expansion, which can be converted to the spin-vorticity
coupling through polarized parton scattering amplitudes encoding the
spin-orbit coupling effect, as we will show shortly. The second term
in Eq. (\ref{eq:oam-vorticity-s}) invloves the symmetric part of
the thermal velocity derivatives in space-time, which is assumed to
vanish in thermal equilibrium for the spin, known as the Killing condition
\cite{Becattini:2013fla,Becattini:2015nva,Becattini:2016gvu,Florkowski:2018ahw}.
In this paper, however, we do not assume the thermal equilibrium for
the spin degree of freedom, so we keep this symmetric term in the
calculation. 

Keeping the first order term in the $y_{c,T}$ expansion and neglecting
the zeroth order term which is irrelevant, Eq. (\ref{eq:rate-pol})
can be simplified as 
\begin{eqnarray}
\frac{d^{4}\mathbf{P}_{AB\rightarrow12}(X)}{dX^{4}} & = & -\frac{1}{(2\pi)^{4}}\int\frac{d^{3}p_{A}}{(2\pi)^{3}2E_{A}}\frac{d^{3}p_{B}}{(2\pi)^{3}2E_{B}}\frac{d^{3}p_{c,1}}{(2\pi)^{3}2E_{c,1}}\frac{d^{3}p_{c,2}}{(2\pi)^{3}2E_{c,2}}\nonumber \\
 &  & \times|v_{c,A}-v_{c,B}|\int d^{3}k_{c,A}d^{3}k_{c,B}d^{3}k_{c,A}^{\prime}d^{3}k_{c,B}^{\prime}\nonumber \\
 &  & \times\phi_{A}(\mathbf{k}_{c,A}-\mathbf{p}_{c,A})\phi_{B}(\mathbf{k}_{c,B}-\mathbf{p}_{c,B})\phi_{A}^{*}(\mathbf{k}_{c,A}^{\prime}-\mathbf{p}_{c,A})\phi_{B}^{*}(\mathbf{k}_{c,B}^{\prime}-\mathbf{p}_{c,B})\nonumber \\
 &  & \times\delta^{(4)}(k_{c,A}^{\prime}+k_{c,B}^{\prime}-p_{c,1}-p_{c,2})\delta^{(4)}(k_{c,A}+k_{c,B}-p_{c,1}-p_{c,2})\nonumber \\
 &  & \times\frac{1}{2}\int d^{2}\mathbf{b}_{c}\exp\left[i(\mathbf{k}_{c,A}^{\prime}-\mathbf{k}_{c,A})\cdot\mathbf{b}_{c}\right]\mathbf{b}_{c,j}[\Lambda^{-1}]_{\; j}^{\nu}\frac{\partial(\beta u_{\rho})}{\partial X^{\nu}}\nonumber \\
 &  & \times\left[p_{A}^{\rho}-p_{B}^{\rho}\right]f_{A}\left(X,p_{A}\right)f_{B}\left(X,p_{B}\right)\Delta I_{M}^{AB\rightarrow12}\mathbf{n}_{c},\label{eq:polarization-1}
\end{eqnarray}
where we have used $d^{3}p_{c,i}/E_{c,i}=d^{3}p_{i}/E_{i}$ for $i=A,B$,
the Lorentz transformation matrix is defined by $\partial X^{\nu}/\partial X_{c}^{\mu}=[\Lambda^{-1}]_{\;\mu}^{\nu}=\Lambda_{\mu}^{\;\nu}$,
the minus sign in the right-hand side comes from $df_{i}\left(X,p_{i}\right)/d(\beta u\cdot p_{i})$
for $i=A,B$, and $\Delta I_{M}^{AB\rightarrow12}$ is defined by
\begin{eqnarray}
\Delta I_{M}^{AB\rightarrow12} & = & \sum_{s_{A},s_{B},s_{1},s_{2}}\sum_{color}2s_{2}\mathcal{M}\left(\{s_{A},k_{c,A};s_{B},k_{c,B}\}\rightarrow\{s_{1},p_{c,1};s_{2},p_{c,2}\}\right)\nonumber \\
 &  & \times\mathcal{M}^{*}\left(\{s_{A},k_{c,A}^{\prime};s_{B},k_{c,B}^{\prime}\}\rightarrow\{s_{1},p_{c,1};s_{2},p_{c,2}\}\right),\label{eq:d-im-ab12}
\end{eqnarray}
where the factor 2 arises from the normalization convention for the
polarization. Note that in the above formula there is a sum over color
degrees of freedom of all incident and outgoing partons. We may write
$\Delta I_{M}^{AB\rightarrow12}\mathbf{n}_{c}$ as 
\begin{eqnarray}
\Delta I_{M}^{AB\rightarrow12}\mathbf{n}_{c} & = & \Delta I_{M}^{AB\rightarrow12}(\hat{\mathbf{b}}_{c}\times\hat{\mathbf{p}}_{c,A})\nonumber \\
 & = & i(\hat{\mathbf{b}}_{c}\cdot\mathbf{I}_{c})\mathbf{e}_{c,i}\epsilon_{ikh}\hat{\mathbf{b}}_{c,k}\hat{\mathbf{p}}_{c,A}^{h}\nonumber \\
 & = & i\mathbf{e}_{c,i}\epsilon_{ikh}\hat{\mathbf{p}}_{c,A}^{h}\mathbf{I}_{c,l}\hat{\mathbf{b}}_{c,l}\hat{\mathbf{b}}_{c,k},\label{eq:deltaM-nc}
\end{eqnarray}
where $\mathbf{e}_{c,i}$ ($i=x,y,z$) are the basis vectors in the
CMS, and $\Delta I_{M}^{AB\rightarrow12}$ can be put into the form
$i\hat{\mathbf{b}}_{c}\cdot\mathbf{I}_{c}$, in this way we can single
out the direction $\hat{\mathbf{b}}_{c}$ out of $\Delta I_{M}^{AB\rightarrow12}$,
see Eq. (\ref{eq:final-result}) for an example of what $\mathbf{I}_{c}$
looks like. 

Substituting Eq. (\ref{eq:deltaM-nc}) into Eq. (\ref{eq:polarization-1}),
completing the integration over $\mathbf{b}_{c}$, and removing delta
functions by integration, we obtain 

\begin{eqnarray}
\frac{d^{4}\mathbf{P}_{AB\rightarrow12}(X)}{dX^{4}} & = & \frac{\pi}{(2\pi)^{4}}\frac{\partial(\beta u_{\rho})}{\partial X^{\nu}}\int\frac{d^{3}p_{A}}{(2\pi)^{3}2E_{A}}\frac{d^{3}p_{B}}{(2\pi)^{3}2E_{B}}\nonumber \\
 &  & \times|v_{c,A}-v_{c,B}|[\Lambda^{-1}]_{\; j}^{\nu}\mathbf{e}_{c,i}\epsilon_{ikh}\hat{\mathbf{p}}_{c,A}^{h}\nonumber \\
 &  & \times f_{A}\left(X,p_{A}\right)f_{B}\left(X,p_{B}\right)\left(p_{A}^{\rho}-p_{B}^{\rho}\right)\nonumber \\
 &  & \times\int\frac{d^{3}p_{c,1}}{(2\pi)^{3}2E_{c,1}}\frac{d^{3}p_{c,2}}{(2\pi)^{3}2E_{c,2}}d^{2}\mathbf{k}_{c,A}^{T}d^{2}\mathbf{k}_{c,A}^{\prime T}\nonumber \\
 &  & \times\sum_{j_{1},j_{2}=1,2}\frac{1}{|\mathrm{Ja}(k_{c,A}^{L}(j_{1}))|}\cdot\frac{1}{|\mathrm{Ja}(k_{c,A}^{\prime L}(j_{2}))|}\nonumber \\
 &  & \times\phi_{A}(\mathbf{k}_{c,A}-\mathbf{p}_{c,A})\phi_{B}(\mathbf{k}_{c,B}-\mathbf{p}_{c,B})\phi_{A}^{*}(\mathbf{k}_{c,A}^{\prime}-\mathbf{p}_{c,A})\phi_{B}^{*}(\mathbf{k}_{c,B}^{\prime}-\mathbf{p}_{c,B})\nonumber \\
 &  & \times\mathbf{I}_{c,l}\frac{1}{a^{3}}\left[Q_{jkl}^{L}\left(-2+2J_{0}(w_{0})+w_{0}J_{1}(w_{0})+w_{0}^{2}J_{2}(w_{0})\right)\right.\nonumber \\
 &  & \left.+Q_{jkl}^{T}\left(2-2J_{0}(w_{0})-w_{0}J_{1}(w_{0})\right)\right].\label{eq:polarization-2}
\end{eqnarray}
Here we have used 
\begin{eqnarray}
Q_{jkl}^{L} & = & \frac{\mathbf{a}_{l}\mathbf{a}_{j}\mathbf{a}_{k}}{a^{3}},\nonumber \\
Q_{jkl}^{T} & = & \frac{1}{a^{3}}\left(a^{2}\mathbf{a}_{k}\delta_{lj}+a^{2}\mathbf{a}_{l}\delta_{jk}+a^{2}\mathbf{a}_{j}\delta_{lk}-3\mathbf{a}_{l}\mathbf{a}_{j}\mathbf{a}_{k}\right),\label{eq:q-tensor}
\end{eqnarray}
with $\mathbf{a}\equiv\mathbf{k}_{c,A}^{\prime}-\mathbf{k}_{c,A}$
and $a=|\mathbf{a}|$, $w_{0}=ab_{0}$ with $b_{0}$ being the upper
limit or cutoff of $b_{c}$, $J_{i}$ for $i=0,1,2$ are Bessel functions,
$\mathbf{k}_{c,B}=\mathbf{p}_{c,1}+\mathbf{p}_{c,2}-\mathbf{k}_{c,A}$,
$\mathbf{k}_{c,B}^{\prime}=\mathbf{p}_{c,1}+\mathbf{p}_{c,2}-\mathbf{k}_{c,A}^{\prime}$,
$\mathrm{Ja}(k_{c,A}^{L})$ and $\mathrm{Ja}(k_{c,A}^{\prime L})$
are Jacobians for the longitudinal momenta $k_{c,A}^{L}$ and $k_{c,A}^{\prime L}$
and are given by 
\begin{eqnarray}
\mathrm{Ja}(k_{c,A}^{L}) & = & k_{c,A}^{L}\left(\frac{1}{E_{c,A}}+\frac{1}{E_{c,B}}\right)-\frac{1}{E_{c,B}}(p_{c,1}^{L}+p_{c,2}^{L}),\nonumber \\
\mathrm{Ja}(k_{c,A}^{\prime L}) & = & k_{c,A}^{\prime L}\left(\frac{1}{E_{c,A}^{\prime}}+\frac{1}{E_{c,B}^{\prime}}\right)-\frac{1}{E_{c,B}^{\prime}}(p_{c,1}^{L}+p_{c,2}^{L}),
\end{eqnarray}
and $k_{c,A}^{L}(j_{1})$ and $k_{c,A}^{\prime L}(j_{2})$ with $j_{1},j_{2}=1,2$
are two roots of the energy conservation equation $E_{c,A}+E_{c,B}-E_{c,1}-E_{c,2}=0$
and $E_{c,A}^{\prime}+E_{c,B}^{\prime}-E_{c,1}-E_{c,2}=0$ respectively.
In (\ref{eq:polarization-2}) and (\ref{eq:q-tensor}) Latin indices
label spatial components in the the CMS. The derivation of (\ref{eq:polarization-2})
is given in Appendix \ref{sec:integration}. 

In a system of gluons and quarks with multi-flavors, there are many
2-to-2 parton scatterings with at least one quark in the final state.
The quark polarization rate for a specific flavor reads 
\begin{equation}
\frac{d^{4}\mathbf{P}_{q}(X)}{dX^{4}}=\sum_{A,B,1=\{q_{a},\bar{q}_{a},g\}}\frac{d^{4}\mathbf{P}_{AB\rightarrow1q}(X)}{dX^{4}},\label{eq:quark-polar}
\end{equation}
where $d^{4}\mathbf{P}_{AB\rightarrow1q}(X)/dX^{4}$ is given by Eq.
(\ref{eq:polarization-2}), and 2-to-2 parton scatterings are listed
in Table \ref{tab:feyn-22-quark}. The antiquark polarization rate
can be similarly obtained.

\subsection{Polarized amplitudes for quarks/antiquarks in 2-to-2 parton scatterings}

In this subsection we will derive the polarized amplitudes for quarks
in 2-to-2 parton scatterings. The Feynman diagrams of all 2-to-2 parton
scatterings at the tree level with at least one quark in the final
state are shown in Table \ref{tab:feyn-22-quark}. For anti-quark
polarization, we can make particle-antiparticle transformation in
all processes listed in Table \ref{tab:feyn-22-quark}, for example,
$q_{a}q_{b}\rightarrow q_{a}q_{b}$ becomes $\bar{q}_{a}\bar{q}_{b}\rightarrow\bar{q}_{a}\bar{q}_{b}$,
$\bar{q}_{a}q_{b}\rightarrow\bar{q}_{a}q_{b}$ becomes $q_{a}\bar{q}_{b}\rightarrow q_{a}\bar{q}_{b}$,
$gg\rightarrow\bar{q}_{a}q_{a}$ becomes $gg\rightarrow q_{a}\bar{q}_{a}$,
etc.. In this subsection, we discuss polarized amplitudes for quarks,
those for antiquarks can be easily obtained. 

In order to obtain the quark polarization, we sum over the spin states
of all partons in the scattering except one quark in the final state.
For simplicity of the calculation, we assume that the quark masses
are equal for all flavors and the external gluon is massless. We introduce
a small mass in the gluon propagator in the t-channel to regulate
the possible divergence. 

In this subsection, all variables are defined in the CMS, for notational
simplicity we will suppress the subscript 'c', for example, $p_{A}$
actually means $p_{cA}$. 

\begin{table}[H]
\caption{The Feynman diagrams of all 2-to-2 parton scatterings at the tree
level with at least one quark in the final state. We calculate the
polarization of the quark (the second parton) in the final state.
Here $a$ and $b$ denote the quark flavor, $s_{i}=\pm1/2$ ($i=A,B,1,2$)
denote the spin states, $k_{i}$ ($i=A,B,1,2$) denote the momenta,
$q,q_{1},q_{2},q_{3}$ denote the momenta in propagators. The processes
for antiquark polarization can be obtained by making a particle-antiparticle
transformation. \label{tab:feyn-22-quark}}

\centering{}%
\begin{tabular}{|c|c|}
\hline 
$q_{a}q_{b}\rightarrow q_{a}q_{b}$ & $\bar{q}_{a}q_{b}\rightarrow\bar{q}_{a}q_{b}$\tabularnewline
\includegraphics[scale=0.25]{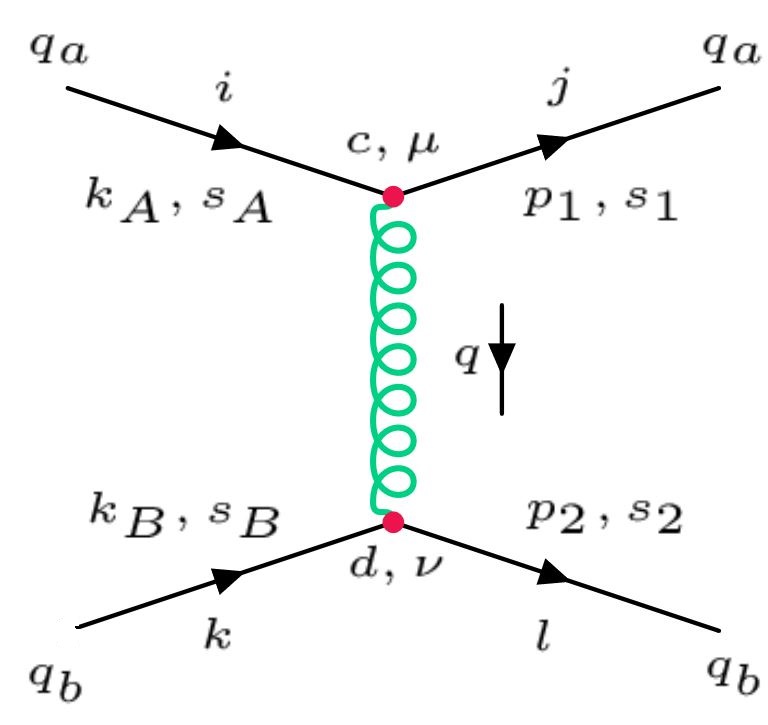} & \includegraphics[scale=0.25]{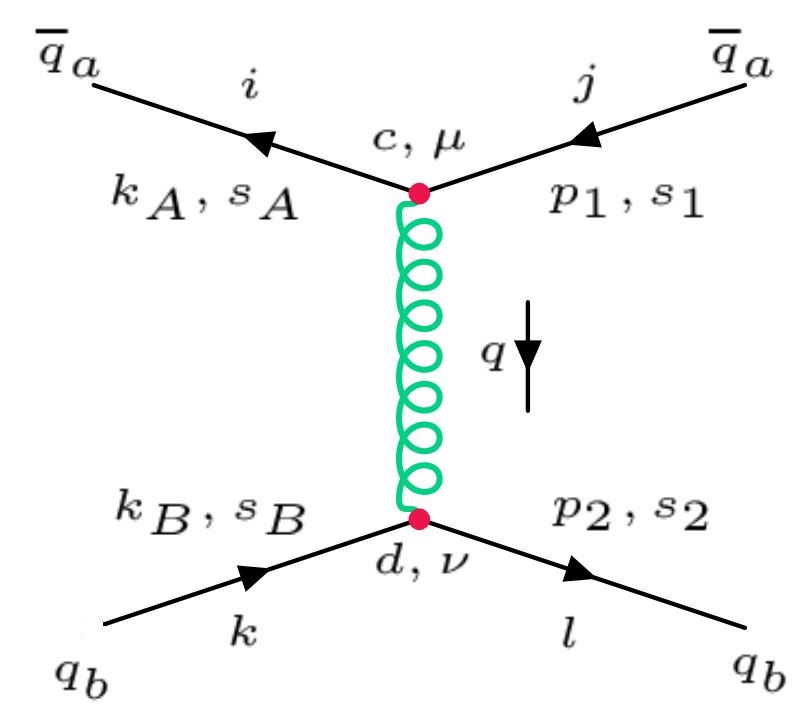}\tabularnewline
\hline 
$\bar{q}_{a}q_{a}\rightarrow\bar{q}_{a}q_{a}$  & $q_{a}q_{a}\rightarrow q_{a}q_{a}$\tabularnewline
\includegraphics[scale=0.2]{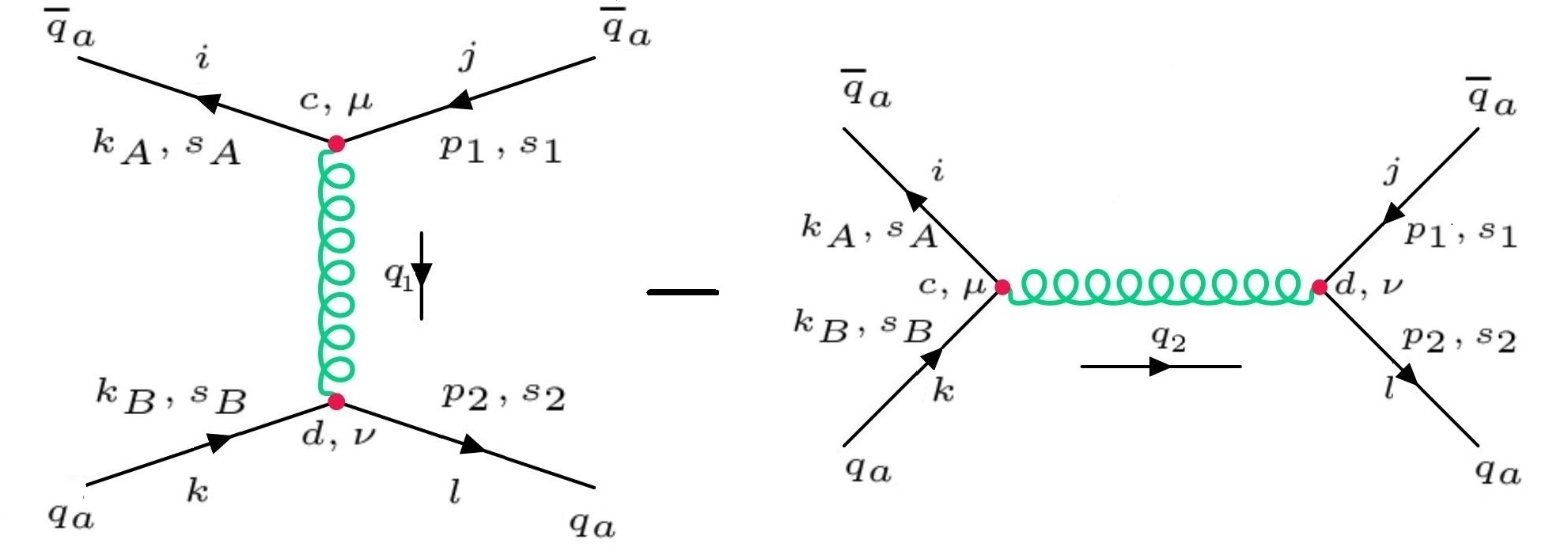} & \includegraphics[scale=0.2]{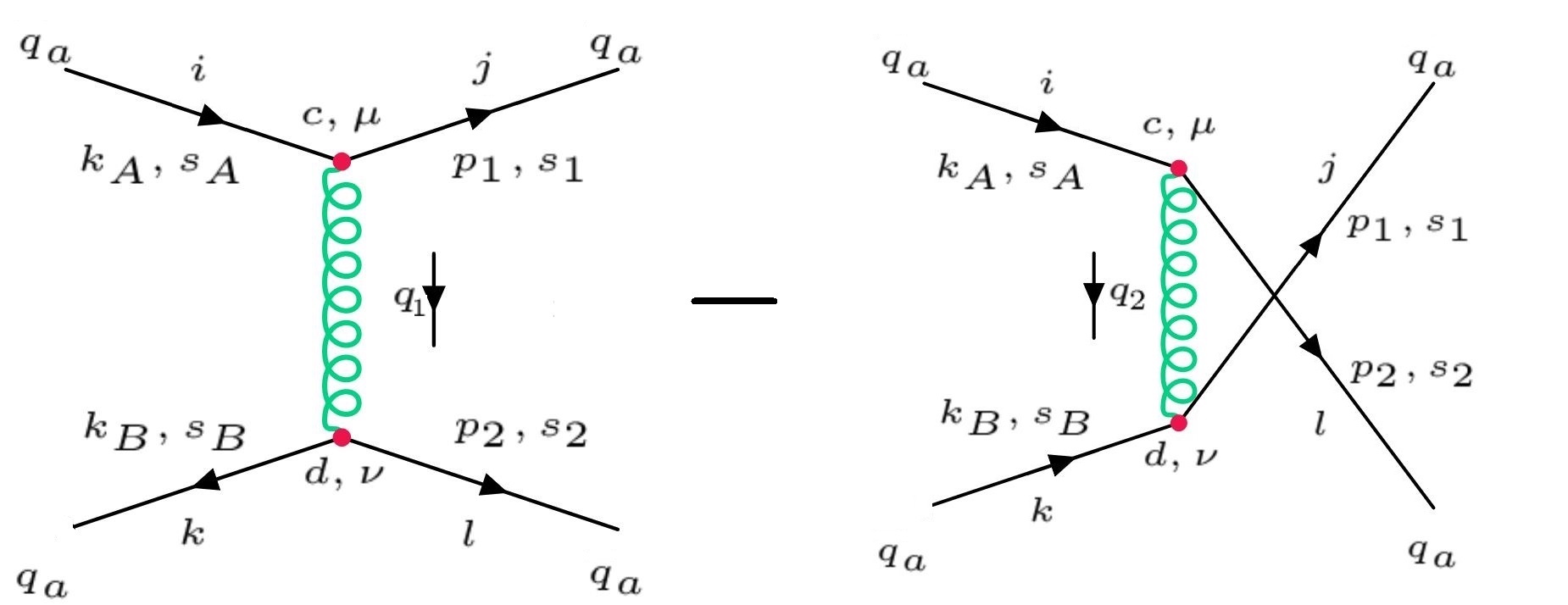}\tabularnewline
\hline 
$gg\rightarrow\bar{q}_{a}q_{a}$ & $gq_{a}\rightarrow gq_{a}$\tabularnewline
\includegraphics[scale=0.15]{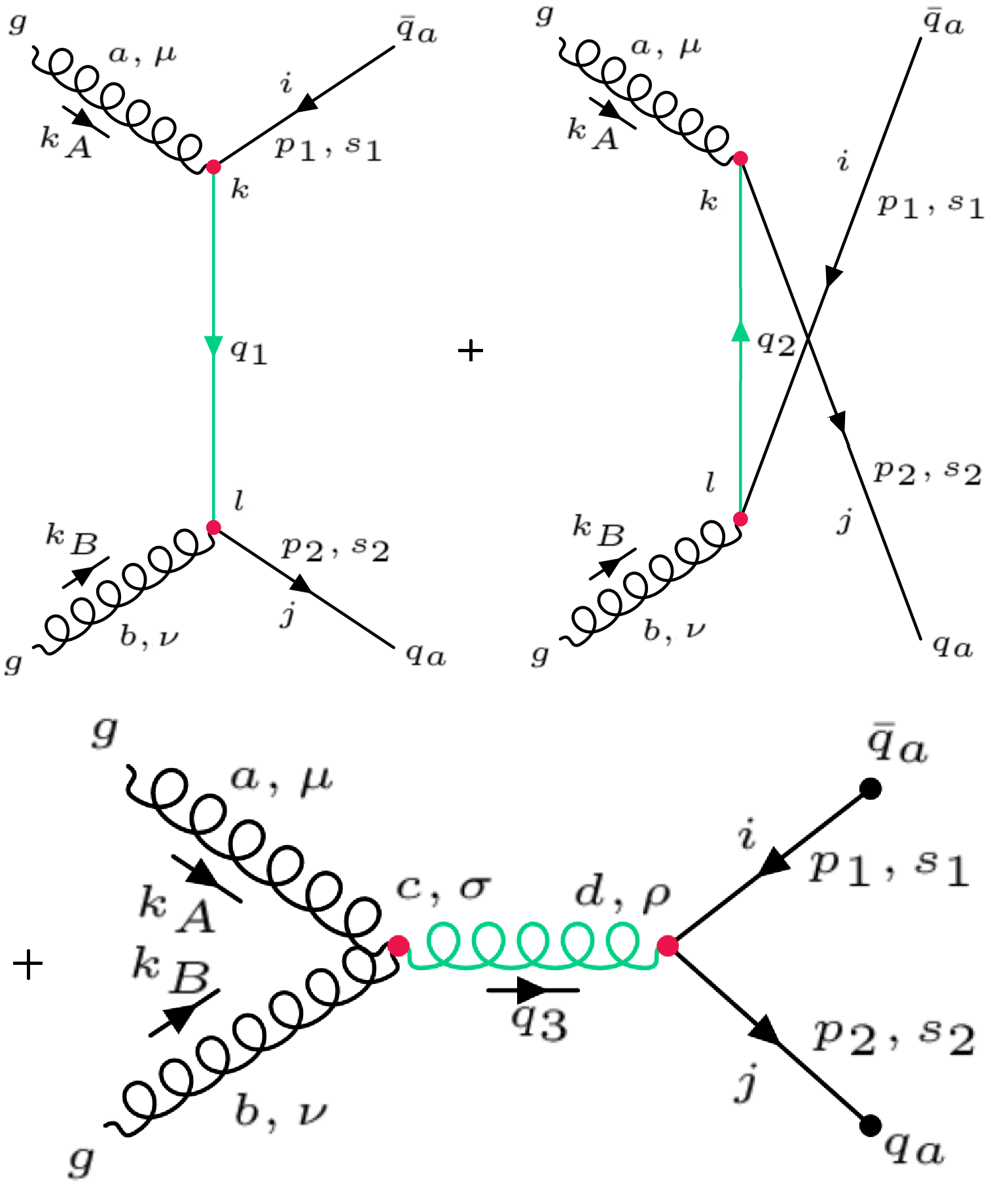} & \includegraphics[scale=0.15]{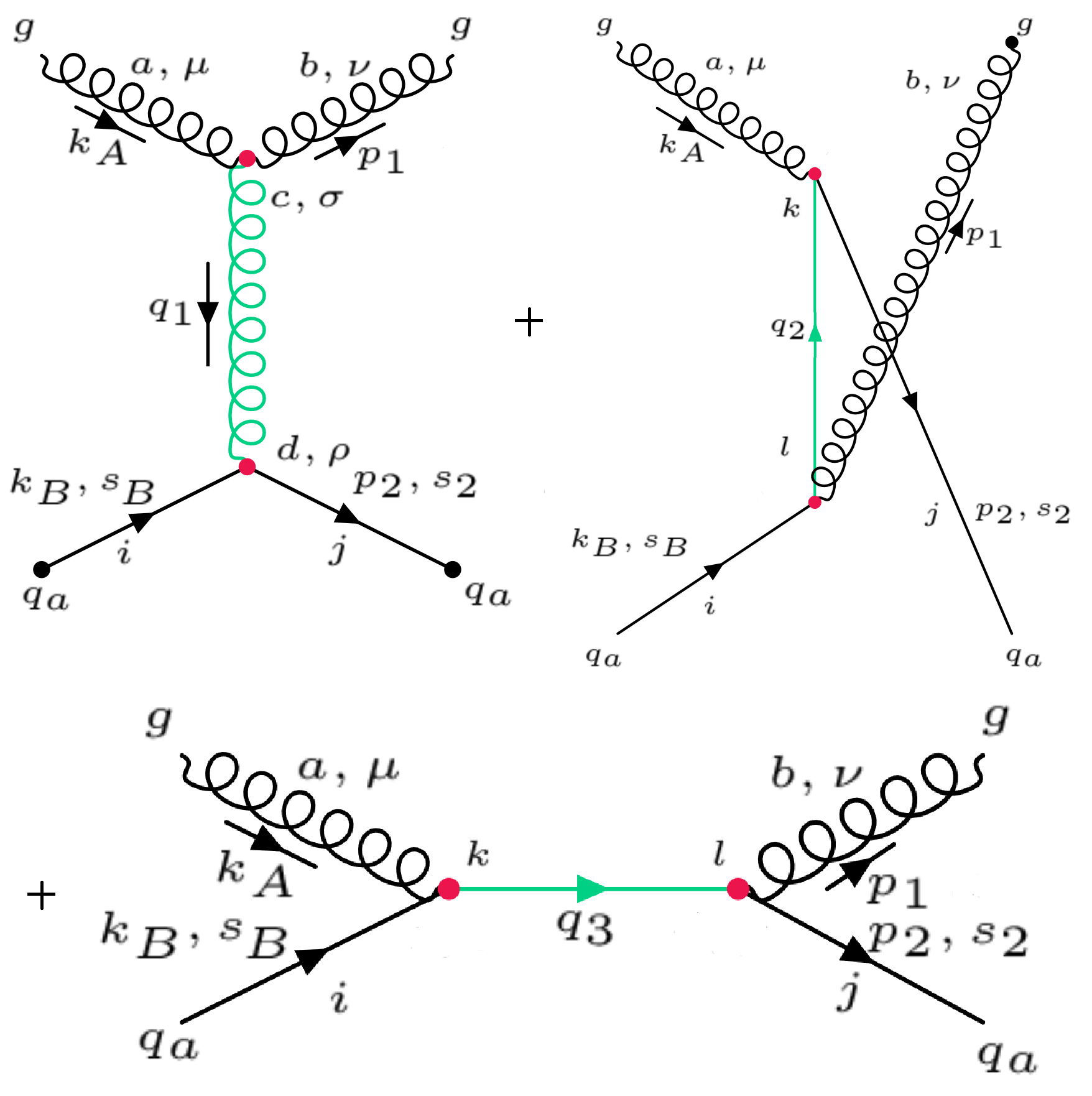}\tabularnewline
\hline 
$\bar{q}_{a}q_{a}\rightarrow\bar{q}_{b}q_{b}$ & \tabularnewline
\includegraphics[scale=0.25]{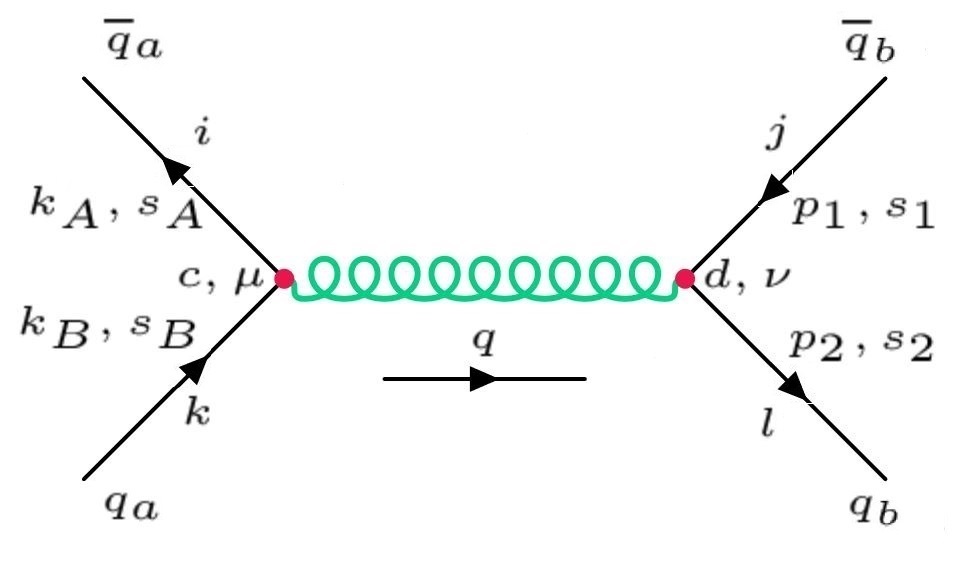} & \tabularnewline
\hline 
\end{tabular}
\end{table}

We take the quark-quark scattering $q_{a}q_{b}\rightarrow q_{a}q_{b}$
with $a\neq b$ (different flavor) as an example to demonstrate how
to derive the polarized scattering amplitude which depends on the
spin state of the quark in the final state. The Feynman diagram of
this process is shown in Table \ref{tab:feyn-22-quark}. The spin-momentum
configurations are shown in the diagram. We can then write down the
corresponding amplitudes following the Feynman rule 
\begin{eqnarray}
I_{1} & = & -i\mathcal{M}\left(\{s_{A},k_{A};s_{B},k_{B}\}\rightarrow\{s_{1},p_{1};s_{2},p_{2}\}\right)\nonumber \\
 & = & ig_{s}^{2}t_{ji}^{c}t_{lk}^{c}\frac{1}{q^{2}}[\bar{u}(s_{1},p_{1})\gamma^{\mu}u(s_{A},k_{A})][\bar{u}(s_{2},p_{2})\gamma_{\mu}u(s_{B},k_{B})],\nonumber \\
I_{2} & = & -i\mathcal{M}\left(\{s_{A},k_{A}^{\prime};s_{B},k_{B}^{\prime}\}\rightarrow\{s_{1},p_{1};s_{2},p_{2}\}\right)\nonumber \\
 & = & ig_{s}^{2}t_{ji}^{d}t_{lk}^{d}\frac{1}{q^{\prime2}}[\bar{u}(s_{1},p_{1})\gamma^{\nu}u(s_{A},k_{A}^{\prime})][\bar{u}(s_{2},p_{2})\gamma_{\nu}u(s_{B},k_{B}^{\prime})].\label{eq:qaqb-qaqb-mm}
\end{eqnarray}
where $g_{s}$ is the strong coupling constant, $i,j,k,l=1,2,3$ denote
the fundamental colors of quarks, $c,d=1,\cdots,8$ denote the adjoint
colors of gluons, $t^{c}$ and $t^{d}$ are generators of $SU(N_{c})$
in fundamental representation satisfying $[t^{a},t^{b}]=if^{abc}t^{c}$,
$q=k_{A}-p_{1}$, and $q^{\prime}=k_{A}^{\prime}-p_{1}$. We obtain
the product $I_{1}I_{2}^{*}$ as 

\begin{eqnarray}
I_{M}^{q_{a}q_{b}\rightarrow q_{a}q_{b}}(s_{2}) & = & \sum_{s_{A},s_{B},s_{1}}\sum_{i,j,k,l}\mathcal{M}\left(\{s_{A},k_{A};s_{B},k_{B}\}\rightarrow\{s_{1},p_{1};s_{2},p_{2}\}\right)\nonumber \\
 &  & \times\mathcal{M}^{*}\left(\{s_{A},k_{A}^{\prime};s_{B},k_{B}^{\prime}\}\rightarrow\{s_{1},p_{1};s_{2},p_{2}\}\right)\nonumber \\
 & = & C_{q_{a}q_{b}\rightarrow q_{a}q_{b}}g_{s}^{4}m^{2}\frac{1}{q^{2}q^{\prime2}}\nonumber \\
 &  & \times\mathrm{Tr}\left[(p_{1}\cdot\gamma+m)\gamma^{\mu}\Lambda_{1/2}(-\mathbf{k}_{A})(\gamma_{0}+1)\Lambda_{1/2}^{-1}(-\mathbf{k}_{A}^{\prime})\gamma^{\nu}\right]\nonumber \\
 &  & \times\mathrm{Tr}\left[\Pi(s_{2},n)(p_{2}\cdot\gamma+m)\gamma_{\mu}\Lambda_{1/2}(-\mathbf{k}_{B})(\gamma_{0}+1)\Lambda_{1/2}^{-1}(-\mathbf{k}_{B}^{\prime})\gamma_{\nu}\right].\label{eq:qaqb-qaqb}
\end{eqnarray}
In Eq. (\ref{eq:qaqb-qaqb}) we have used the notation $p\cdot\gamma\equiv p_{\rho}\gamma^{\rho}$,
a sum over all spins except $s_{2}$ and over all colors of quarks
and gluons have been taken, and $C_{q_{a}q_{b}\rightarrow q_{a}q_{b}}$
is the color factor for this process given in Table \ref{tab:color-factor}.
In the last two lines of Eq. (\ref{eq:qaqb-qaqb}), $\Lambda_{1/2}$
and $\Lambda_{1/2}^{-1}$ are the Lorentz transformation matrices
for spinors defined in Eq. (\ref{eq:lorentz-spinor}), $\Pi(s_{2},n)=(1+s_{2}\gamma_{5}n^{\sigma}\gamma_{\sigma})/2$
is the spin projector where $n^{\sigma}=(0,\mathbf{n})$ is the spin
quantization four-vector in the CMS with $\mathbf{n}=\hat{\mathbf{b}}\times\hat{\mathbf{p}}_{A}$,
and we have applied Eq. (\ref{eq:lorentz-trans}) and Eq. (\ref{eq:proj-spinor}).
From Eq. (\ref{eq:qaqb-qaqb}), we obtain the difference of $I_{M}^{q_{a}q_{b}\rightarrow q_{a}q_{b}}$
between the spin state $s_{2}=1/2$ and $s_{2}=-1/2$ for $q_{b}$,
\begin{eqnarray}
\Delta I_{M}^{q_{a}q_{b}\rightarrow q_{a}q_{b}} & = & I_{M}^{q_{a}q_{b}\rightarrow q_{a}q_{b}}(s_{2}=1/2)-I_{M}^{q_{a}q_{b}\rightarrow q_{a}q_{b}}(s_{2}=-1/2)\nonumber \\
 & = & C_{q_{a}q_{b}\rightarrow q_{a}q_{b}}g_{s}^{4}m^{2}\frac{1}{q^{2}q^{\prime2}}\nonumber \\
 &  & \times\mathrm{Tr}\left[(p_{1}\cdot\gamma+m)\gamma^{\mu}\Lambda_{1/2}(-\mathbf{k}_{A})(\gamma_{0}+1)\Lambda_{1/2}^{-1}(-\mathbf{k}_{A}^{\prime})\gamma^{\nu}\right]\nonumber \\
 &  & \times\mathrm{Tr}\left[\gamma_{5}(n\cdot\gamma)(p_{2}\cdot\gamma+m)\gamma_{\mu}\Lambda_{1/2}(-\mathbf{k}_{B})(\gamma_{0}+1)\Lambda_{1/2}^{-1}(-\mathbf{k}_{B}^{\prime})\gamma_{\nu}\right].\label{eq:delta-M-qaqb-qaqb}
\end{eqnarray}
The expansion of $\Delta I_{M}^{q_{a}q_{b}\rightarrow q_{a}q_{b}}$
gives about 200 terms. In accordance with Eq. (\ref{eq:lorentz-spinor}),
$\Lambda_{1/2}(\mathbf{p})$ depends on the repidity $\eta_{p}$ and
the momentum direction $\hat{\mathbf{p}}$, where $\eta_{p}$ is related
to the energy-momentum by $E_{p}=m\cosh(\eta_{p})$ and $|\mathbf{p}|=m\sinh(\eta_{p})$.
So the contracted trace part of $\Delta I_{M}^{q_{a}q_{b}\rightarrow q_{a}q_{b}}$
can be expressed as a function of $(\hat{\mathbf{k}}_{A},\hat{\mathbf{k}}_{A}^{\prime},\hat{\mathbf{k}}_{B},\hat{\mathbf{k}}_{B}^{\prime})$
and $(\eta_{kA},\eta_{kA}^{\prime},\eta_{kB},\eta_{kB}^{\prime})$. 

\begin{table}[H]
\caption{Color factors for all 2-to-2 processes with at least one final quark.
The constants which appear in color factors are: $d_{F}=N_{c}$, $d_{A}=N_{c}^{2}-1$,
$C_{F}=(N_{c}^{2}-1)/(2N_{c})$, and $C_{A}=3$ with $N_{c}=3$. \label{tab:color-factor}}

\centering{}%
\begin{tabular}{|c|c|}
\hline 
color factors & Color factors in scattering processes\tabularnewline
\hline 
$d_{F}^{2}C_{F}^{2}/d_{A}$ & $C_{q_{a}q_{b}\rightarrow q_{a}q_{b}}$, $C_{\bar{q}_{a}q_{b}\rightarrow\bar{q}_{a}q_{b}}$,
$C_{\bar{q}_{a}q_{a}\rightarrow\bar{q}_{a}q_{a}}^{(1)}$, $C_{q_{a}q_{a}\rightarrow q_{a}q_{a}}^{(1)}$
, $C_{\bar{q}_{a}q_{a}\rightarrow\bar{q}_{b}q_{b}}$\tabularnewline
\hline 
$d_{F}C_{F}^{2}$ & $C_{gg\rightarrow\bar{q}_{a}q_{a}}^{(1)}$, $C_{gq_{a}\rightarrow gq_{a}}^{(3)}$\tabularnewline
\hline 
$(C_{F}-C_{A}/2)d_{F}C_{F}$ & $C_{\bar{q}_{a}q_{a}\rightarrow\bar{q}_{a}q_{a}}^{(2)}$, $C_{q_{a}q_{a}\rightarrow q_{a}q_{a}}^{(2)}$,
$C_{gg\rightarrow\bar{q}_{a}q_{a}}^{(2)}$, $C_{gq_{a}\rightarrow gq_{a}}^{(4)}$\tabularnewline
\hline 
$\frac{1}{4}d_{A}C_{A}$ & $C_{gq_{a}\rightarrow gq_{a}}^{(2)}$, $C_{gg\rightarrow\bar{q}_{a}q_{a}}^{(3)}$ \tabularnewline
\hline 
$d_{F}C_{F}C_{A}$ & $C_{gq_{a}\rightarrow gq_{a}}^{(1)}$, $C_{gg\rightarrow\bar{q}_{a}q_{a}}^{(4)}$ \tabularnewline
\hline 
\end{tabular}
\end{table}

The polarized amplitudes for quarks in all 2-to-2 parton scatterings
listed in Table \ref{tab:feyn-22-quark} are given in Appendix \ref{sec:all-22-amp},
which results in more than 5000 terms. Here we give an estimate of
how many terms there are in each process: $\Delta I_{M}^{gg\rightarrow\bar{q}_{a}q_{a}}$
gives 136 terms, $\Delta I_{M}^{gq_{a}\rightarrow gq_{a}}$ gives
2442 terms, $\Delta I_{M}^{\bar{q}_{a}q_{a}\rightarrow\bar{q}_{a}q_{a}}$
gives 874 terms, $\Delta I_{M}^{\bar{q}_{a}q_{a}\rightarrow\bar{q}_{b}q_{b}}$
gives 40 terms, $\Delta I_{M}^{\bar{q}_{a}q_{b}\rightarrow\bar{q}_{a}q_{b}}$
gives 210 terms, $\Delta I_{M}^{q_{a}q_{b}\rightarrow q_{a}q_{b}}$
gives 210 terms, $\Delta I_{M}^{q_{a}q_{a}\rightarrow q_{a}q_{a}}$
gives 1156 terms. It is hard to see the physics behind such huge number
of terms unless we make an appropriate approximation.

\subsection{Evaluation of polarized amplitudes for quarks/antiquarks}

\label{sub:Approximation-of-Amplitude}The evaluation of contracted
traces of quark polarized amplitudes are very complicated. This has
been done with the help of FeynCalc \cite{Mertig:1990an,Shtabovenko:2016sxi}.
There are about $10^{4}$ terms in the expansion of contracted traces
for 2-to-2 parton scatterings. 

In this subsection, all variables are defined in the CMS, for notational
simplicity we will suppress the subscript 'c' if not explicitly specified,
for example, $p_{A}$ actually means $p_{cA}$. 

In order to show the physics in the midst of the huge number of terms,
we have to make an appropriate approximation. As we know that the
incident particles are treated as wave packets in order to describe
scatterings displaced by impact parameters. A realistic approximation
is that the wave packets are assumed to be narrow, i.e. the width
is much smaller than the center momenta of the wave packet in Eq.
(\ref{eq:wave-packet-gs}). In the extreme case that the width of
the wave packet is zero, we recover the normal scattering of plane
waves. Since the positions of incident particles can be anywhere in
plane waves, in average the relative OAM of two incident particles
is zero, leading to the vanishing polarization of final state particles.
This fact can be verified by setting 
\begin{eqnarray}
\hat{\mathbf{k}}_{A} & = & \hat{\mathbf{k}}_{A}^{\prime}=\hat{\mathbf{p}}_{A},\nonumber \\
\hat{\mathbf{k}}_{B} & = & \hat{\mathbf{k}}_{B}^{\prime}=-\hat{\mathbf{p}}_{A},\nonumber \\
\mathbf{p}_{1} & = & -\mathbf{p}_{2},\nonumber \\
\eta_{A} & = & \eta_{B}=\eta_{A}^{\prime}=\eta_{B}^{\prime},\label{eq:zero order replacement}
\end{eqnarray}
in the trace part in Eq. (\ref{eq:delta-M-qaqb-qaqb}), then we have
$\Delta I_{M}^{q_{a}q_{b}\rightarrow q_{a}q_{b}}=0$. 

The above result is of the zeroth order, now we turn to the first
order in the deviation from momenta in (\ref{eq:zero order replacement}).
We expand $(\hat{\mathbf{k}}_{A},\hat{\mathbf{k}}_{A}^{\prime},\hat{\mathbf{k}}_{B},\hat{\mathbf{k}}_{B}^{\prime})$
about their central values $(\hat{\mathbf{p}}_{A},\hat{\mathbf{p}}_{A},-\hat{\mathbf{p}}_{A},-\hat{\mathbf{p}}_{A})$
and $(\eta_{kA},\eta_{kA}^{\prime},\eta_{kB},\eta_{kB}^{\prime})$
about their central values $(\eta_{pA},\eta_{pA},\eta_{pA},\eta_{pA})$
to the first order in the differences, 
\begin{eqnarray}
\hat{\mathbf{k}}_{A} & \to & \hat{\mathbf{p}}_{A}+\boldsymbol{\Delta}_{A},\ \hat{\mathbf{k}}_{B}\to-\hat{\mathbf{p}}_{A}+\boldsymbol{\Delta}_{B},\nonumber \\
\hat{\mathbf{k}}_{A}^{\prime} & \to & \hat{\mathbf{p}}_{A}+\boldsymbol{\Delta}_{A}^{\prime},\ \hat{\mathbf{k}}_{B}^{\prime}\to-\hat{\mathbf{p}}_{A}+\boldsymbol{\Delta}_{B}^{\prime},\nonumber \\
\eta_{kA} & = & \eta_{pA}+\Delta\eta_{kA},\nonumber \\
\eta_{kA}^{\prime} & = & \eta_{pA}+\Delta\eta_{kA}^{\prime},\nonumber \\
\eta_{kB} & = & \eta_{pA}+\Delta\eta_{kB},\nonumber \\
\eta_{kB}^{\prime} & = & \eta_{pA}+\Delta\eta_{kB}^{\prime},\label{eq:1st-order-exp}
\end{eqnarray}
where the first order quantities are denoted with $\Delta$ (for example,
$\boldsymbol{\Delta}_{A}$, $\Delta\eta_{kA}$). We also expand $(E_{1},\mathbf{p}_{1},E_{2},\mathbf{p}_{2})$
at $(E_{0},\mathbf{p}_{0},E_{0},-\mathbf{p}_{0})$, 
\begin{eqnarray}
E_{1} & \to & E_{0}+\Delta_{1},\ E_{2}\to E_{0}+\Delta_{2},\nonumber \\
\mathbf{p}_{1} & \to & \mathbf{p}_{0}+\boldsymbol{\Delta}_{1},\ \mathbf{p}_{2}\to-\mathbf{p}_{0}+\boldsymbol{\Delta}_{1}.\label{eq:e1-p1-e2-p2}
\end{eqnarray}
The delta functions in Eq. (\ref{eq:polarization-1}) lead to 
\begin{eqnarray}
\mathbf{k}_{A}+\mathbf{k}_{B} & = & \mathbf{k}_{A}^{\prime}+\mathbf{k}_{B}^{\prime}=\mathbf{p}_{1}+\mathbf{p}_{2}.
\end{eqnarray}
So $\boldsymbol{\Delta}_{1}$ in (\ref{eq:e1-p1-e2-p2}) can be determined
by 
\begin{equation}
\boldsymbol{\Delta}_{1}=\frac{1}{2}(\mathbf{k}_{A}+\mathbf{k}_{B}),
\end{equation}
and $\mathbf{p}_{0}$ determined by 
\begin{equation}
\mathbf{p}_{0}=\frac{1}{2}(\mathbf{p}_{1}-\mathbf{p}_{2}).
\end{equation}
Note that once $\mathbf{p}_{0}$ and $\boldsymbol{\Delta}_{1}$ are
given, $E_{0},\Delta_{1},\Delta_{2}$ satisfy 
\begin{eqnarray}
(E_{0}+\Delta_{1})^{2} & = & (\mathbf{p}_{0}+\boldsymbol{\Delta}_{1})^{2}+m_{1}^{2},\nonumber \\
(E_{0}+\Delta_{2})^{2} & = & (-\mathbf{p}_{0}+\boldsymbol{\Delta}_{1})^{2}+m_{2}^{2}.
\end{eqnarray}
So we have a freedom to choose the value of $E_{0}$. Then we use
(\ref{eq:1st-order-exp}) and (\ref{eq:e1-p1-e2-p2}) in the contracted
trace part in Eq. (\ref{eq:delta-M-qaqb-qaqb}) and expand it to the
first order in $\Delta$-quantities. Still, the final result has many
terms but all terms of $\Delta_{1},\Delta_{2}$ and $\boldsymbol{\Delta}_{1}$
cancel out. 

In order to further simplify the contracted trace part in Eq. (\ref{eq:delta-M-qaqb-qaqb}),
we use the property that the first order contributions do not have
terms of $\Delta_{1},\Delta_{2},\boldsymbol{\Delta}_{1}$ by setting
\begin{eqnarray}
\mathbf{p}_{1} & = & \mathbf{p}_{0},\nonumber \\
\mathbf{p}_{2} & = & -\mathbf{p}_{0},\label{eq:replacement-p1-p2}
\end{eqnarray}
which leads to $\mathbf{k}_{A}+\mathbf{k}_{B}=\mathbf{k}_{A}^{\prime}+\mathbf{k}_{B}^{\prime}=0$
and then 
\begin{eqnarray}
\hat{\mathbf{k}}_{A} & = & -\hat{\mathbf{k}}_{B},\ \hat{\mathbf{k}}_{A}^{\prime}=-\hat{\mathbf{k}}_{B}^{\prime},\nonumber \\
\eta_{kA} & = & \eta_{kB},\ \eta_{kA}^{\prime}=\eta_{kB}^{\prime}.\label{eq:new-replacement}
\end{eqnarray}
Using (\ref{eq:replacement-p1-p2}) and (\ref{eq:new-replacement})
in the contracted trace part in Eq. (\ref{eq:delta-M-qaqb-qaqb})
for $q_{a}q_{b}\rightarrow q_{a}q_{b}$, we obtain a shorter series
of 31 terms 
\begin{eqnarray}
\mathrm{Tr_{q_{a}q_{b}\rightarrow q_{a}q_{b}}^{2}} & = & 16i(\mathbf{n}\times\mathbf{p}_{1})\cdot\hat{\mathbf{k}}_{A}\nonumber \\
 &  & \times\left[5c_{A}s_{A}c_{A}^{\prime}s_{A}^{\prime}\mathbf{p}_{1}\cdot\hat{\mathbf{k}}_{A}^{\prime}+7E_{1}s_{A}^{2}c_{A}^{\prime}s_{A}^{\prime}\hat{\mathbf{k}}_{A}\cdot\hat{\mathbf{k}}_{A}^{\prime}+2mc_{A}s_{A}c_{A}^{\prime2}\right.\nonumber \\
 &  & \left.-2mc_{A}s_{A}s_{A}^{\prime2}+4E_{1}c_{A}s_{A}c_{A}^{\prime2}+E_{1}c_{A}s_{A}s_{A}^{\prime2}-s_{A}^{2}s_{A}^{\prime2}\mathbf{p}_{1}\cdot\hat{\mathbf{k}}_{A}\right]\nonumber \\
 &  & +16i(\mathbf{n}\times\mathbf{p}_{1})\cdot\hat{\mathbf{k}}_{A}^{\prime}\left[4mc_{A}s_{A}s_{A}^{\prime2}\hat{\mathbf{k}}_{A}\cdot\hat{\mathbf{k}}_{A}^{\prime}-5E_{1}c_{A}s_{A}s_{A}^{\prime2}\hat{\mathbf{k}}_{A}\cdot\hat{\mathbf{k}}_{A}^{\prime}\right.\nonumber \\
 &  & -2mc_{A}^{2}c_{A}^{\prime}s_{A}^{\prime}-4E_{1}c_{A}^{2}c_{A}^{\prime}s_{A}^{\prime}-5c_{A}s_{A}c_{A}^{\prime}s_{A}^{\prime}\mathbf{p}_{1}\cdot\hat{\mathbf{k}}_{A}-2ms_{A}^{2}c_{A}^{\prime}s_{A}^{\prime}\nonumber \\
 &  & \left.-3E_{1}s_{A}^{2}c_{A}^{\prime}s_{A}^{\prime}-s_{A}^{2}s_{A}^{\prime2}\mathbf{p}_{1}\cdot\hat{\mathbf{k}}_{A}^{\prime}+2s_{A}^{2}s_{A}^{\prime2}(\mathbf{p}_{1}\cdot\hat{\mathbf{k}}_{A})(\hat{\mathbf{k}}_{A}\cdot\hat{\mathbf{k}}_{A}^{\prime})\right]\nonumber \\
 &  & +16i(\mathbf{n}\times\hat{\mathbf{k}}_{A})\cdot\hat{\mathbf{k}}_{A}^{\prime}\left[4ms_{A}^{2}c_{A}^{\prime}s_{A}^{\prime}\mathbf{p}_{1}\cdot\hat{\mathbf{k}}_{A}+8m^{2}c_{A}s_{A}c_{A}^{\prime}s_{A}^{\prime}\right.\nonumber \\
 &  & +4E_{1}ms_{A}^{2}s_{A}^{\prime2}\hat{\mathbf{k}}_{A}\cdot\hat{\mathbf{k}}_{A}^{\prime}-s_{A}^{2}s_{A}^{\prime2}(\mathbf{p}_{1}\cdot\mathbf{p}_{1})(\hat{\mathbf{k}}_{A}\cdot\hat{\mathbf{k}}_{A}^{\prime})\nonumber \\
 &  & -3E_{1}c_{A}s_{A}s_{A}^{\prime2}\mathbf{p}_{1}\cdot\hat{\mathbf{k}}_{A}^{\prime}-E_{1}s_{A}^{2}c_{A}^{\prime}s_{A}^{\prime}\mathbf{p}_{1}\cdot\hat{\mathbf{k}}_{A}\nonumber \\
 &  & \left.-3c_{A}s_{A}c_{A}^{\prime}s_{A}^{\prime}\mathbf{p}_{1}\cdot\mathbf{p}_{1}-8E_{1}^{2}c_{A}s_{A}c_{A}^{\prime}s_{A}^{\prime}\right]\nonumber \\
 &  & +16i(\mathbf{p}_{1}\times\hat{\mathbf{k}}_{A})\cdot\hat{\mathbf{k}}_{A}^{\prime}\left[s_{A}^{2}s_{A}^{\prime2}(\mathbf{p}_{1}\cdot\hat{\mathbf{k}}_{A})(\mathbf{n}\cdot\hat{\mathbf{k}}_{A}^{\prime})-s_{A}^{2}s_{A}^{\prime2}(\mathbf{n}\cdot\hat{\mathbf{k}}_{A})(\mathbf{p}_{1}\cdot\hat{\mathbf{k}}_{A}^{\prime})\right.\nonumber \\
 &  & +s_{A}^{2}s_{A}^{\prime2}(\mathbf{n}\cdot\mathbf{p}_{1})(\hat{\mathbf{k}}_{A}\cdot\hat{\mathbf{k}}_{A}^{\prime})+4mc_{A}s_{A}s_{A}^{\prime2}\mathbf{n}\cdot\hat{\mathbf{k}}_{A}^{\prime}+E_{1}s_{A}^{2}c_{A}^{\prime}s_{A}^{\prime R}\mathbf{n}\cdot\hat{\mathbf{k}}_{A}\nonumber \\
 &  & \left.+3E_{1}c_{A}s_{A}s_{A}^{\prime2}\mathbf{n}\cdot\hat{\mathbf{k}}_{A}^{\prime}+3c_{A}s_{A}c_{A}^{\prime}s_{A}^{\prime}\mathbf{n}\cdot\mathbf{p}_{1}\right],\label{eq:final-result}
\end{eqnarray}
where we denote the contracted trace part for $q_{a}q_{b}\rightarrow q_{a}q_{b}$
as $\mathrm{Tr_{q_{a}q_{b}\rightarrow q_{a}q_{b}}^{2}}$, $c_{A}\equiv\cosh(\eta_{kA}/2)$,
$c_{A}^{\prime}\equiv\cosh(\eta_{kA}^{\prime}/2)$, $s_{A}\equiv\sinh(\eta_{kA}^{\prime}/2)$,
and $s_{A}^{\prime}\equiv\sinh(\eta_{kA}^{\prime}/2)$. We see in
(\ref{eq:final-result}) that there are four typical terms proportional
to $(\mathbf{n}\times\mathbf{p}_{1})\cdot\hat{\mathbf{k}}_{A}$, $(\mathbf{n}\times\mathbf{p}_{1})\cdot\hat{\mathbf{k}}_{A}^{\prime}$,
$(\mathbf{n}\times\hat{\mathbf{k}}_{A})\cdot\hat{\mathbf{k}}_{A}^{\prime}$
and $(\mathbf{p}_{1}\times\hat{\mathbf{k}}_{A})\cdot\hat{\mathbf{k}}_{A}^{\prime}$,
in which the first three terms are from the spin-orbit coupling and
the last one corresponds to the non-coplanar part of $\mathbf{p}_{1}$,
$\hat{\mathbf{k}}_{A}$ and $\hat{\mathbf{k}}_{A}^{\prime}$. We will
show in the next section that (\ref{eq:final-result}) is a good approximation
for the contracted trace part to the exact result. 

It can be proved that $\Delta I_{M}^{AB\rightarrow12}$ for all 2-to-2
parton scatterings in Table \ref{tab:feyn-22-quark} have the same
structure as in (\ref{eq:final-result}) for $q_{a}q_{b}\rightarrow q_{a}q_{b}$
under the approximation in (\ref{eq:replacement-p1-p2},\ref{eq:new-replacement}). 

Note that $\Delta I_{M}^{AB\rightarrow12}$ depends linearly on the
direction of the scattering plane $\mathbf{n}=\hat{\mathbf{b}}\times\hat{\mathbf{p}}_{A}$,
we can write the contracted trace part in the form of $\hat{\mathbf{b}}\cdot\mathbf{I}$,
as is done in Eq. (\ref{eq:deltaM-nc}). We take the term $(\mathbf{n}\times\mathbf{p}_{1})\cdot\hat{\mathbf{k}}_{A}$
in (\ref{eq:final-result}) as an example, which can be rewritten
as 
\begin{equation}
[(\hat{\mathbf{b}}\times\hat{\mathbf{p}}_{A})\times\mathbf{p}_{1}]\cdot\hat{\mathbf{k}}_{A}=\hat{\mathbf{b}}\cdot[(\hat{\mathbf{p}}_{A}\cdot\hat{\mathbf{k}}_{A})\mathbf{p}_{1}-(\hat{\mathbf{p}}_{A}\cdot\mathbf{p}_{1})\hat{\mathbf{k}}_{A}].\label{eq:example-Ic}
\end{equation}
Therefore $\mathbf{I}$ contains the term inside the square brackets
on the right-hand side of Eq. (\ref{eq:example-Ic}). Another example
is the term proportional to $(\mathbf{p}_{1}\times\hat{\mathbf{k}}_{A})\cdot\hat{\mathbf{k}}_{A}^{\prime}$,
we see that all terms have factors of the form $\mathbf{n}\cdot\mathbf{V}$
($\mathbf{V}=\hat{\mathbf{k}}_{A},\hat{\mathbf{k}}_{A}^{\prime},\mathbf{p}_{1}$)
inside the square brackets, these terms can be rewritten as $\mathbf{n}\cdot\mathbf{V}=\hat{\mathbf{b}}\cdot(\hat{\mathbf{p}}_{A}\times\mathbf{V})$,
so $\mathbf{I}$ contains the term $\hat{\mathbf{p}}_{A}\times\mathbf{V}$.

\section{Numerical method to calculate quark/antiquark polarization rate}

In this section we will calculate the polarization rate for quarks
in a QGP from Eq. (\ref{eq:polarization-2}). Here we assume a local
equilibrium in particle momentum but not in spin. We will consider
two cases: the approximation as in (\ref{eq:replacement-p1-p2},\ref{eq:new-replacement})
and the exact result without any appoximation. The main parameters
are set to following values: the quark mass $m_{q}=0.2$ GeV for quarks
of all flavors ($u,d,s,\bar{u},\bar{d},\bar{s}$), the gluon mass
$m_{g}=0$ for the external gluon, the internal gluon mass (Debye
screening mass) $m_{g}=m_{D}=0.2$ GeV in gluon propagators in the
t and u channel to regulate the possible divergence, the width $\alpha=0.28$
GeV of the Gaussian wave packet, and the temperature $T=0.3$ GeV. 

Although the 2-to-2 processes for anti-qaurk polarization are different
from those for quarks, it can be shown that the polarization rate
for anti-quarks is the same as that for quarks, because all 2-to-2
scatterings for anti-quark polarization can be obtained from those
in Table \ref{tab:feyn-22-quark} by making a particle-antiparticle
transformation. In the following we discuss only the quark polarization.
The same discussion can also be applied to the antiquark polarization. 

The local polarization rate in Eq. (\ref{eq:polarization-2}) for
quarks involves a 16-dimensional integration, which is a major challenge
in the numerical calculation. In the Monte Carlo integration, the
number of sample points grows exponentially with the dimension, so
even a very rough calculation in high dimensions would need huge number
of sample points. 

To overcome this difficulty, we split the integration into two parts:
a 10-dimension (10D) integration over $(\mathbf{p}_{c,1},\mathbf{p}_{c,2},\mathbf{k}_{c,A}^{T},\mathbf{k}_{c,A}^{\prime T})$
and a 6-dimension (6D) integration over $(\mathbf{p}_{A},\mathbf{p}_{B})$.
We carry out the 10D integration and store the result as a function
of $\mathbf{p}_{c,A}$ (and $\mathbf{p}_{c,B}=-\mathbf{p}_{c,A}$).
Then we carry out the 6D integration using the pre-calculated 10D
integral. 

The 10D integral, the last five lines of Eq. (\ref{eq:polarization-2}),
depends on $\mathbf{p}_{c,A}$ and $\mathbf{p}_{c,B}=-\mathbf{p}_{c,A}$
which appear in the wave packet function $\phi_{A}$ and $\phi_{B}$
respectively. So we denote the 10D integral as $\Theta_{jk}(\mathbf{p}_{c,A})$,
from Eq. (\ref{eq:quark-polar}) the polarization rate per unit volume
for one quark flavor can be rewritten as 
\begin{eqnarray}
\frac{d^{4}\mathbf{P}_{q}(X)}{dX^{4}} & = & \frac{\pi}{(2\pi)^{4}}\frac{\partial(\beta u_{\rho})}{\partial X^{\nu}}\sum_{A,B,1}\int\frac{d^{3}p_{A}}{(2\pi)^{3}2E_{A}}\frac{d^{3}p_{B}}{(2\pi)^{3}2E_{B}}\nonumber \\
 &  & \times|v_{c,A}-v_{c,B}|[\Lambda^{-1}]_{\; j}^{\nu}\mathbf{e}_{c,i}\epsilon_{ikh}\hat{\mathbf{p}}_{c,A}^{h}\nonumber \\
 &  & \times f_{A}\left(X,p_{A}\right)f_{B}\left(X,p_{B}\right)\left(p_{A}^{\rho}-p_{B}^{\rho}\right)\Theta_{jk}(\mathbf{p}_{c,A})\nonumber \\
 & \equiv & \frac{\partial(\beta u_{\rho})}{\partial X^{\nu}}\mathbf{W}^{\rho\nu},\label{eq:diff-rate-1}
\end{eqnarray}
where the second equality defines $\mathbf{W}^{\rho\nu}$ and the
sum of $A,B,1$ is over all 2-to-2 processes in Table \ref{tab:feyn-22-quark}.

\subsection{The 10D integration}

The 10D integral $\Theta_{jk}(\mathbf{p}_{c,A}^{(z)})$ is calculated
in the CMS by assuming $\mathbf{p}_{c,A}^{(z)}=(0,0,|\mathbf{p}_{c,A}|)$
and $\mathbf{p}_{c,B}^{(z)}=(0,0,-|\mathbf{p}_{c,A}|)$, where $|\mathbf{p}_{c,A}|$
is determined by the momenta of two incident particles in the lab
frame as in Eq. (\ref{eq:cmf-lorentz}). We can obtain $\Theta_{jk}(\mathbf{p}_{c,A})$
by carrying out the rotation operation on the tensor $\Theta_{jk}(\mathbf{p}_{c,A}^{(z)})$
in accordance with the rotation matrix from $\mathbf{p}_{c,A}^{(z)}$
to $\mathbf{p}_{c,A}$. 

For the Monte Carlo integration we have to sample $\mathbf{k}_{c,A}^{T}$,
$\mathbf{k}_{c,A}^{\prime T}$, $\mathbf{p}_{c,1}$, and $\mathbf{p}_{c,2}$.
First we sample $\mathbf{k}_{c,A}^{T}$ and $\mathbf{k}_{c,A}^{\prime T}$,
where the main contribution comes from the Gaussian distribution (\ref{eq:wave-packet-gs}).
Here we draw samples of $\mathbf{k}_{c,A}^{T}=(k_{c,A,x},k_{c,A,y},0)$
and $\mathbf{k}_{c,A}^{\prime T}=(k_{c,A,x}^{\prime},k_{c,A,y}^{\prime},0)$
inside the $3\sigma$ ($\sigma=\alpha/\sqrt{2}$) region of the Gaussian
distribution around the center point $\mathbf{p}_{c,A}^{(z)}$. The
longitudinal momentum $k_{c,A,z}$ and $k_{c,A,z}^{\prime}$ can be
determined by the energy conservation once $\mathbf{p}_{c,1}$ and
$\mathbf{p}_{c,2}$ are given. 

Then we sample $\mathbf{p}_{c,1}$ and $\mathbf{p}_{c,2}$. In order
to increase the efficiency of the sampling, we should determine the
range of $\mathbf{p}_{c,1}$ and $\mathbf{p}_{c,2}$. We can first
determine the ranges of lengths $|\mathbf{p}_{c,1}|$ and $|\mathbf{p}_{c,2}|$
by a numerical search. Then we determine the ranges of directions
$\hat{\mathbf{p}}_{c,1}$ and $\hat{\mathbf{p}}_{c,2}$. For a given
$\hat{\mathbf{p}}_{c,1}$, which can be randomly chosen, we find that
the largest value of $\theta\equiv\arccos(-\mathbf{\hat{p}}_{c,1}\cdot\mathbf{\hat{p}}_{c,2})$
between $\hat{\mathbf{p}}_{c,2}$ and $-\hat{\mathbf{p}}_{c,1}$ occurs
when 
\begin{eqnarray}
\left|\mathbf{k}_{c,A}\right|=\left|\mathbf{k}_{c,B}\right| & = & \left|\mathbf{p}_{c,1}\right|=\left|\mathbf{p}_{c,2}\right|\nonumber \\
 & = & \sqrt{p_{c,A}^{2}+(3\sigma)^{2}}.\label{eq:-1}
\end{eqnarray}
Hence we obtain the range of $\theta$ as 
\begin{eqnarray}
\theta\equiv\arccos(-\mathbf{\hat{p}}_{c,1}\cdot\mathbf{\hat{p}}_{c,2}) & \in & \left[0,\pi-2\text{arccos}\left(\frac{3\sigma}{\sqrt{p_{c,A}^{2}+(3\sigma)^{2}}}\right)\right].
\end{eqnarray}
The azimuthal angle $\varphi$ of $\hat{\mathbf{p}}_{c,2}$ around
$-\hat{\mathbf{p}}_{c,1}$ is in the range $\left[0,2\pi\right]$. 

With the given values of $\mathbf{p}_{c,1}$ and $\mathbf{p}_{c,2}$,
the values of $k_{c,A,z}$ and $k_{c,A,z}^{\prime}$ can be obtained
by solving Eq. (\ref{eq:root-kal-kal1}). Then $\mathbf{k}_{c,B}$
and $\mathbf{k}_{c,B}^{\prime}$ can be determined by $\mathbf{k}_{c,B}=\mathbf{p}_{c,1}+\mathbf{p}_{c,2}-\mathbf{k}_{c,A}$
and $\mathbf{k}_{c,B}^{\prime}=\mathbf{p}_{c,1}+\mathbf{p}_{c,2}-\mathbf{k}_{c,A}^{\prime}$
respectively. 

The 10D integral is done by ZMCintegral-3.0, a Monte Carlo integration
package, that we have newly developed and runs on multi-GPUs \cite{Wu:2019tsf}.
The ZMCintegral package is able to evaluate $15^{10}$ sample points
within a couple of hours depending on the complexity of the integrand.
For our integrand with all 2-to-2 processes for quarks of all flavors
and gluons, it takes about 5 hours on one Tesla v100 card. We scan
the values of $|\mathbf{p}_{c,A}|$ from 0.1 to 2.2 GeV and those
of $b_{0}$ from 0.1 to 3.5 fm, then we store the integration results
of $\Theta_{jk}(\mathbf{p}_{c,A}^{(z)})$ for later use. It takes
a couple of days to finish the calculation. We find that when $|\mathbf{p}_{c,A}|>2.5$
GeV, the 10D integral is almost zero. This is due to the fact that
if $\alpha\ll|\mathbf{p}_{c,A}|$, the incident wave packets can be
almost regarded as plane waves which give vanishing polarization.

\subsection{The 6D integration}

Now we carry out the remaining 6D integration over $\mathbf{p}_{A}$
and $\mathbf{p}_{B}$ in (\ref{eq:diff-rate-1}). As we have mentioned
in Section \ref{sec:quark-rate} that we assume partons with $p_{A}^{\mu}=(E_{A},\mathbf{p}_{A})$
and $p_{B}^{\mu}=(E_{B},\mathbf{p}_{B})$ in the lab frame follow
the Boltzmann distribution, $f_{i}(X,p_{i})=\exp[-\beta(X)p_{i}\cdot u(X)]$
for $i=A,B$. 

The energy-momentum $p_{c,A}^{\mu}=(E_{c,A},\mathbf{p}_{c,A})$ and
$p_{c,B}^{\mu}=(E_{c,B},\mathbf{p}_{c,B})$ in the CMS of two scattering
particles are given by Eq. (\ref{eq:cmf-lorentz}), where the boost
velocity and the Lorentz contraction factor are given by Eq. (\ref{eq:boost-v})
and (\ref{eq:lorentz-factor}) respectively. The impact parameter
$\mathbf{b}_{c}$ in the CMS is given by Eq. (\ref{eq:impact-para}). 

In the preceding subsection, we calculated the 10D integral $\Theta_{jk}(\mathbf{p}_{c,A}^{(z)})$
where $\mathbf{p}_{c,A}^{(z)}$ is in the z direction. We have to
transform the tensor $\Theta_{jk}(\mathbf{p}_{c,A}^{(z)})$ to $\Theta_{jk}(\mathbf{p}_{c,A})$
so that $\mathbf{p}_{c,A}^{(z)}$ is rotated to the real direction
of $\mathbf{p}_{c,A}$ determined by Eq. (\ref{eq:cmf-lorentz}).
The rotation matrix $R_{ij}$ is defined by $\mathbf{p}_{c,A,i}=R_{ij}\mathbf{p}_{c,A,j}^{(z)}$,
with which we define the transformation for the tensor $\Theta_{jk}(\mathbf{p}_{c,A})=R_{jj^{\prime}}R_{kk^{\prime}}\Theta_{j^{\prime}k^{\prime}}(\mathbf{p}_{c,A}^{(z)})$. 

Our numerical results show that the tensor $\mathbf{W}^{\rho\nu}$
has the form 
\begin{equation}
\mathbf{W}^{\rho\nu}=W\epsilon^{0\rho\nu j}\mathbf{e}_{j},\label{eq:w-vector}
\end{equation}
where we see that $\rho$ and $\nu$ should be spatial indices or
$\mathbf{W}^{0\nu}=\mathbf{W}^{\rho0}=\mathbf{0}$. The form of (\ref{eq:w-vector})
will be verified in the numerical results in Section \ref{sec:result}.
Then from (\ref{eq:diff-rate-1}) we obtain the polarization rate
per unit volume for one quark flavor 
\begin{eqnarray}
\frac{d^{4}\mathbf{P}_{q}(X)}{dX^{4}} & = & \epsilon^{0j\rho\nu}\frac{\partial(\beta u_{\rho})}{\partial X^{\nu}}W\mathbf{e}_{j}=2\epsilon_{jkl}\omega_{kl}W\mathbf{e}_{j}\nonumber \\
 & = & 2W\nabla_{X}\times(\beta\mathbf{u}),\label{eq:w31}
\end{eqnarray}
where $\omega_{\rho\nu}=-(1/2)[\partial_{\rho}^{X}(\beta u_{\nu})-\partial_{\nu}^{X}(\beta u_{\rho})]$,
and for spatial indices we have the 3D form $\omega_{kl}=(1/2)[\nabla_{k}^{X}(\beta\mathbf{u}_{l})-\nabla_{l}^{X}(\beta\mathbf{u}_{k})]$
with $\mathbf{u}$ being the spatial part of the four-velocity $u^{\rho}$.

\section{Numerical results}

\label{sec:result}
\begin{figure}[H]
\caption{Comparison of the results of the symmetric tensor $\Theta_{jk}(\mathbf{p}_{c,A}^{(z)})$
for $q(\bar{q})+q\rightarrow q(\bar{q})+q$ and $g+q\rightarrow g+q$
in two cases: (1) with the approximation in (\ref{eq:replacement-p1-p2},\ref{eq:new-replacement})
and (2) exact calculation of the integral without any approximation.
The results for $g+g\rightarrow q+\bar{q}$ are not shown because
they are negligibly small (almost zero). Here we choose $b_{0}=0.5$
fm and $|\mathbf{p}_{c,A}^{(z)}|=0,\,0.5,\,1.0,\,1.5,\,2.0$ GeV.
The solid symbols are the exact results without any approximation,
while the dashed symbols are the results with approximation in (\ref{eq:replacement-p1-p2},\ref{eq:new-replacement}).
The unit of $\Theta_{jk}(\mathbf{p}_{c,A}^{(z)})$ is $\text{GeV}^{-1}$.
\label{fig:component}}

\begin{center}\includegraphics[scale=0.35]{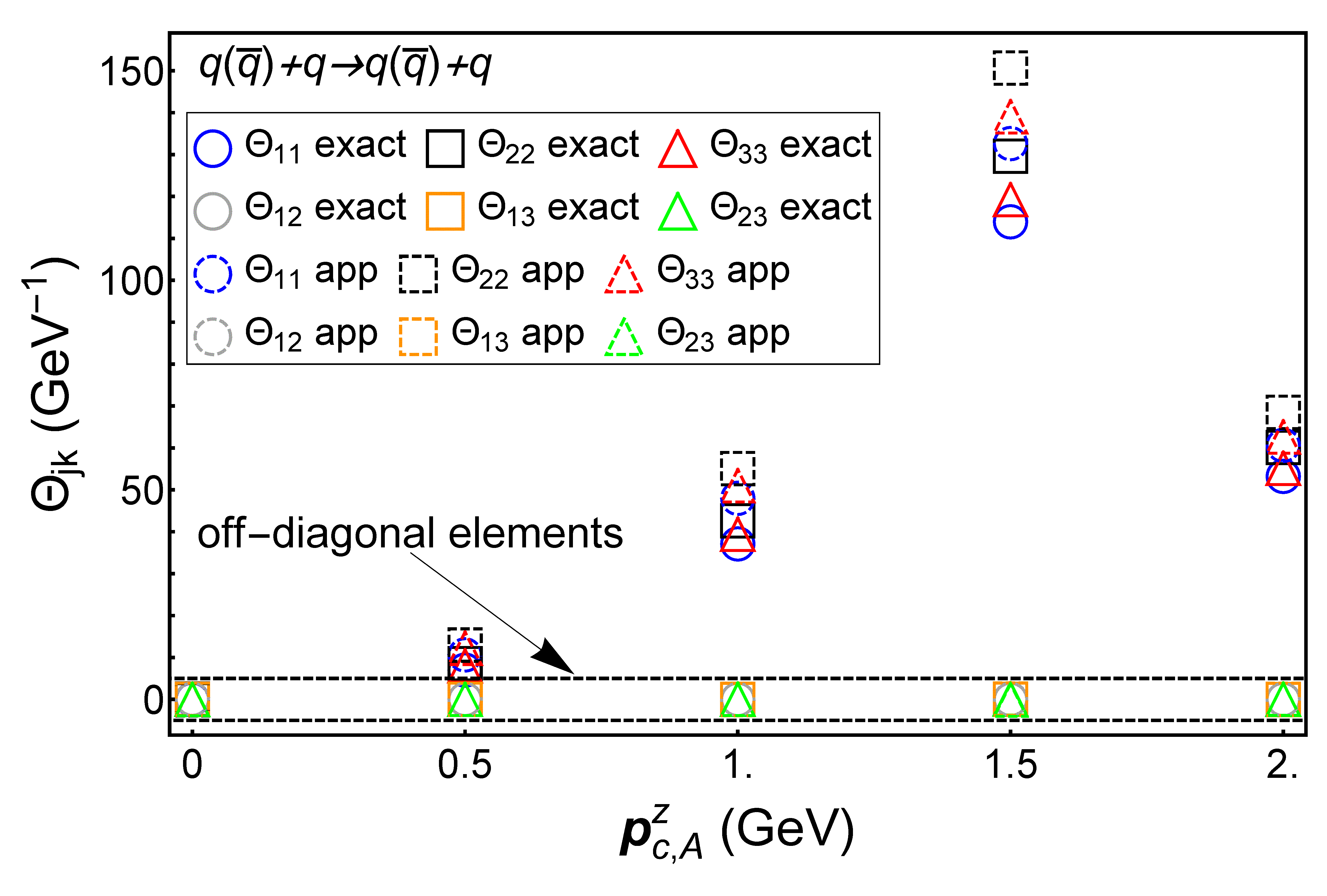} \includegraphics[scale=0.35]{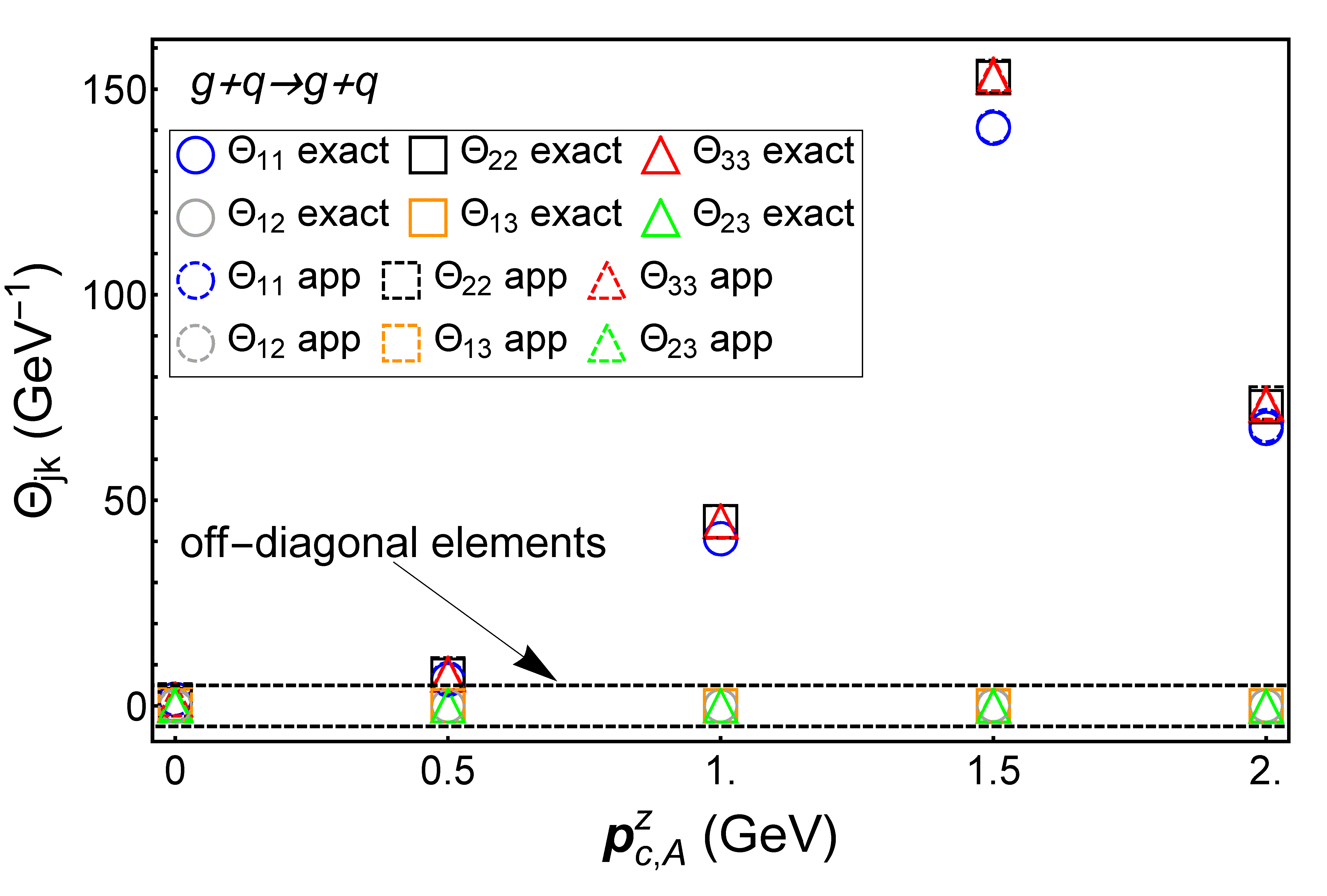}\end{center}
\end{figure}

In this section we will present our numerical results.
The approximation in (\ref{eq:replacement-p1-p2},\ref{eq:new-replacement})
is inspired by the first order contribution in the narrow wave packet
approximation. In order to see how effective the approximation is,
we compare in Fig. \ref{fig:component} the results of the 10D integral
$\Theta_{jk}(\mathbf{p}_{c,A}^{(z)})$ for the scattering processes
$q(\bar{q})+q\rightarrow q(\bar{q})+q$ and $g+q\rightarrow g+q$
in two cases: with and without the approximation. Here the process
$q(\bar{q})+q\rightarrow q(\bar{q})+q$ stands for a sum over 5 different
processes in Table \ref{tab:feyn-22-quark}. Note that we do not show
the results for $g+g\rightarrow q+\bar{q}$ for which all elements
of $\Theta_{jk}(\mathbf{p}_{c,A}^{(z)})$ are almost zero in contrast
to processes with at least one incident quark. We see in the figure
that the results with the approximation are in agreement with the
exact ones in 20\% precision. In the figure we see that all elements
of $\Theta(\mathbf{p}_{c,A}^{(z)})$ fluctuate around zero for $|\mathbf{p}_{c,A}^{(z)}|=0$,
which leads to vanishing polarization. When $|\mathbf{p}_{c,A}^{(z)}|$
is non-vanishing, the off-diagonal elements of $\Theta(\mathbf{p}_{c,A}^{(z)})$
are still zero within errors, but all diagonal elements take positive
values which are almost equal to each other. 

\begin{figure}[H]
\caption{The dependence of the results of $\mathbf{W}_{y}^{31}$ on the integral
ranges $|\mathbf{p}_{A}|_{max}=|\mathbf{p}_{B}|_{max}$ for $q(\bar{q})+q\rightarrow q(\bar{q})+q$.
We choose $b_{0}=2.2$ fm, $z=0$ fm, $T=0.3$ GeV. \label{fig:result-w31y}}

\begin{center}\includegraphics[scale=0.5]{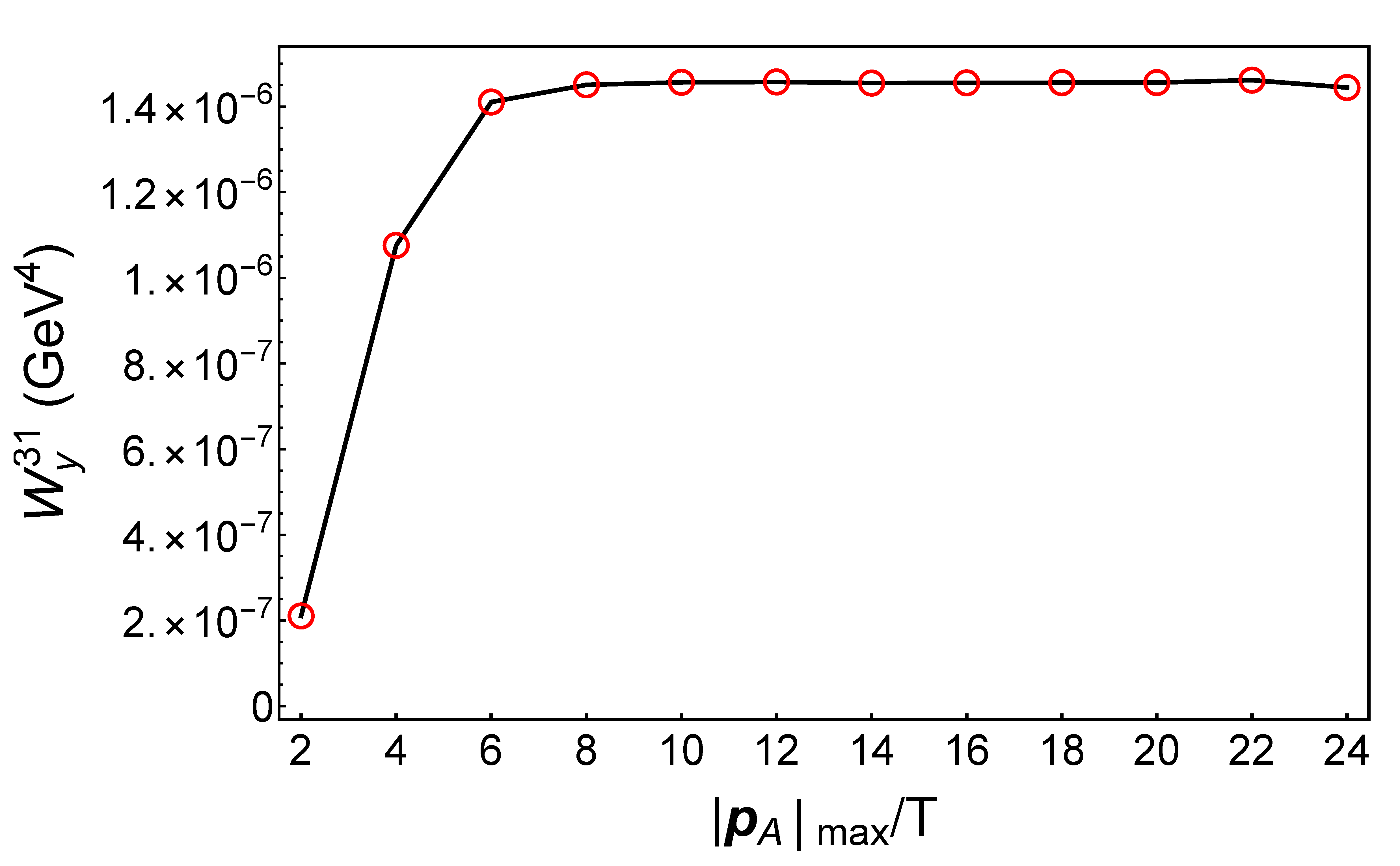}\end{center}
\end{figure}

We then work out the rest 6D integral and obtain $\mathbf{W}^{\rho\nu}$
in Eq. (\ref{eq:w-vector}). In the 6D integration we have to determine
the maximum value of $|\mathbf{p}_{A}|$ and $|\mathbf{p}_{B}|$ or
the integration range of $|\mathbf{p}_{A}|$ and $|\mathbf{p}_{B}|$.
In Fig. \ref{fig:result-w31y}, as an example, we show the dependence
of $\mathbf{W}_{y}^{31}$ on $|\mathbf{p}_{A}|_{max}=|\mathbf{p}_{B}|_{max}$
for $q(\bar{q})+q\rightarrow q(\bar{q})+q$, where we choose $b_{0}=2.2$
fm, $z=0$ fm and $T=0.3$ GeV. We see in the figure that the value
of $\mathbf{W}_{y}^{31}$ is very stable when $|\mathbf{p}_{A}|_{max}=|\mathbf{p}_{B}|_{max}>8T$. 

\begin{figure}[H]
\caption{Results for $\mathbf{W}_{x}^{31}$, $\mathbf{W}_{y}^{31}$ and $\mathbf{W}_{z}^{31}$
as functions of the cutoff $b_{0}$ in fm. There are large fluctuations
in $\mathbf{W}_{x}^{31}$ and $\mathbf{W}_{z}^{31}$ above $b_{0}=1.5$
fm due to the strong oscillation of Bessel functions. \label{fig:w31-b0}}

\begin{center}

\includegraphics[scale=0.5]{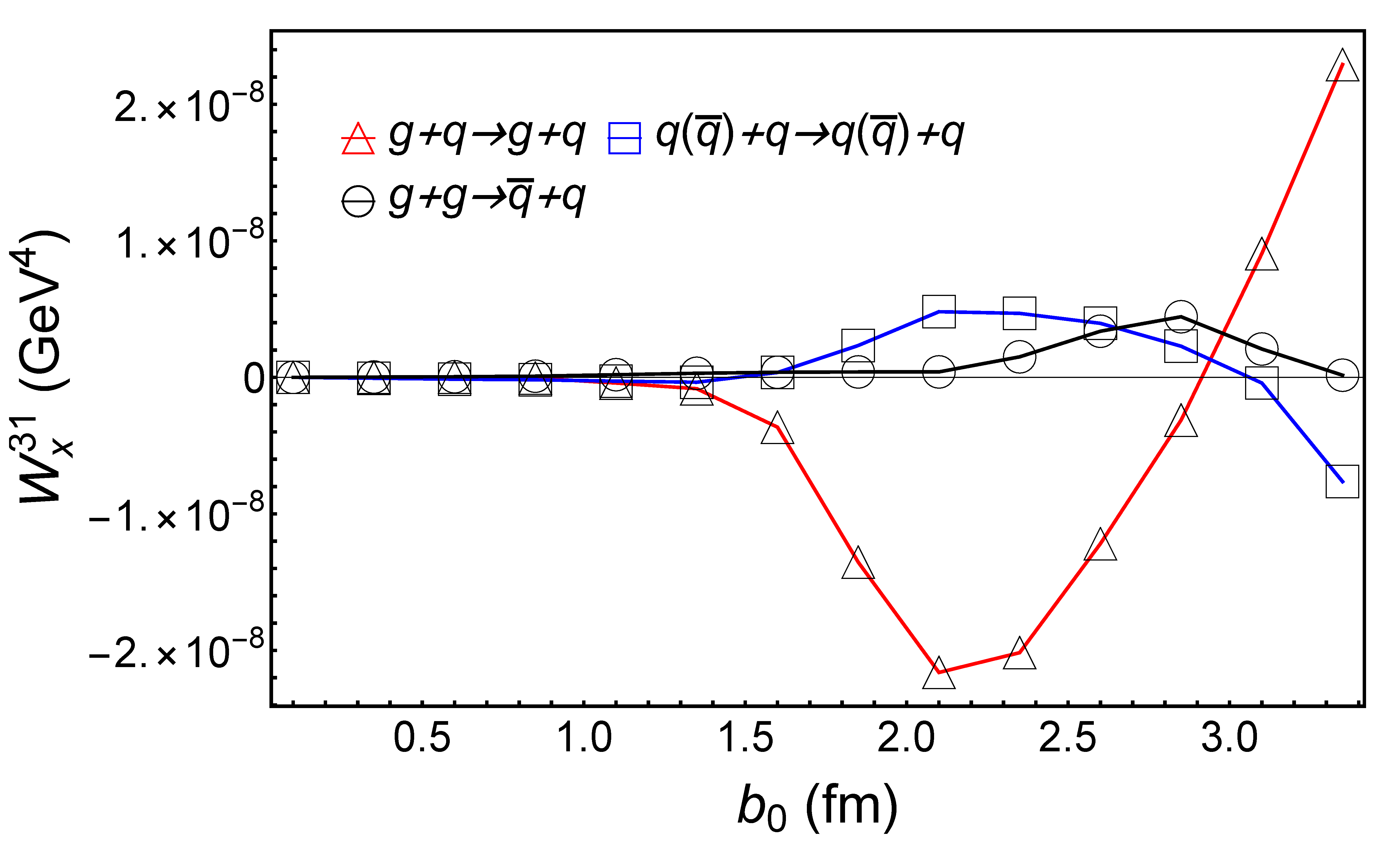}

\includegraphics[scale=0.5]{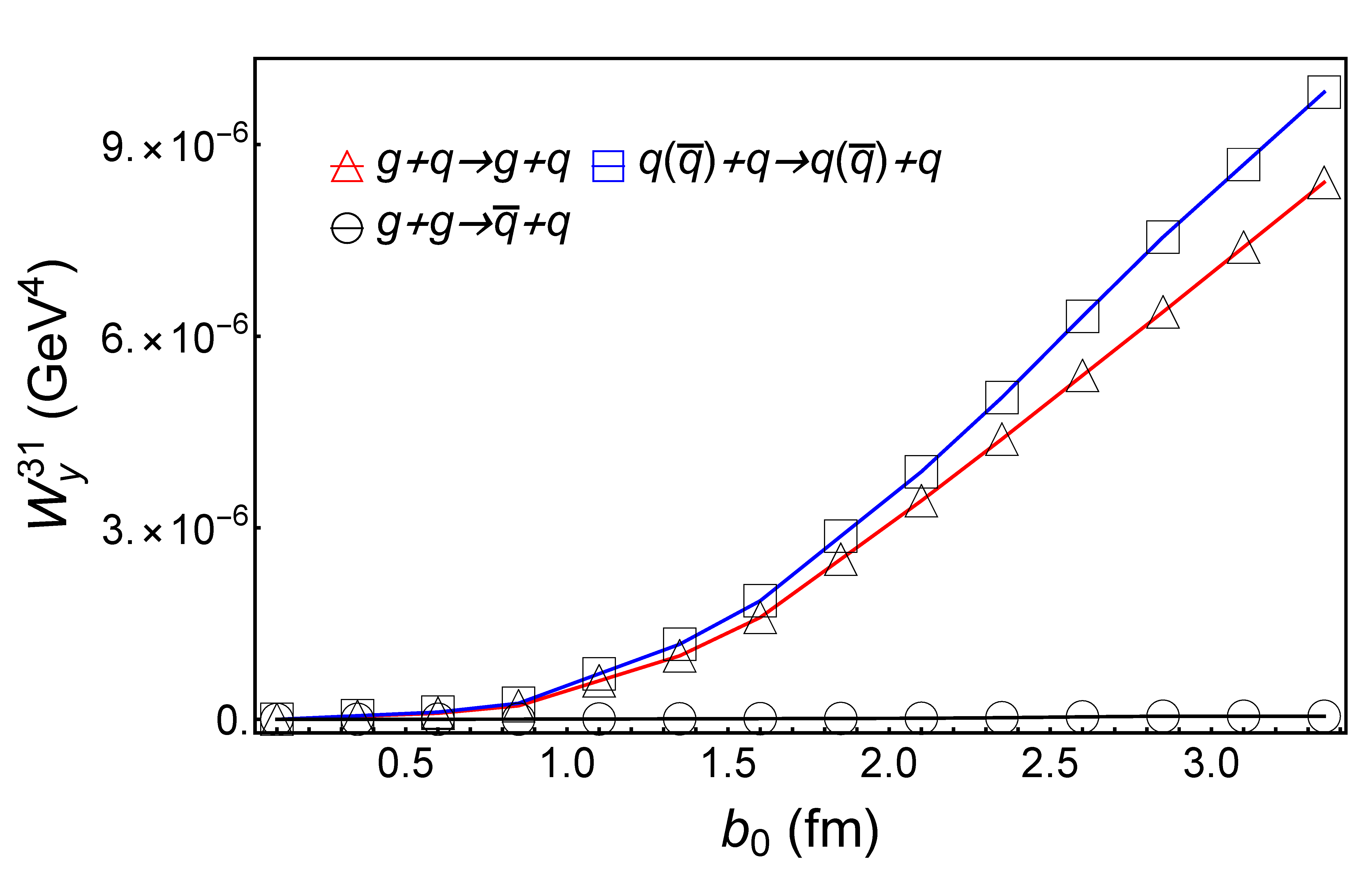}

\includegraphics[scale=0.5]{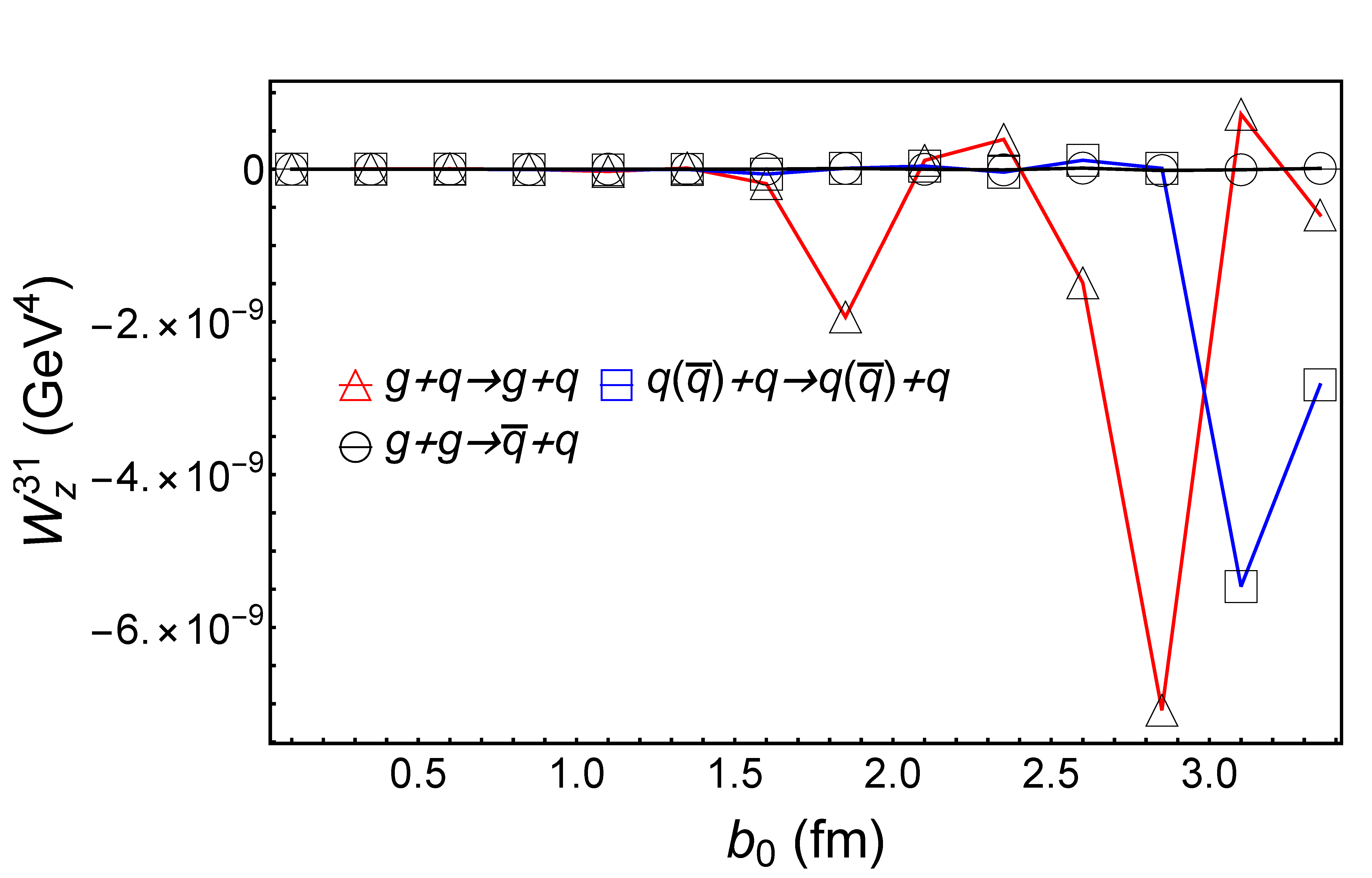}

\end{center}
\end{figure}

The numerical results for $\mathbf{W}^{\rho\nu}$ show the structure
of (\ref{eq:w-vector}). We can write $\mathbf{W}^{\rho\nu}$ in an
explicit matrix form
\begin{equation}
\mathbf{W}^{\rho\nu}=\left(\begin{array}{cccc}
0 & 0 & 0 & 0\\
0 & 0 & W\mathbf{e}_{z} & -W\mathbf{e}_{y}\\
0 & -W\mathbf{e}_{z} & 0 & W\mathbf{e}_{x}\\
0 & W\mathbf{e}_{y} & -W\mathbf{e}_{x} & 0
\end{array}\right)\label{eq:w-vector-exp}
\end{equation}
As an example, we show in Fig. \ref{fig:w31-b0} the results for all
components of $\mathbf{W}^{31}$ as functions of the cutoff $b_{0}$
for the quark polarization. We see in the figure that $\mathbf{W}_{x}^{31}$
and $\mathbf{W}_{z}^{31}$ are two or three orders of magnitude smaller
than the positive values of $\mathbf{W}_{y}^{31}$, which gives the
polarization in the y direction. As we can see in the figure that
$\mathbf{W}_{y}^{31}$ increases with the cutoff $b_{0}$. The reason
for such a rising behavior is due to the Taylor expansion of $f_{A}(x_{c,A},p_{c,A})f_{B}(x_{c,B},p_{c,B})$
to the linear order in $y_{c,T}=(0,\mathbf{b}_{c})$ as in App. \ref{sec:Derivation-of-Expansion}.
There should exist an upper limit for $b_{0}$ above which the coherence
of the incident wave packets is broken and the results are not physical.
Such an upper limit can be set to be the order of the hydrodynamical
length scale $\sim1/\partial_{X}^{\mu}u^{\nu}$ and should be larger
than the interaction length scale $1/m_{D}$. 

It can be proved that $\mathbf{W}^{31}$ for the anti-quark polarization
is the same as that for the quark one. The numerical results show
that the magnitude of all element $\mathbf{W}^{\rho\nu}$ are equal
so we denote it as $W$. 

\section{Discussions}

We have constructed a microscopic model for the global polarization
from particle scatterings in a many body system. The core of the idea
is the scattering of particles as wave packets so that the orbital
angular momentum is present in scatterings and can be converted to
spin polarization. As an illustrative example, we have calculated
the quark/antiquark polarization in a QGP. The quarks and gluons are
assumed to obey the Boltzmann distribution which simplifies the heavy
numerical calculation. There is no essential difficulty to treat quarks
and gluons as fermions and bosons respectively. 

To simplify the calculation, we also assume that the quark distributions
are the same for all flavors and spin states. As a consequence, the
inverse processes that one polarized quark is scattered by a parton
to two final state partons as wave packets are absent. So the relaxation
of polarization cannot be described without inverse processes and
polarized distributions. We will extend our model by including the
inverse processes in the future.

\section{Summary and conclusions}

The global polarization in heavy ion collisions arises from scattering
processes of partons or hadrons with spin-orbit couplings. However
it is hard to implement this microscopic picture consistently to describe
particle scatterings at specified impact parameters in a thermal medium
with a shear flow. On the other hand the statistic-hydro model or
Wigner function method are widely used to calculate the global polarization
in heavy ion collisions. These models are based on the assumption
that the spin degrees of freedom have reached a local equilibrium.
So there should be a spin-vorticity coupling term in the distribution
function to give the global polarization proportional to the vorticity
when it is small. However it is unknown if particle spins are really
in a local equilibrium. In this paper we aim to construct a microscopic
model for the global polarization from particle collisions without
the assumption of local equilibrium for spins. The polarization effect
is incorporated into particle scatterings at specified impact parameters
with spin-orbit couplings encoded. The spin-vorticity coupling naturally
emerges from particle collisions if we assume a local equilibrium
in particle momenta instead of particle spins. This provides a microscopic
mechanism for the global polarization from the first principle through
particle collisions in non-equilibrium. 

As an illustrative example, we have calculated the quark polarization
rate per unit volume from all 2-to-2 parton (quark or gluon) scatterings
in a locally thermalized quark-gluon plasma in momentum. Although
the processes for anti-quark polarization are different from those
for quarks, it can be shown that the polarization rate for anti-quarks
is the same as that for quarks because they are connected by the charge
conjugate transformation. This is consistent with the fact that the
rotation does not distinguish particles and antiparticles. The spin-orbit
coupling is hidden in the polarized scattering amplitude at specified
impact parameters. The polarization rate involves an integral of 16
dimensions, which is far beyond the capability of the current numerical
algorithm. We have developed a new Monte-Carlo integration algorithm
ZMCintegral on multi-GPUs to make such a heavy task feasible. We have
shown that the polarization rate per unit volume is proportional to
the vorticity as the result of particle scatterings, a non-equilibrium
senario for the global polarization. So we can see in this example
how the spin-vorticity coupling emerges naturally from particle scatterings. 
\begin{acknowledgments}
QW thanks F. Becattini and M. Lisa for insightful discussions. QW
is supported in part by the National Natural Science Foundation of
China (NSFC) under Grant No. 11535012 and No. 11890713, the 973 program
under Grant No. 2015CB856902, and the Key Research Program of the
Chinese Academy of Sciences under the Grant No. XDPB09. XNW is supported
in part by the National Natural Science Foundation of China (NSFC)
under Grant No. 11890714 and No. 11861131009, and by the Director,
Office of Energy Research, Office of High Energy and Nuclear Physics,
Division of Nuclear Physics, of the U.S. Department of Energy under
Contract Nos. DE- AC02-05CH11231. 
\end{acknowledgments}
\appendix

\section{Single particle state as a wave packet in relativistic quantum mechanics}

\label{sec:one-part-wave-pack}In this appendix, we will give definitions
and conventions for the single particle state in coordinate and momentum
space and those for the wave packet.

\subsection{Single particle state in coordinate and momentum space}

\label{sub:one-particle}For simplicity we first consider the single
particle state of spin-0 particles, then we generalize it to spin-1/2
particles. 

A position eigenstate is denoted as $\left|\mathbf{x}\right\rangle $
and satisfies following orthogonality and completeness conditions
\begin{eqnarray}
\left\langle \mathbf{x}^{\prime}|\mathbf{x}\right\rangle  & = & \delta^{(3)}(\mathbf{x}^{\prime}-\mathbf{x}),\nonumber \\
1 & = & \int d^{3}x\left|\mathbf{x}\right\rangle \left\langle \mathbf{x}\right|.\label{eq:orth-x}
\end{eqnarray}
The normalization of the state $|\mathbf{x}\rangle$ is then 
\begin{equation}
\left\langle \mathbf{x}|\mathbf{x}\right\rangle =\delta^{(3)}(\mathbf{x}-\mathbf{x})=\int\frac{d^{3}p}{(2\pi)^{3}}=\frac{1}{\Omega}\sum_{\mathbf{p}},
\end{equation}
where $\Omega$ is the space volume. 

A momentum eigenstate is denoted as $\left|\mathbf{p}\right\rangle $
and satisfies following orthogonality and completeness conditions
\begin{eqnarray}
\left\langle \mathbf{p}^{\prime}|\mathbf{p}\right\rangle  & = & 2E_{p}(2\pi)^{3}\delta^{(3)}(\mathbf{p}-\mathbf{p}^{\prime}),\nonumber \\
1 & = & \int\frac{d^{3}p}{(2\pi)^{3}}\frac{1}{2E_{p}}\left|\mathbf{p}\right\rangle \left\langle \mathbf{p}\right|,\label{eq:ortho-p}
\end{eqnarray}
where $E_{p}=\sqrt{|\mathbf{p}|^{2}+m^{2}}$ is the energy of the
particle. Note that $\left\langle \mathbf{p}^{\prime}|\mathbf{p}\right\rangle $
is Lorentz invariant. The normalization of $|\mathbf{p}\rangle$ is
then 
\begin{equation}
\left\langle \mathbf{p}|\mathbf{p}\right\rangle =2E_{p}(2\pi)^{3}\delta^{(3)}(\mathbf{p}-\mathbf{p})=2E_{p}\Omega.
\end{equation}
From Eq. (\ref{eq:orth-x}) and (\ref{eq:ortho-p}) we can define
the inner product $\left\langle \mathbf{x}|\mathbf{p}\right\rangle $
as 
\begin{equation}
\left\langle \mathbf{x}|\mathbf{p}\right\rangle =\sqrt{2E_{p}}e^{i\mathbf{p}\cdot\mathbf{x}}.\label{eq:x-p-overlap}
\end{equation}
With the above relation we can check
\begin{eqnarray}
\delta^{(3)}(\mathbf{x}-\mathbf{x}^{\prime}) & = & \left\langle \mathbf{x}^{\prime}|\mathbf{x}\right\rangle =\int\frac{d^{3}p}{(2\pi)^{3}}\frac{1}{2E_{p}}\left\langle \mathbf{x}^{\prime}|\mathbf{p}\right\rangle \left\langle \mathbf{p}|\mathbf{x}\right\rangle \nonumber \\
 & = & \int\frac{d^{3}p}{(2\pi)^{3}}e^{i\mathbf{p}\cdot(\mathbf{x}^{\prime}-\mathbf{x})},
\end{eqnarray}
where we have inserted the completeness relation in (\ref{eq:ortho-p}).
We can express $\left|\mathbf{x}\right\rangle $ in terms of $\left|\mathbf{p}\right\rangle $
and vice versa, 

\begin{eqnarray}
\left|\mathbf{x}\right\rangle  & = & \int\frac{d^{3}p}{(2\pi)^{3}}\frac{1}{2E_{p}}\left|\mathbf{p}\right\rangle \left\langle \mathbf{p}|\mathbf{x}\right\rangle =\int\frac{d^{3}p}{(2\pi)^{3}}\frac{1}{\sqrt{2E_{p}}}e^{-i\mathbf{p}\cdot\mathbf{x}}\left|\mathbf{p}\right\rangle ,\nonumber \\
\left|\mathbf{p}\right\rangle  & = & \int d^{3}x\left|\mathbf{x}\right\rangle \left\langle \mathbf{x}|\mathbf{p}\right\rangle =\sqrt{2E_{p}}\int d^{3}xe^{i\mathbf{p}\cdot\mathbf{x}}\left|\mathbf{x}\right\rangle .
\end{eqnarray}

\subsection{Single particle state as a wavepacket}

\label{sub:wave-pack}In the real world a particle is always localized
in some finite region, so its state can be represented by a wavepacket
$\left|\phi\right\rangle $ which is a superposition of plane wave
states, 
\begin{equation}
\left|\phi\right\rangle =\int\frac{d^{3}k}{(2\pi)^{3}}\frac{1}{\sqrt{2E_{k}}}\phi(\mathbf{k})\left|\mathbf{k}\right\rangle ,
\end{equation}
and $\phi(\mathbf{k})$ is the amplitude and can be normalized to
unity,
\begin{equation}
\left\langle \phi|\phi\right\rangle =\int\frac{d^{3}k}{(2\pi)^{3}}|\phi(\mathbf{k})|^{2}=1.\label{eq:wavepack}
\end{equation}
The energy dimension of $\left|\phi\right\rangle $ is $0$. A typical
form for $\phi(\mathbf{p})$ satisfying Eq. (\ref{eq:wavepack}) is
the Gaussian wavepacket 

\begin{equation}
\phi(\mathbf{p}-\mathbf{p}_{0})=\frac{(8\pi)^{3/4}}{\alpha^{3/2}}\exp\left[-\frac{(\mathbf{p}-\mathbf{p}_{0})^{2}}{\alpha^{2}}\right],\label{eq:width-wavepacket}
\end{equation}
which is centered at $\mathbf{p}_{0}$. The wavepacket function in
coordinate space is 
\begin{eqnarray}
\phi(\mathbf{x}) & = & \left\langle \mathbf{x}|\phi\right\rangle =\int\frac{d^{3}k}{(2\pi)^{3}}\phi(\mathbf{k})e^{i\mathbf{k}\cdot\mathbf{x}},
\end{eqnarray}
where we have used Eq. (\ref{eq:x-p-overlap}). 

If we displace the particle state by $\mathbf{b}$ in coordinate space,
the new wavepacket function is given by 
\begin{equation}
\phi^{\prime}(\mathbf{x})=\phi(\mathbf{x}-\mathbf{b})=\int\frac{d^{3}k}{(2\pi)^{3}}\phi(\mathbf{k})e^{i\mathbf{k}\cdot(\mathbf{x}-\mathbf{b})}=\left\langle \mathbf{x}|\phi^{\prime}\right\rangle ,
\end{equation}
where the new wavepacket state is 
\begin{equation}
\left|\phi^{\prime}\right\rangle =\int\frac{d^{3}k}{(2\pi)^{3}}\frac{1}{\sqrt{2E_{k}}}\phi(\mathbf{k})e^{-i\mathbf{k}\cdot\mathbf{b}}\left|\mathbf{k}\right\rangle .
\end{equation}

For spin-1/2 particles, the single particle state $\left|\mathbf{k},\lambda\right\rangle $
has a spin index $\lambda$ which is the spin along a quantization
direction. The orthogonality and completeness conditions in (\ref{eq:ortho-p})
now become 
\begin{eqnarray}
\left\langle \mathbf{k}^{\prime},\lambda^{\prime}|\mathbf{k},\lambda\right\rangle  & = & 2E_{k}(2\pi)^{3}\delta^{(3)}(\mathbf{k}-\mathbf{k}^{\prime})\delta_{\lambda,\lambda^{\prime}},\nonumber \\
1 & = & \int\frac{d^{3}p}{(2\pi)^{3}}\frac{1}{2E_{p}}\sum_{\lambda}\left|\mathbf{p},\lambda\right\rangle \left\langle \mathbf{p},\lambda\right|.
\end{eqnarray}
The wavepacket has the form 
\begin{equation}
\left|\phi,\lambda\right\rangle =\int\frac{d^{3}k}{(2\pi)^{3}}\frac{1}{\sqrt{2E_{k}}}\phi(\mathbf{k})\left|\mathbf{k},\lambda\right\rangle ,
\end{equation}
and satisfies the normalization condition $\left\langle \phi,\lambda|\phi,\lambda\right\rangle =1$
similar to Eq. (\ref{eq:wavepack}).

\section{Expansion of $f_{A}$ and $f_{B}$ in impact parameter}

\label{sec:Derivation-of-Expansion}We can make an expansion of $f_{A}\left(X_{c}+y_{c,T}/2,p_{c,A}\right)f_{B}\left(X_{c}-y_{c,T}/2,p_{c,B}\right)$
in $y_{c,T}=(0,\mathbf{b}_{c})$ if $|\mathbf{b}_{c}|$ is small compared
with the range in which $f_{A}$ and $f_{B}$ change slowly. The variables
with the subscript 'c' are defined in the CMS of the scattering, while
those without 'c' are defined in the lab frame. We assume that the
system has reached local equilibrium in momentum and the phase space
distributions depend on the space-time through the fluid velocity
$u^{\mu}(x)$ and temperature $T(x)$ in the form $f(x,p)=f[\beta(x)p\cdot u(x)]$. 

To the linear order in $y_{c,T}$, we have 

\begin{eqnarray}
 &  & f_{A}\left(X_{c}+\frac{y_{c,T}}{2},p_{c,A}\right)f_{B}\left(X_{c}-\frac{y_{c,T}}{2},p_{c,B}\right)\nonumber \\
 & \approx & f_{A}\left(X_{c},p_{c,A}\right)f_{B}\left(X_{c},p_{c,B}\right)\nonumber \\
 &  & +\frac{1}{2}y_{c,T}^{\mu}\left[\frac{\partial f_{A}\left(X_{c},p_{c,A}\right)}{\partial X_{c}^{\mu}}f_{B}\left(X_{c},p_{c,B}\right)-f_{A}\left(X_{c},p_{c,A}\right)\frac{\partial f_{B}\left(X_{c},p_{c,B}\right)}{\partial X_{c}^{\mu}}\right]\nonumber \\
 & = & f_{A}\left(X_{c},p_{c,A}\right)f_{B}\left(X_{c},p_{c,B}\right)+\frac{1}{2}y_{c,T}^{\mu}\frac{\partial(\beta u_{c,\rho})}{\partial X_{c}^{\nu}}\nonumber \\
 &  & \times\left[p_{c,A}^{\rho}f_{B}\left(X_{c},p_{c,B}\right)\frac{df_{A}\left(X_{c},p_{c,A}\right)}{d(\beta u_{c}\cdot p_{c,A})}-p_{c,B}^{\rho}f_{A}\left(X_{c},p_{c,A}\right)\frac{df_{B}\left(X_{c},p_{c,B}\right)}{d(\beta u_{c}\cdot p_{c,B})}\right]\nonumber \\
 & = & f_{A}\left(X,p_{A}\right)f_{B}\left(X,p_{B}\right)+\frac{1}{2}y_{c,T}^{\mu}\frac{\partial X^{\nu}}{\partial X_{c}^{\mu}}\frac{\partial(\beta u_{\rho})}{\partial X^{\nu}}\nonumber \\
 &  & \times\left[p_{A}^{\rho}f_{B}\left(X,p_{B}\right)\frac{df_{A}\left(X,p_{A}\right)}{d(\beta u\cdot p_{A})}-p_{B}^{\rho}f_{A}\left(X,p_{A}\right)\frac{df_{B}\left(X,p_{B}\right)}{d(\beta u\cdot p_{B})}\right],\label{eq:expansion-ff}
\end{eqnarray}
where in the second equality we have boosted to the lab frame using
$f_{A}\left(X,p_{A}\right)=f_{A}\left(X_{c},p_{c,A}\right)$ and $f_{B}\left(X,p_{B}\right)=f_{B}\left(X_{c},p_{c,B}\right)$.
We look closely at the term $y_{c,T}^{\mu}[\partial(\beta u_{c,\rho})/\partial X_{c}^{\mu}]p_{c,A}^{\rho}$,
\begin{eqnarray}
y_{c,T}^{\mu}p_{c,A}^{\rho}\frac{\partial(\beta u_{\rho})}{\partial X_{c}^{\mu}} & = & \frac{1}{4}y_{c,T}^{[\mu}p_{c,A}^{\rho]}\left[\frac{\partial(\beta u_{c,\rho})}{\partial X_{c}^{\mu}}-\frac{\partial(\beta u_{c,\mu})}{\partial X_{c}^{\rho}}\right]\nonumber \\
 &  & +\frac{1}{4}y_{c,T}^{\{\mu}p_{c,A}^{\rho\}}\left[\frac{\partial(\beta u_{c,\rho})}{\partial X_{c}^{\mu}}+\frac{\partial(\beta u_{c,\mu})}{\partial X_{c}^{\rho}}\right]\nonumber \\
 & = & -\frac{1}{2}y_{c,T}^{[\mu}p_{c,A}^{\rho]}\varpi_{\mu\rho}^{(c)}+\frac{1}{4}y_{c,T}^{\{\mu}p_{c,A}^{\rho\}}\left[\frac{\partial(\beta u_{c,\rho})}{\partial X_{c}^{\mu}}+\frac{\partial(\beta u_{c,\mu})}{\partial X_{c}^{\rho}}\right]\nonumber \\
 & = & -\frac{1}{2}L_{(c)}^{\mu\rho}\varpi_{\mu\rho}^{(c)}+\frac{1}{4}y_{c,T}^{\{\mu}p_{c,A}^{\rho\}}\left[\frac{\partial(\beta u_{c,\rho})}{\partial X_{c}^{\mu}}+\frac{\partial(\beta u_{c,\mu})}{\partial X_{c}^{\rho}}\right],\label{eq:oam-vorticity}
\end{eqnarray}
where $[\mu\rho]$ and $\{\mu\rho\}$ denote the anti-symmetrization
and symmetrization of two indices respectively, $L_{(c)}^{\mu\rho}\equiv y_{c,T}^{[\mu}p_{c,A}^{\rho]}$
is the OAM tensor, and $\omega_{\mu\rho}^{(c)}\equiv-(1/2)[\partial_{\mu}^{X_{c}}(\beta u_{c,\rho})-\partial_{\rho}^{X_{c}}(\beta u_{c,\mu})]$
is the thermal vorticity. We see that the coupling term of the OAM
and vorticity appear in Eq. (\ref{eq:expansion-ff}). The second term
in last line of Eq. (\ref{eq:oam-vorticity}) is related to the Killing
condition required by the thermal equilibrium of the spin. 

Using $X_{c}^{\mu}=\Lambda_{\;\nu}^{\mu}X^{\nu}$ and $X^{\mu}=[\Lambda^{-1}]_{\;\nu}^{\mu}X_{c}^{\nu}$,
so we have $\frac{\partial X^{\nu}}{\partial X_{c}^{\mu}}=[\Lambda^{-1}]_{\;\mu}^{\nu}=\Lambda_{\mu}^{\;\nu}$
and then Eq. (\ref{eq:expansion-ff}) becomes  
\begin{eqnarray}
 &  & f_{A}\left(X_{c}+\frac{y_{c,T}}{2},p_{c,A}\right)f_{B}\left(X_{c}-\frac{y_{c,T}}{2},p_{c,B}\right)\nonumber \\
 & = & f_{A}\left(X,p_{A}\right)f_{B}\left(X,p_{B}\right)+\frac{1}{2}y_{c,T}^{\mu}[\Lambda^{-1}]_{\;\mu}^{\nu}\frac{\partial(\beta u_{\rho})}{\partial X^{\nu}}\nonumber \\
 &  & \times\left[p_{A}^{\rho}f_{B}\left(X,p_{B}\right)\frac{df_{A}\left(X,p_{A}\right)}{d(\beta u\cdot p_{A})}-p_{B}^{\rho}f_{A}\left(X,p_{A}\right)\frac{df_{B}\left(X,p_{B}\right)}{d(\beta u\cdot p_{B})}\right].\label{eq:expansion-ff-1}
\end{eqnarray}
In Appendix \ref{sec:lorentz} we give the exact form of $\Lambda_{\;\nu}^{\mu}$
and $[\Lambda^{-1}]_{\;\nu}^{\mu}$.

\section{Lorentz transformation}

\label{sec:lorentz}In the lab frame two colliding particles have
on-shell momenta $p_{A}=(E_{A},\mathbf{p}_{A})$ and $p_{B}=(E_{B},\mathbf{p}_{B})$.
The Lorentz transformation for the energy-momentum from the lab frame
to the CMS of two colliding particles is 
\begin{eqnarray}
\mathbf{p}_{c,i} & = & \mathbf{p}_{i}+(\gamma_{\mathrm{bst}}-1)\hat{\mathbf{v}}_{\mathrm{bst}}(\hat{\mathbf{v}}_{\mathrm{bst}}\cdot\mathbf{p}_{i})-\gamma_{\mathrm{bst}}\mathbf{v}_{\mathrm{bst}}E_{i},\nonumber \\
E_{c,i} & = & \gamma_{\mathrm{bst}}(E_{i}-\mathbf{v}_{\mathrm{bst}}\cdot\mathbf{p}_{i}).\label{eq:cmf-lorentz}
\end{eqnarray}
where $i=A,B$, $\mathbf{v}_{\mathrm{bst}}$ is the boost velocity
or the velocity of CMS in the lab frame and is given by 
\begin{equation}
\mathbf{v}_{\mathrm{bst}}=\frac{\mathbf{p}_{A}+\mathbf{p}_{B}}{E_{A}+E_{B}},\label{eq:boost-v}
\end{equation}
and 
\begin{equation}
\gamma_{\mathrm{bst}}=(1-|\mathbf{v}_{\mathrm{bst}}|^{2})^{-1/2},\label{eq:lorentz-factor}
\end{equation}
is the Lorentz contraction facror corresponding to $\mathbf{v}_{\mathrm{bst}}$.
Equation (\ref{eq:cmf-lorentz}) defines the Lorentz transformation
matrix $\Lambda_{\;\nu}^{\mu}$. The reverse transformation to (\ref{eq:cmf-lorentz})
from the CMS to the lab frame can be obtained by flipping the sign
of $\hat{\mathbf{v}}_{\mathrm{bst}}$, 
\begin{eqnarray}
\mathbf{p}_{i} & = & \mathbf{p}_{c,i}+(\gamma_{\mathrm{bst}}-1)\hat{\mathbf{v}}_{\mathrm{bst}}(\hat{\mathbf{v}}_{\mathrm{bst}}\cdot\mathbf{p}_{c,i})+\gamma_{\mathrm{bst}}\mathbf{v}_{\mathrm{bst}}E_{c,i},\nonumber \\
E_{i} & = & \gamma_{\mathrm{bst}}(E_{c,i}+\mathbf{v}_{\mathrm{bst}}\cdot\mathbf{p}_{c,i}).\label{eq:lab-to-cmf}
\end{eqnarray}
The above defines the Lorentz transformation matrix $[\Lambda^{-1}]_{\;\nu}^{\mu}$. 

The Lorentz transformation for $x_{A}=(t_{A},\mathbf{x}_{A})$ and
$x_{B}=(t_{B},\mathbf{x}_{B})$ is 
\begin{eqnarray}
\mathbf{x}_{c,i} & = & \mathbf{x}_{i}+(\gamma_{\mathrm{bst}}-1)\hat{\mathbf{v}}_{\mathrm{bst}}(\hat{\mathbf{v}}_{\mathrm{bst}}\cdot\mathbf{x}_{i})-\gamma_{\mathrm{bst}}\mathbf{v}_{\mathrm{bst}}t_{i},\nonumber \\
t_{c,i} & = & \gamma_{\mathrm{bst}}(t_{i}-\mathbf{v}_{\mathrm{bst}}\cdot\mathbf{x}_{i}).
\end{eqnarray}
The difference of two space-time points in the CMS are expressed in
lab frame variables, 
\begin{eqnarray}
\Delta t_{c} & = & t_{c,A}-t_{c,B}=\gamma_{\mathrm{bst}}(\Delta t-\mathbf{v}_{\mathrm{bst}}\cdot\Delta\mathbf{x}),\nonumber \\
\Delta\mathbf{x}_{c} & = & \Delta\mathbf{x}+(\gamma_{\mathrm{bst}}-1)\hat{\mathbf{v}}_{\mathrm{bst}}(\hat{\mathbf{v}}_{\mathrm{bst}}\cdot\Delta\mathbf{x})-\gamma_{\mathrm{bst}}\mathbf{v}_{\mathrm{bst}}\Delta t,\label{eq:xt-lorentz}
\end{eqnarray}
where $\Delta t=t_{A}-t_{B}$ and $\Delta\mathbf{x}=\mathbf{x}_{A}-\mathbf{x}_{B}$.
We then express the impact parameter as 
\begin{equation}
\mathbf{b}_{c}=\Delta\mathbf{x}_{c}\cdot(1-\mathbf{\hat{p}}_{c,A}\mathbf{\hat{p}}_{c,A}).\label{eq:impact-para}
\end{equation}
Let us look at the CMS constraint $\delta(\Delta t_{c})\delta(\Delta x_{c,L})$
in Eq. (\ref{eq:delta-sigma}) (we have recovered the subscript 'c').
The condition $\Delta t_{c}=0$ leads to 
\begin{equation}
\Delta t=\mathbf{v}_{\mathrm{bst}}\cdot\Delta\mathbf{x},\label{eq:lorentz-t-cond}
\end{equation}
while the condition $\mathbf{\hat{p}}_{c,A}\cdot\Delta\mathbf{x}_{c}=0$
leads to 
\begin{equation}
(\mathbf{v}_{A}-\mathbf{v}_{B})\cdot\Delta\mathbf{x}=0,\label{eq:lorentz-x-cond}
\end{equation}
where we have used 
\begin{equation}
\Delta\mathbf{x}_{c}=\Delta\mathbf{x}+(\gamma_{\mathrm{bst}}^{-1}-1)\hat{\mathbf{v}}_{\mathrm{bst}}(\hat{\mathbf{v}}_{\mathrm{bst}}\cdot\Delta\mathbf{x}),
\end{equation}
which is the result of Eqs. (\ref{eq:xt-lorentz},\ref{eq:lorentz-t-cond}).
The condition in Eq. (\ref{eq:lorentz-x-cond}) means that $(\mathbf{x}_{A}-\mathbf{x}_{B})\perp(\mathbf{v}_{A}-\mathbf{v}_{B})$.
Equation (\ref{eq:lorentz-t-cond}) and (\ref{eq:lorentz-x-cond})
are the lab frame version of the constraint $\delta(\Delta t_{c})\delta(\Delta x_{c,L})$.

\section{Integration over impact parameter and Delta Functions in Eq. (\ref{eq:polarization-1})}

\label{sec:integration}We carry out the integration over the impact
parameter and show how to remove the delta functions by integration
in Eq. (\ref{eq:polarization-1}). 

Substitute Eq. \ref{eq:deltaM-nc} into Eq. \ref{eq:polarization-1},
we have the integration of $\mathbf{b}_{c}$ in the following form
\begin{eqnarray}
I(\mathbf{b}_{c}) & = & i\int d^{2}\mathbf{b}_{c}\exp\left(i\mathbf{a}\cdot\mathbf{b}_{c}\right)\frac{1}{b_{c}^{2}}\mathbf{b}_{c,j}\mathbf{b}_{c,k}\mathbf{b}_{c,l}\nonumber \\
 & = & -\frac{\partial}{\partial\mathbf{a}_{l}}\frac{\partial}{\partial\mathbf{a}_{j}}\frac{\partial}{\partial\mathbf{a}_{k}}\int d^{2}\mathbf{b}_{c}\exp\left(i\mathbf{a}\cdot\mathbf{b}_{c}\right)\frac{1}{b_{c}^{2}}\nonumber \\
 & = & -2\pi\frac{\partial}{\partial\mathbf{a}_{l}}\frac{\partial}{\partial\mathbf{a}_{j}}\frac{\partial}{\partial\mathbf{a}_{k}}\int_{0}^{b_{0}}db_{c}\frac{1}{b_{c}}J_{0}(ab_{c}),\label{eq:b-integral}
\end{eqnarray}
where $b_{c}\equiv|\mathbf{b}_{c}|$, $b_{0}$ is the cutoff of $b_{c}$,
$\mathbf{a}=\mathbf{k}_{c,A}^{\prime}-\mathbf{k}_{c,A}$, and 
\begin{equation}
J_{0}(ab_{c})=\frac{1}{2\pi}\int_{0}^{2\pi}d\phi\exp\left(iab_{c}\cos\phi\right).
\end{equation}
Then we carry out the derivatives on $\mathbf{a}_{j}$, $\mathbf{a}_{k}$
and $\mathbf{a}_{l}$, 
\begin{eqnarray}
I(\mathbf{b}_{c}) & = & -2\pi\frac{1}{a^{3}}Q_{jkl}^{L}\int_{0}^{w_{0}}dww^{2}J_{0}^{\prime\prime\prime}(w)\nonumber \\
 &  & -2\pi\frac{1}{a^{3}}Q_{jkl}^{T}\int_{0}^{w_{0}}dw\left[wJ_{0}^{\prime\prime}(w)+J_{1}(w)\right],\label{eq:b-integral-1}
\end{eqnarray}
where we have used $w_{0}=ab_{0}$ with $b_{0}$ being the upper limit
or cutoff of $b_{c}$, $J_{i}$ ($i=0,1,2$) are Bessel functions,
and 
\begin{eqnarray}
Q_{jkl}^{L} & = & \frac{\mathbf{a}_{l}\mathbf{a}_{j}\mathbf{a}_{k}}{a^{3}},\nonumber \\
Q_{jkl}^{T} & = & \frac{1}{a^{3}}\left(a^{2}\mathbf{a}_{k}\delta_{lj}+a^{2}\mathbf{a}_{l}\delta_{jk}+a^{2}\mathbf{a}_{j}\delta_{lk}-3\mathbf{a}_{l}\mathbf{a}_{j}\mathbf{a}_{k}\right).
\end{eqnarray}
Note that the overall minus sign of Eq. (\ref{eq:b-integral-1}) cancels
the one in Eq. (\ref{eq:polarization-1}). 

We carry out the integration to remove the delta functions. First
we integrate over $\mathbf{k}_{c,B}$ and $\mathbf{k}_{c,B}^{\prime}$
to remove six delta functions in three momenta, the result is to make
following replacement in the integrand 
\begin{eqnarray}
\mathbf{k}_{c,B} & = & \mathbf{p}_{c,1}+\mathbf{p}_{c,2}-\mathbf{k}_{c,A},\nonumber \\
\mathbf{k}_{c,B}^{\prime} & = & \mathbf{p}_{c,1}+\mathbf{p}_{c,2}-\mathbf{k}_{c,A}^{\prime}.
\end{eqnarray}
We are left with two delta functions for energy conservation which
can be removed by the integration over $k_{c,A}^{L}$ and $k_{c,A}^{\prime L}$,
where 'L' means the longitudinal direction along $\mathbf{p}_{c,A}$.
To this purpose, we express the energies in terms of longitudinal
and transverse momenta 
\begin{eqnarray}
E_{c,A} & = & \sqrt{(k_{c,A}^{L})^{2}+(\mathbf{k}_{c,A}^{T})^{2}+m^{2}},\nonumber \\
E_{c,B} & = & \sqrt{(\mathbf{p}_{c,1}^{T}+\mathbf{p}_{c,2}^{T}-\mathbf{k}_{c,A}^{T})^{2}+(p_{c,1}^{L}+p_{c,2}^{L}-k_{c,A}^{L})^{2}+m^{2}},\nonumber \\
E_{c,A}^{\prime} & = & \sqrt{(k_{c,A}^{\prime L})^{2}+(\mathbf{k}_{c,A}^{\prime T})^{2}+m^{2}},\nonumber \\
E_{c,B}^{\prime} & = & \sqrt{(\mathbf{p}_{c,1}^{T}+\mathbf{p}_{c,2}^{T}-\mathbf{k}_{c,A}^{\prime T})^{2}+(p_{c,1}^{L}+p_{c,2}^{L}-k_{c,A}^{\prime L})^{2}+m^{2}}.
\end{eqnarray}
So two delta functions for energy conservation become 
\begin{eqnarray}
I(\delta E) & = & \delta(E_{c,A}+E_{c,B}-E_{c,1}-E_{c,2})\nonumber \\
 & = & \frac{1}{|\mathrm{Ja}(k_{c,A}^{L}(1))|}\delta[k_{c,A}^{L}-k_{c,A}^{L}(1)]+\frac{1}{|\mathrm{Ja}(k_{c,A}^{L}(2))|}\delta[k_{c,A}^{L}-k_{c,A}^{L}(2)]\nonumber \\
I(\delta E^{\prime}) & = & \delta(E_{c,A}^{\prime}+E_{c,B}^{\prime}-E_{c,1}-E_{c,2})\nonumber \\
 & = & \frac{1}{|\mathrm{Ja}(k_{c,A}^{\prime L}(1))|}\delta[k_{c,A}^{\prime L}-k_{c,A}^{\prime L}(1)]+\frac{1}{|\mathrm{Ja}(k_{c,A}^{\prime L}(2))|}\delta[k_{c,A}^{\prime L}-k_{c,A}^{\prime L}(2)]
\end{eqnarray}
where the Jacobians of two delta functions are given by 
\begin{eqnarray}
\mathrm{Ja}(k_{c,A}^{L}) & = & \frac{\partial}{\partial k_{c,A}^{L}}(E_{c,A}+E_{c,B}-E_{c,1}-E_{c,2})\nonumber \\
 & = & k_{c,A}^{L}\left(\frac{1}{E_{c,A}}+\frac{1}{E_{c,B}}\right)-\frac{1}{E_{c,B}}(p_{c,1}^{L}+p_{c,2}^{L}),\nonumber \\
\mathrm{Ja}(k_{c,A}^{\prime L}) & = & \frac{\partial}{\partial k_{c,A}^{\prime L}}(E_{c,A}^{\prime}+E_{c,B}^{\prime}-E_{c,1}-E_{c,2})\nonumber \\
 & = & k_{c,A}^{\prime L}\left(\frac{1}{E_{c,A}^{\prime}}+\frac{1}{E_{c,B}^{\prime}}\right)-\frac{1}{E_{c,B}^{\prime}}(p_{c,1}^{L}+p_{c,2}^{L}),
\end{eqnarray}
and $k_{c,A}^{L}(i=1,2)$ and $k_{c,A}^{\prime L}(i=1,2)$ are two
roots of the energy conservation equation $E_{c,A}+E_{c,B}-E_{c,1}-E_{c,2}=0$
and $E_{c,A}^{\prime}+E_{c,B}^{\prime}-E_{c,1}-E_{c,2}=0$, respectively.
The explicit forms of $k_{c,A}^{L}(i=1,2)$ and $k_{c,A}^{\prime L}(i=1,2)$
are 
\begin{eqnarray}
k_{c,A}^{L}(1,2) & = & C_{1}\pm C_{2},\nonumber \\
k_{c,A}^{\prime L}(1,2) & = & k_{c,A}^{L}(1,2)[\mathbf{k}_{c,A}^{T}\rightarrow\mathbf{k}_{c,A}^{\prime T}],\label{eq:root-kal-kal1}
\end{eqnarray}
where $C_{1}$ and $C_{2}$ are given by 
\begin{eqnarray}
C_{1} & = & \frac{1}{2}\cdot\frac{p_{c,1}^{L}+p_{c,2}^{L}}{(E_{c,1}+E_{c,2})^{2}-(p_{c,1}^{L}+p_{c,2}^{L})^{2}}\nonumber \\
 &  & \times\left[(E_{c,1}+E_{c,2})^{2}-(p_{c,1}^{L}+p_{c,2}^{L})^{2}\right.\nonumber \\
 &  & \left.+2(\mathbf{p}_{c,1}^{T}+\mathbf{p}_{c,2}^{T})\cdot\mathbf{k}_{c,A}^{T}-(\mathbf{p}_{c,1}^{T}+\mathbf{p}_{c,2}^{T})^{2}\right],\nonumber \\
C_{2} & = & -\frac{1}{2}\cdot\frac{E_{c,1}+E_{c,2}}{(E_{c,1}+E_{c,2})^{2}-(p_{c,1}^{L}+p_{c,2}^{L})^{2}}\sqrt{H},
\end{eqnarray}
with $H$ being defined by 
\begin{eqnarray}
H & = & (E_{c,1}+E_{c,2})^{4}+4m^{2}(p_{c,1}^{L}+p_{c,2}^{L})^{2}+(\mathbf{p}_{c,1}+\mathbf{p}_{c,2})^{4}\nonumber \\
 &  & +4(\mathbf{k}_{c,A}^{T})^{2}(\mathbf{p}_{c,1}+\mathbf{p}_{c,2})^{2}-4(\mathbf{p}_{c,1}+\mathbf{p}_{c,2})^{2}[\mathbf{k}_{c,A}^{T}\cdot(\mathbf{p}_{c,1}^{T}+\mathbf{p}_{c,2}^{T})]\nonumber \\
 &  & -2(E_{c,1}+E_{c,2})^{2}\nonumber \\
 &  & \times[2m^{2}+2(\mathbf{k}_{c,A}^{T})^{2}-2\mathbf{k}_{c,A}^{T}\cdot(\mathbf{p}_{c,1}^{T}+\mathbf{p}_{c,2}^{T})+(\mathbf{p}_{c,1}+\mathbf{p}_{c,2})^{2}].
\end{eqnarray}

\section{Some formula for Dirac spinors}

The Hamiltonian for a Dirac fermion with the mass $m$ is given by
\begin{eqnarray}
H & = & \boldsymbol{\alpha}\cdot\mathbf{p}+\gamma_{0}m\nonumber \\
 & = & \left(\begin{array}{cc}
m & \boldsymbol{\sigma}\cdot\mathbf{p}\\
\boldsymbol{\sigma}\cdot\mathbf{p} & -m
\end{array}\right),\label{eq:dirac-mass}
\end{eqnarray}
where $\gamma^{\mu}=(\gamma_{0},\boldsymbol{\gamma})$ are Dirac gamma-matrices,
$\boldsymbol{\alpha}\equiv\gamma_{0}\boldsymbol{\gamma}$, and $\boldsymbol{\sigma}=(\sigma_{1},\sigma_{2},\sigma_{3})$
are Pauli matrices. The energy eigenstate can be found from the equation
\begin{equation}
H\left(\begin{array}{c}
\chi\\
\phi
\end{array}\right)=\pm E_{p}\left(\begin{array}{c}
\chi\\
\phi
\end{array}\right),
\end{equation}
where $E_{p}=\sqrt{\mathbf{p}^{2}+m^{2}}$, the sign $\pm$ in the
right-hand side corresponds to positive/negative energy state, $\chi$
and $\phi$ are Pauli spinors which form a Dirac spinor $(\chi,\phi)$.
We can express $\chi$ in terms of $\phi$ and vice versa, 
\begin{eqnarray}
\chi & = & \frac{\boldsymbol{\sigma}\cdot\mathbf{p}}{\eta E_{p}-m}\phi,\nonumber \\
\phi & = & \frac{\boldsymbol{\sigma}\cdot\mathbf{p}}{\eta E_{p}+m}\chi,
\end{eqnarray}
where $\eta=\pm1$ correspond to the positive and negative energy
state respectively. So the positive energy solution becomes 
\begin{equation}
u(s,\mathbf{p})=\sqrt{E_{p}+m}\left(\begin{array}{c}
\chi_{s}\\
\frac{\boldsymbol{\sigma}\cdot\mathbf{p}}{E_{p}+m}\chi_{s}
\end{array}\right),\label{eq:positive-en}
\end{equation}
where $s=\pm1$ is the spin orientation of the Pauli spinor and $\mathbf{n}=(\sin\theta\cos\phi,\sin\theta\sin\phi,\cos\theta)$
is the spin quantization direction. The spin eigenstates along $\mathbf{n}$
are given by 
\begin{eqnarray}
\chi_{+} & = & \left(\begin{array}{c}
e^{-i\phi}\cos\frac{\theta}{2}\\
\sin\frac{\theta}{2}
\end{array}\right),\nonumber \\
\chi_{-} & = & \left(\begin{array}{c}
-e^{-i\phi}\sin\frac{\theta}{2}\\
\cos\frac{\theta}{2}
\end{array}\right),\label{eq:pauli-spinor}
\end{eqnarray}
which satisfy 
\begin{eqnarray}
\sigma\cdot\mathbf{n} & = & \left(\begin{array}{cc}
\cos\theta & e^{-i\phi}\sin\theta\\
e^{i\phi}\sin\theta & -\cos\theta
\end{array}\right),\nonumber \\
(\sigma\cdot\mathbf{n})\chi_{s} & = & s\chi_{s}.
\end{eqnarray}
The negative energy solution can be put into the form 
\begin{eqnarray}
\tilde{v}(s,\mathbf{p}) & = & \sqrt{E_{p}+m}\left(\begin{array}{c}
-\frac{\boldsymbol{\sigma}\cdot\mathbf{p}}{E_{p}+m}\chi_{s}\\
\chi_{s}
\end{array}\right),\label{eq:negative-en}
\end{eqnarray}
The Dirac spinor for anti-particles can be defined by 
\begin{equation}
v(s,\mathbf{p})=\tilde{v}(-s,-\mathbf{p})=\sqrt{E_{p}+m}\left(\begin{array}{c}
\frac{\boldsymbol{\sigma}\cdot\mathbf{p}}{E_{p}+m}\chi_{-s}\\
\chi_{-s}
\end{array}\right),\label{eq:anti-spinor-1}
\end{equation}
or defined in terms of the positive energy solution, 
\begin{equation}
v(s,\mathbf{p})=i\gamma^{2}u^{*}(s,\mathbf{p})=-i\sqrt{E_{p}+m}\left(\begin{array}{c}
\frac{\boldsymbol{\sigma}\cdot\mathbf{p}}{E_{p}+m}\sigma_{2}\chi_{s}^{*}\\
\sigma_{2}\chi_{s}^{*}
\end{array}\right).\label{eq:anti-spinor-2}
\end{equation}
The two Dirac spinors in (\ref{eq:anti-spinor-1}) and (\ref{eq:anti-spinor-2})
are actually the same up to a sign. 

Now we rewrite the Dirac spinor of a moving particle in the way of
a Lorentz transformation of the one in the particle's rest frame.
The Lorentz transformation matrix for the Dirac spinor is given by
\begin{eqnarray}
\Lambda_{1/2}(\mathbf{p}) & = & \exp\left(-\frac{1}{2}\eta_{p}\boldsymbol{\alpha}\cdot\hat{\mathbf{p}}\right)\nonumber \\
 & = & \cosh\left(\frac{1}{2}\eta_{p}\right)-(\boldsymbol{\alpha}\cdot\hat{\mathbf{p}})\sinh\left(\frac{1}{2}\eta_{p}\right),\nonumber \\
\Lambda_{1/2}^{-1}(\mathbf{p}) & = & \Lambda_{1/2}(-\mathbf{p})=\exp\left(\frac{1}{2}\eta_{p}\boldsymbol{\alpha}\cdot\hat{\mathbf{p}}\right),\label{eq:lorentz-spinor}
\end{eqnarray}
where $\hat{\mathbf{p}}\equiv\mathbf{p}/|\mathbf{p}|$ is the momentum
direction, $\eta_{p}$ is the rapidity satisfying $E_{p}=m\cosh(\eta_{p})$,
$|\mathbf{p}|=m\sinh(\eta_{p})$, $v_{p}=\tanh(\eta_{p})$, $E_{p}+m=2m\cosh^{2}\left(\frac{1}{2}\eta_{p}\right)$,
$E_{p}-m=2m\sinh^{2}\left(\frac{1}{2}\eta_{p}\right)$. So $u(s,\mathbf{p})$
can be expressed by a Lorentz boost of $u(s,\mathbf{0})$ for the
particle at rest, 
\begin{eqnarray}
u(s,\mathbf{p}) & = & \sqrt{E_{p}+m}\left(\begin{array}{c}
\chi_{s}\\
\frac{\boldsymbol{\sigma}\cdot\mathbf{p}}{E_{p}+m}\chi_{s}
\end{array}\right)=\Lambda_{1/2}(-\mathbf{p})u(s,\mathbf{0})\nonumber \\
 & = & \sqrt{2m}\left(\begin{array}{c}
\cosh\left(\frac{1}{2}\eta_{p}\right)\chi_{s}\\
(\boldsymbol{\sigma}\cdot\hat{\mathbf{p}})\sinh\left(\frac{1}{2}\eta_{p}\right)\chi_{s}
\end{array}\right).\label{eq:spinor-p}
\end{eqnarray}
In the same way we can rewrite $v(s,\mathbf{p})$ as 
\begin{eqnarray}
v(s,\mathbf{p}) & = & \sqrt{E_{p}+m}\left(\begin{array}{c}
\frac{\boldsymbol{\sigma}\cdot\mathbf{p}}{E_{p}+m}\chi_{-s}\\
\chi_{-s}
\end{array}\right)=\Lambda_{1/2}(-\mathbf{p})v(s,\mathbf{0})\nonumber \\
 & = & \sqrt{2m}\left(\begin{array}{c}
(\boldsymbol{\sigma}\cdot\hat{\mathbf{p}})\sinh\left(\frac{1}{2}\eta_{p}\right)\chi_{-s}\\
\cosh\left(\frac{1}{2}\eta_{p}\right)\chi_{-s}
\end{array}\right).\label{eq:spinor-ap}
\end{eqnarray}
With Eqs. (\ref{eq:spinor-p},\ref{eq:spinor-ap}) we have following
formula 
\begin{eqnarray}
\sum_{s}u(s,\mathbf{p})\bar{u}(s,\mathbf{q}) & = & \Lambda_{1/2}(-\mathbf{p})\left[\sum_{s}u(s,\mathbf{0})\bar{u}(s,\mathbf{0})\right]\Lambda_{1/2}^{-1}(-\mathbf{q})\nonumber \\
 & = & m\Lambda_{1/2}(-\mathbf{p})(1+\gamma_{0})\Lambda_{1/2}^{-1}(-\mathbf{q}),\nonumber \\
\sum_{s}v(s,\mathbf{p})\bar{v}(s,\mathbf{q}) & = & \Lambda_{1/2}(-\mathbf{p})\left[\sum_{s}v(s,\mathbf{0})\bar{v}(s,\mathbf{0})\right]\Lambda_{1/2}^{-1}(-\mathbf{q})\nonumber \\
 & = & m\Lambda_{1/2}(-\mathbf{p})(\gamma_{0}-1)\Lambda_{1/2}^{-1}(-\mathbf{q}),\label{eq:lorentz-trans}
\end{eqnarray}
where we have used $\bar{u}(s,\mathbf{q})=\bar{u}(s,\mathbf{0})\Lambda_{1/2}^{-1}(-\mathbf{q})$,
$\bar{v}(s,\mathbf{q})=\bar{v}(s,\mathbf{0})\Lambda_{1/2}^{-1}(-\mathbf{q})$,
$\sum_{s}u(s,\mathbf{0})\bar{u}(s,\mathbf{0})=m(1+\gamma_{0})$ and
$\sum_{s}v(s,\mathbf{0})\bar{v}(s,\mathbf{0})=m(-1+\gamma_{0})$. 

The spin projector is defined by 
\begin{equation}
\Pi(s,n)=\frac{1}{2}(1+s\gamma_{5}n^{\sigma}\gamma_{\sigma})\label{eq:spin-projector}
\end{equation}
where $n^{\sigma}$ is the Lorentz boost of the polarization vector
$(0,\mathbf{n})$ in the particle's rest frame satisfying $n\cdot p=0$
and $n^{2}=-1$. In the particle's rest frame, we have 
\begin{eqnarray}
\Pi_{\mathrm{rest}}(s,n) & = & \frac{1}{2}(1+s\mathbf{n}\cdot\boldsymbol{\Sigma})\nonumber \\
 & \equiv & \frac{1}{2}\left(\begin{array}{cc}
1+s\mathbf{n}\cdot\boldsymbol{\sigma} & 0\\
0 & 1-s\mathbf{n}\cdot\boldsymbol{\sigma}
\end{array}\right).
\end{eqnarray}
We have following properties for the spin projector 
\begin{eqnarray}
\Pi(s,n)u(s,\mathbf{p}) & = & u(s,\mathbf{p}),\nonumber \\
\Pi(s,n)v(s,\mathbf{p}) & = & v(s,\mathbf{p}),\nonumber \\
\Pi(s,n)u(-s,\mathbf{p}) & = & 0,\nonumber \\
\Pi(s,n)v(-s,\mathbf{p}) & = & 0.
\end{eqnarray}
As an example, we can explicitly verify the first one as 
\begin{eqnarray}
\Pi(s,n)u(s,\mathbf{p}) & = & \frac{1}{2}\Lambda_{1/2}(-\mathbf{p})u(s,\mathbf{0})\nonumber \\
 &  & +\frac{1}{2}sn^{\sigma}\gamma_{5}\Lambda_{1/2}(-\mathbf{p})\Lambda_{1/2}^{-1}(-\mathbf{p})\gamma_{\sigma}\Lambda_{1/2}(-\mathbf{p})u(s,\mathbf{0})\nonumber \\
 & = & \frac{1}{2}\Lambda_{1/2}(-\mathbf{p})u(s,\mathbf{0})\nonumber \\
 &  & +\frac{1}{2}s\Lambda_{1/2}(-\mathbf{p})\gamma_{5}n^{\sigma}\varLambda_{\sigma}^{\:\nu}(-\mathbf{p})\gamma_{\nu}u(s,\mathbf{0})\nonumber \\
 & = & \frac{1}{2}\Lambda_{1/2}(-\mathbf{p})u(s,\mathbf{0})\nonumber \\
 &  & +\frac{1}{2}s\Lambda_{1/2}(-\mathbf{p})\gamma_{0}(\mathbf{n}\cdot\boldsymbol{\Sigma})u(s,\mathbf{0})\nonumber \\
 & = & \Lambda_{1/2}(-\mathbf{p})\Pi_{\mathrm{rest}}(s,n)u(s,\mathbf{0})\nonumber \\
 & = & u(s,\mathbf{p}),
\end{eqnarray}
where we have used $\Lambda_{1/2}^{-1}(-\mathbf{p})\gamma_{\sigma}\Lambda_{1/2}(-\mathbf{p})=\varLambda_{\sigma}^{\:\nu}(-\mathbf{p})\gamma_{\nu}$
and $\varLambda_{\sigma}^{\:\nu}(-\mathbf{p})=\varLambda_{\:\sigma}^{\nu}(\mathbf{p})$.
Using the spin projector, we have the following relation 
\begin{eqnarray}
\Pi(s_{0},n)\sum_{s}u(s,\mathbf{p})\bar{u}(s,\mathbf{p}) & = & \Pi(s_{0},n)\left.(p\cdot\gamma+m)\right|_{p^{\mu}=(E_{p},\mathbf{p})}\nonumber \\
 & = & u(s_{0},\mathbf{p})\bar{u}(s_{0},\mathbf{p}),\label{eq:proj-spinor}
\end{eqnarray}
where $p\cdot\gamma\equiv p_{\mu}\gamma^{\mu}$.

\section{Polarized amplitudes for quarks in 2-to-2 parton scatterings}

\label{sec:all-22-amp}In this appendix, we give polarized amplitudes
for quarks in all 2-to-2 parton scatterings listed in Table \ref{tab:feyn-22-quark}.
We assume the same quark mass $m$ for all flavors and that the external
gluon is massless. We introduce a mass into internal gluons or gluon
propagators in the t and u channel to regulate the possible divergence. 

All kinematic variables are defined in the CMS in this appendix, for
notational simplicity we will suppress the subscript 'c' for all variables,
for example, $p_{A}$ actually means $p_{cA}$. The values of color
factors, denoted as $C_{AB\rightarrow CD}$ for the process $A+B\rightarrow C+D$,
are given in Table \ref{tab:color-factor}.

\subsection{$q_{a}q_{b}\rightarrow q_{a}q_{b}$ with $a\protect\neq b$}

Following the Feynman diagram in Table \ref{tab:feyn-22-quark}, we
obtain the difference in the squared amplitude between the spin state
$s_{2}=1/2$ and $s_{2}=-1/2$ for $q_{b}$ in the final state,

\begin{eqnarray}
\Delta I_{M}^{q_{a}q_{b}\rightarrow q_{a}q_{b}} & = & I_{M}^{q_{a}q_{b}\rightarrow q_{a}q_{b}}(s_{2}=1/2)-I_{M}^{q_{a}q_{b}\rightarrow q_{a}q_{b}}(s_{2}=-1/2)\nonumber \\
 & = & C_{q_{a}q_{b}\rightarrow q_{a}q_{b}}g_{s}^{4}m^{2}\frac{1}{q^{2}q^{\prime2}}\nonumber \\
 &  & \times\mathrm{Tr}\left[(p_{1}\cdot\gamma+m)\gamma^{\mu}\Lambda_{1/2}(-\mathbf{k}_{A})(\gamma_{0}+1)\Lambda_{1/2}^{-1}(-\mathbf{k}_{A}^{\prime})\gamma^{\nu}\right]\nonumber \\
 &  & \times\mathrm{Tr}\left[\gamma_{5}(n\cdot\gamma)(p_{2}\cdot\gamma+m)\gamma_{\mu}\Lambda_{1/2}(-\mathbf{k}_{B})(\gamma_{0}+1)\Lambda_{1/2}^{-1}(-\mathbf{k}_{B}^{\prime})\gamma_{\nu}\right],\label{eq:delta-M-qaqb-qaqb-1}
\end{eqnarray}
where $q=k_{A}-p_{1}$ and $q^{\prime}=k_{A}^{\prime}-p_{1}$ are
momenta in the propagators.

\subsection{$\bar{q}_{a}q_{b}\rightarrow\bar{q}_{a}q_{b}$ with $a\protect\neq b$}

For the polarization of $q_{b}$, we obtain 
\begin{eqnarray}
\Delta I_{M}^{\bar{q}_{a}q_{b}\rightarrow\bar{q}_{a}q_{b}} & = & C_{\bar{q}_{a}q_{b}\rightarrow\bar{q}_{a}q_{b}}g_{s}^{4}m^{2}\frac{1}{q^{2}q^{\prime2}}\nonumber \\
 &  & \times\mathrm{Tr}\left[\gamma^{\mu}(p_{1}\cdot\gamma-m)\gamma^{\nu}\Lambda_{1/2}(-\mathbf{k}_{A}^{\prime})(\gamma_{0}-1)\Lambda_{1/2}^{-1}(-\mathbf{k}_{A})\right]\nonumber \\
 &  & \times\mathrm{Tr}\left[\gamma_{5}(n\cdot\gamma)(p_{2}\cdot\gamma+m)\gamma_{\mu}\Lambda_{1/2}(-\mathbf{k}_{B})(\gamma_{0}+1)\Lambda_{1/2}^{-1}(-\mathbf{k}_{B}^{\prime})\gamma_{\nu}\right],
\end{eqnarray}
where $q=k_{A}-p_{1}$ and $q^{\prime}=k_{A}^{\prime}-p_{1}$ are
momenta in the propagators.

\subsection{$\bar{q}_{a}q_{a}\rightarrow\bar{q}_{a}q_{a}$}

For the polarization of $q_{a}$ in the final state, we obtain 
\begin{eqnarray}
\Delta I_{M}^{\bar{q}_{a}q_{a}\rightarrow\bar{q}_{a}q_{a}} & = & C_{\bar{q}_{a}q_{a}\rightarrow\bar{q}_{a}q_{a}}^{(1)}g_{s}^{4}m^{2}\frac{1}{q_{1}^{2}q_{1}^{\prime2}}\nonumber \\
 &  & \times\mathrm{Tr}\left[\gamma_{5}(n\cdot\gamma)(p_{2}\cdot\gamma+m)\gamma^{\mu}\Lambda_{1/2}(-\mathbf{k}_{B})(\gamma_{0}+1)\Lambda_{1/2}^{-1}(-\mathbf{k}_{B}^{\prime})\gamma^{\nu}\right]\nonumber \\
 &  & \times\mathrm{Tr}\left[(p_{1}\cdot\gamma-m)\gamma_{\nu}\Lambda_{1/2}(-\mathbf{k}_{A}^{\prime})(\gamma_{0}-1)\Lambda_{1/2}^{-1}(-\mathbf{k}_{A})\gamma_{\mu}\right]\nonumber \\
 &  & -C_{\bar{q}_{a}q_{a}\rightarrow\bar{q}_{a}q_{a}}^{(2)}g_{s}^{4}m^{2}\frac{1}{q_{1}^{2}q_{2}^{\prime2}}\nonumber \\
 &  & \times\mathrm{Tr}\left[\gamma_{5}(n\cdot\gamma)(p_{2}\cdot\gamma+m)\gamma^{\mu}\Lambda_{1/2}(-\mathbf{k}_{B})(\gamma_{0}+1)\Lambda_{1/2}^{-1}(-\mathbf{k}_{B}^{\prime})\gamma^{\nu}\right.\nonumber \\
 &  & \times\left.\Lambda_{1/2}(-\mathbf{k}_{A}^{\prime})(\gamma_{0}-1)\Lambda_{1/2}^{-1}(-\mathbf{k}_{A})\gamma_{\mu}(p_{1}\cdot\gamma-m)\gamma_{\nu}\right]\nonumber \\
 &  & -C_{\bar{q}_{a}q_{a}\rightarrow\bar{q}_{a}q_{a}}^{(2)}g_{s}^{4}m^{2}\frac{1}{q_{2}^{2}q_{1}^{\prime2}}\nonumber \\
 &  & \times\mathrm{Tr}\left[\gamma_{5}(n\cdot\gamma)(p_{2}\cdot\gamma+m)\gamma_{\mu}(p_{1}\cdot\gamma-m)\gamma_{\nu}\right.\nonumber \\
 &  & \times\left.\Lambda_{1/2}(-\mathbf{k}_{A}^{\prime})(\gamma_{0}-1)\Lambda_{1/2}^{-1}(-\mathbf{k}_{A})\gamma^{\mu}\Lambda_{1/2}(-\mathbf{k}_{B})(\gamma_{0}+1)\Lambda_{1/2}^{-1}(-\mathbf{k}_{B}^{\prime})\gamma^{\nu}\right]\nonumber \\
 &  & +C_{\bar{q}_{a}q_{a}\rightarrow\bar{q}_{a}q_{a}}^{(1)}g_{s}^{4}m^{2}\frac{1}{q_{2}^{2}q_{2}^{\prime2}}\nonumber \\
 &  & \times\mathrm{Tr}\left[\Lambda_{1/2}(-\mathbf{k}_{A}^{\prime})(\gamma_{0}-1)\Lambda_{1/2}^{-1}(-\mathbf{k}_{A})\gamma^{\mu}\Lambda_{1/2}(-\mathbf{k}_{B})(\gamma_{0}+1)\Lambda_{1/2}^{-1}(-\mathbf{k}_{B}^{\prime})\gamma^{\nu}\right]\nonumber \\
 &  & \times\mathrm{Tr}\left[\gamma_{5}(n\cdot\gamma)(p_{2}\cdot\gamma+m)\gamma_{\mu}(p_{1}\cdot\gamma-m)\gamma_{\nu}\right],
\end{eqnarray}
where $q_{1}=k_{A}-p_{1}$, $q_{2}=k_{A}+k_{B}$, $q_{1}^{\prime}=k_{A}^{\prime}-p_{1}$
and $q_{2}^{\prime}=k_{A}^{\prime}+k_{B}^{\prime}$ are momenta in
the propagators.

\subsection{$q_{a}q_{a}\rightarrow q_{a}q_{a}$}

For the polarization of $q_{a}$ in the final state, we obtain 

\begin{eqnarray}
\Delta I_{M}^{q_{a}q_{a}\rightarrow q_{a}q_{a}} & = & C_{q_{a}q_{a}\rightarrow q_{a}q_{a}}^{(1)}g_{s}^{4}m^{2}\frac{1}{q_{1}^{2}q_{1}^{\prime2}}\nonumber \\
 &  & \times\mathrm{Tr}\left[(p_{1}\cdot\gamma+m)\gamma^{\mu}\Lambda_{1/2}(-\mathbf{k}_{A})(\gamma_{0}+1)\Lambda_{1/2}^{-1}(-\mathbf{k}_{A}^{\prime})\gamma^{\nu}\right]\nonumber \\
 &  & \times\mathrm{Tr}\left[\gamma_{5}(n\cdot\gamma)(p_{2}\cdot\gamma+m)\gamma_{\mu}\Lambda_{1/2}(-\mathbf{k}_{B})(\gamma_{0}+1)\Lambda_{1/2}^{-1}(-\mathbf{k}_{B}^{\prime})\gamma_{\nu}\right]\nonumber \\
 &  & -C_{q_{a}q_{a}\rightarrow q_{a}q_{a}}^{(2)}g_{s}^{4}m^{2}\frac{1}{q_{1}^{2}q_{2}^{\prime2}}\nonumber \\
 &  & \times\mathrm{Tr}\left[(p_{1}\cdot\gamma+m)\gamma^{\mu}\Lambda_{1/2}(-\mathbf{k}_{A})(\gamma_{0}+1)\Lambda_{1/2}^{-1}(-\mathbf{k}_{A}^{\prime})\gamma^{\nu}\right.\nonumber \\
 &  & \times\left.\gamma_{5}(n\cdot\gamma)(p_{2}\cdot\gamma+m)\gamma_{\mu}\Lambda_{1/2}(-\mathbf{k}_{B})(\gamma_{0}+1)\Lambda_{1/2}^{-1}(-\mathbf{k}_{B}^{\prime})\gamma_{\nu}\right]\nonumber \\
 &  & -C_{q_{a}q_{a}\rightarrow q_{a}q_{a}}^{(2)}g_{s}^{4}m^{2}\frac{1}{q_{1}^{\prime2}q_{2}^{2}}\nonumber \\
 &  & \times\mathrm{Tr}\left[\gamma^{\mu}\Lambda_{1/2}(-\mathbf{k}_{A})(\gamma_{0}+1)\Lambda_{1/2}^{-1}(-\mathbf{k}_{A}^{\prime})\gamma^{\nu}(p_{1}\cdot\gamma+m)\right.\nonumber \\
 &  & \times\left.\gamma_{\mu}\Lambda_{1/2}(-\mathbf{k}_{B})(\gamma_{0}+1)\Lambda_{1/2}^{-1}(-\mathbf{k}_{B}^{\prime})\gamma_{\nu}\gamma_{5}(n\cdot\gamma)(p_{2}\cdot\gamma+m)\right]\nonumber \\
 &  & +C_{q_{a}q_{a}\rightarrow q_{a}q_{a}}^{(1)}g_{s}^{4}m^{2}\frac{1}{q_{2}^{\prime2}q_{2}^{2}}\nonumber \\
 &  & \times\mathrm{Tr}\left[\gamma_{5}(n\cdot\gamma)(p_{2}\cdot\gamma+m)\gamma^{\mu}\Lambda_{1/2}(-\mathbf{k}_{A})(\gamma_{0}+1)\Lambda_{1/2}^{-1}(-\mathbf{k}_{A}^{\prime})\gamma^{\nu}\right]\nonumber \\
 &  & \times\mathrm{Tr}\left[\Lambda_{1/2}(-\mathbf{k}_{B})(\gamma_{0}+1)\Lambda_{1/2}^{-1}(-\mathbf{k}_{B}^{\prime})\gamma_{\nu}(p_{1}\cdot\gamma+m)\gamma_{\mu}\right],
\end{eqnarray}
where $q_{1}=k_{A}-p_{1}$, $q_{2}=k_{A}-p_{2}$, $q_{1}^{\prime}=k_{A}^{\prime}-p_{1}$
and $q_{2}^{\prime}=k_{A}^{\prime}-p_{2}$ are momenta in propagators.

\subsection{$gg\rightarrow\bar{q}_{a}q_{a}$}

In principle, the ghost diagrams should also contribute. However,
its contribution is canceled when we calculate $\Delta I_{M}^{gg\rightarrow\bar{q}_{a}q_{a}}$.
For the polarization of $q_{a}$ in the final state, we obtain 
\begin{eqnarray}
\Delta I_{M}^{gg\rightarrow\bar{q}_{a}q_{a}} & = & C_{gg\rightarrow\bar{q}_{a}q_{a}}^{(1)}g_{s}^{4}\frac{1}{(q_{1}^{2}-m^{2})(q_{1}^{\prime2}-m^{2})}I_{1}\nonumber \\
 &  & +C_{gg\rightarrow\bar{q}_{a}q_{a}}^{(2)}g_{s}^{4}\frac{1}{(q_{1}^{2}-m^{2})(q_{2}^{\prime2}-m^{2})}I_{2}\nonumber \\
 &  & -C_{gg\rightarrow\bar{q}_{a}q_{a}}^{(3)}g_{s}^{4}\frac{1}{(q_{1}^{2}-m^{2})q_{3}^{\prime2}}I_{3}\nonumber \\
 &  & +C_{gg\rightarrow\bar{q}_{a}q_{a}}^{(2)}g_{s}^{4}\frac{1}{(q_{1}^{\prime2}-m^{2})(q_{2}^{2}-m^{2})}I_{4}\nonumber \\
 &  & +C_{gg\rightarrow\bar{q}_{a}q_{a}}^{(1)}g_{s}^{4}\frac{1}{(q_{2}^{2}-m^{2})(q_{2}^{\prime2}-m^{2})}I_{5}\nonumber \\
 &  & +C_{gg\rightarrow\bar{q}_{a}q_{a}}^{(3)}g_{s}^{4}\frac{1}{(q_{2}^{2}-m^{2})q_{3}^{\prime2}}I_{6}\nonumber \\
 &  & -C_{gg\rightarrow\bar{q}_{a}q_{a}}^{(3)}g_{s}^{4}\frac{1}{(q_{1}^{\prime2}-m^{2})q_{3}^{2}}I_{7}\nonumber \\
 &  & +C_{gg\rightarrow\bar{q}_{a}q_{a}}^{(3)}g_{s}^{4}\frac{1}{(q_{2}^{\prime2}-m^{2})q_{3}^{2}}I_{8}\nonumber \\
 &  & +C_{gg\rightarrow\bar{q}_{a}q_{a}}^{(4)}g_{s}^{4}\frac{1}{q_{3}^{2}q_{3}^{\prime2}}I_{9},\label{eq:gg-q-bar-q}
\end{eqnarray}
where $q_{1}=k_{A}-p_{1}$, $q_{2}=p_{2}-k_{A}$, $q_{3}=k_{A}+k_{B}$
, $q_{1}^{\prime}=k_{A}^{\prime}-p_{1}$, $q_{2}^{\prime}=p_{2}-k_{A}^{\prime}$
and $q_{3}^{\prime}=k_{A}^{\prime}+k_{B}^{\prime}$ are momenta in
propagators, and the terms $I_{i}^{\rho}$ for $i=1,2,\cdots,9$ are
given by 

\begin{eqnarray}
I_{1} & = & \mathrm{Tr}[\gamma_{5}(n\cdot\gamma)(p_{2}\cdot\gamma+m)\gamma^{\nu}(q_{1}\cdot\gamma+m)\gamma^{\mu}\nonumber \\
 &  & \times(p_{1}\cdot\gamma-m)\gamma^{\mu^{\prime}}(q_{1}^{\prime}\cdot\gamma+m)\gamma^{\nu^{\prime}}]g_{\mu\mu^{\prime}}g_{\nu\nu^{\prime}}
\end{eqnarray}

\begin{eqnarray}
I_{2} & = & \mathrm{Tr}[\gamma_{5}(n\cdot\gamma)(p_{2}\cdot\gamma+m)\gamma^{\nu}(q_{1}\cdot\gamma+m)\gamma^{\mu}\nonumber \\
 &  & \times(p_{1}\cdot\gamma-m)\gamma^{\nu^{\prime}}(q_{2}^{\prime}\cdot\gamma+m)\gamma^{\mu^{\prime}}]g_{\mu\mu^{\prime}}g_{\nu\nu^{\prime}}
\end{eqnarray}
\begin{eqnarray}
I_{3} & = & \mathrm{Tr}[\gamma_{5}(n\cdot\gamma)(p_{2}\cdot\gamma+m)\gamma^{\nu}(q_{1}\cdot\gamma+m)\gamma^{\mu}(p_{1}\cdot\gamma-m)\gamma_{\sigma^{\prime}}]g_{\mu\mu^{\prime}}g_{\nu\nu^{\prime}}\nonumber \\
 &  & \times[g^{\sigma^{\prime}\mu^{\prime}}(-q_{3}^{\prime}-k_{A}^{\prime})^{\nu^{\prime}}+g^{\mu^{\prime}\nu^{\prime}}(k_{A}^{\prime}-k_{B}^{\prime})^{\sigma^{\prime}}+g^{\nu^{\prime}\sigma^{\prime}}(k_{B}^{\prime}+q_{3}^{\prime})^{\mu^{\prime}}]
\end{eqnarray}
\begin{eqnarray}
I_{4} & = & \mathrm{Tr}[\gamma_{5}(n\cdot\gamma)(p_{2}\cdot\gamma+m)\gamma^{\mu}(q_{2}\cdot\gamma+m)\gamma^{\nu}\nonumber \\
 &  & \times(p_{1}\cdot\gamma-m)\gamma^{\mu^{\prime}}(q_{1}^{\prime}\cdot\gamma+m)\gamma^{\nu^{\prime}}]g_{\mu\mu^{\prime}}g_{\nu\nu^{\prime}}
\end{eqnarray}
\begin{eqnarray}
I_{5} & = & \mathrm{Tr}[\gamma_{5}(n\cdot\gamma)(p_{2}\cdot\gamma+m)\gamma^{\mu}(q_{2}\cdot\gamma+m)\gamma^{\nu}\nonumber \\
 &  & \times(p_{1}\cdot\gamma-m)\gamma^{\nu^{\prime}}(q_{2}^{\prime}\cdot\gamma+m)\gamma^{\mu^{\prime}}]g_{\mu\mu^{\prime}}g_{\nu\nu^{\prime}}
\end{eqnarray}
\begin{eqnarray}
I_{6} & = & \mathrm{Tr}[\gamma_{5}(n\cdot\gamma)(p_{2}\cdot\gamma+m)\gamma^{\mu}(q_{2}\cdot\gamma+m)\gamma^{\nu}(p_{1}\cdot\gamma-m)\gamma_{\sigma^{\prime}}]g_{\mu\mu^{\prime}}g_{\nu\nu^{\prime}}\nonumber \\
 &  & \times[g^{\sigma^{\prime}\mu^{\prime}}(-q_{3}^{\prime}-k_{A}^{\prime})^{\nu^{\prime}}+g^{\mu^{\prime}\nu^{\prime}}(k_{A}^{\prime}-k_{B}^{\prime})^{\sigma^{\prime}}+g^{\nu^{\prime}\sigma^{\prime}}(k_{B}^{\prime}+q_{3}^{\prime})^{\mu^{\prime}}]
\end{eqnarray}
\begin{eqnarray}
I_{7} & = & \mathrm{Tr}[\gamma_{5}(n\cdot\gamma)(p_{2}\cdot\gamma+m)\gamma_{\sigma}(p_{1}\cdot\gamma-m)\gamma^{\mu^{\prime}}(q_{1}^{\prime}\cdot\gamma+m)\gamma^{\nu^{\prime}}]g_{\mu\mu^{\prime}}g_{\nu\nu^{\prime}}\nonumber \\
 &  & \times[g^{\sigma\mu}(-q_{3}-k_{A})^{\nu}+g^{\mu\nu}(k_{A}-k_{B})^{\sigma}+g^{\nu\sigma}(k_{B}+q_{3})^{\mu}]
\end{eqnarray}
\begin{eqnarray}
I_{8} & = & \mathrm{Tr}[\gamma_{5}(n\cdot\gamma)(p_{2}\cdot\gamma+m)\gamma_{\sigma}(p_{1}\cdot\gamma-m)\gamma^{\nu^{\prime}}(q_{2}^{\prime}\cdot\gamma+m)\gamma^{\mu^{\prime}}]g_{\mu\mu^{\prime}}g_{\nu\nu^{\prime}}\nonumber \\
 &  & \times[g^{\sigma\mu}(-q_{3}-k_{A})^{\nu}+g^{\mu\nu}(k_{A}-k_{B})^{\sigma}+g^{\nu\sigma}(k_{B}+q_{3})^{\mu}]
\end{eqnarray}
\begin{eqnarray}
I_{9} & = & \mathrm{Tr}[\gamma_{5}(n\cdot\gamma)(p_{2}\cdot\gamma+m)\gamma_{\sigma}(p_{1}\cdot\gamma-m)\gamma_{\sigma^{\prime}}]\nonumber \\
 &  & \times[g^{\sigma\mu}(-q_{3}-k_{A})^{\nu}+g^{\mu\nu}(k_{A}-k_{B})^{\sigma}+g^{\nu\sigma}(k_{B}+q_{3})^{\mu}]\nonumber \\
 &  & \times[g^{\sigma^{\prime}\mu^{\prime}}(-q_{3}^{\prime}-k_{A}^{\prime})^{\nu^{\prime}}+g^{\mu^{\prime}\nu^{\prime}}(k_{A}^{\prime}-k_{B}^{\prime})^{\sigma^{\prime}}+g^{\nu^{\prime}\sigma^{\prime}}(k_{B}^{\prime}+q_{3}^{\prime})^{\mu^{\prime}}]\nonumber \\
 &  & \times g_{\mu\mu^{\prime}}g_{\nu\nu^{\prime}}
\end{eqnarray}

\subsection{$gq_{a}\rightarrow gq_{a}$}

In principle, the ghost diagram should also contribute. However, its
contribution is canceled when we calculate $\Delta I_{M}^{gq_{a}\rightarrow gq_{a}}$.
For the polarization of $q_{a}$ in the final state, we obtain 
\begin{eqnarray}
\Delta I_{M}^{gq_{a}\rightarrow gq_{a}} & = & C_{gq_{a}\rightarrow gq_{a}}^{(1)}g_{s}^{4}m\frac{1}{q_{1}^{2}q_{1}^{\prime2}}I_{1}\nonumber \\
 &  & +C_{gq_{a}\rightarrow gq_{a}}^{(2)}g_{s}^{4}m\frac{1}{q_{1}^{2}(q_{2}^{\prime2}-m^{2})}I_{2}\nonumber \\
 &  & -C_{gq_{a}\rightarrow gq_{a}}^{(2)}g_{s}^{4}m\frac{1}{q_{1}^{2}(q_{3}^{\prime2}-m^{2})}I_{3}\nonumber \\
 &  & +C_{gq_{a}\rightarrow gq_{a}}^{(2)}g_{s}^{4}m\frac{1}{q_{1}^{\prime2}(q_{2}^{2}-m^{2})}I_{4}\nonumber \\
 &  & +C_{gq_{a}\rightarrow gq_{a}}^{(3)}g_{s}^{4}m\frac{1}{(q_{2}^{2}-m^{2})(q_{2}^{\prime2}-m^{2})}I_{5}\nonumber \\
 &  & +C_{gq_{a}\rightarrow gq_{a}}^{(4)}g_{s}^{4}m\frac{1}{(q_{2}^{2}-m^{2})(q_{3}^{\prime2}-m^{2})}I_{6}\nonumber \\
 &  & -C_{gq_{a}\rightarrow gq_{a}}^{(2)}g_{s}^{4}m\frac{1}{q_{1}^{\prime2}(q_{3}^{2}-m^{2})}I_{7}\nonumber \\
 &  & +C_{gq_{a}\rightarrow gq_{a}}^{(4)}g_{s}^{4}m\frac{1}{(q_{2}^{\prime2}-m^{2})(q_{3}^{2}-m^{2})}I_{8}\nonumber \\
 &  & +C_{gq_{a}\rightarrow gq_{a}}^{(3)}g_{s}^{4}m\frac{1}{(q_{3}^{2}-m^{2})(q_{3}^{\prime2}-m^{2})}I_{9}
\end{eqnarray}
where $q_{1}=k_{A}-p_{1}$, $q_{2}=p_{2}-k_{A}$, $q_{3}=k_{A}+k_{B}$,
$q_{1}^{\prime}=k_{A}^{\prime}-p_{1}$, $q_{2}^{\prime}=p_{2}-k_{A}^{\prime}$
and $q_{3}^{\prime}=k_{A}^{\prime}+k_{B}^{\prime}$ are momenta in
propagators, and the terms $I_{i}^{\rho}$ for $i=1,2,\cdots,9$ are
given by 

\begin{eqnarray}
I_{1} & = & \mathrm{Tr}[\gamma_{5}(n\cdot\gamma)(p_{2}\cdot\gamma+m)\gamma_{\sigma}\Lambda_{1/2}(-\mathbf{k}_{B})(\gamma_{0}+1)\Lambda_{1/2}^{-1}(-\mathbf{k}_{B}^{\prime})\gamma_{\sigma^{\prime}}]\nonumber \\
 &  & \times g_{\mu\mu^{\prime}}g_{\nu\nu^{\prime}}[g^{\mu\nu}(k_{A}+p_{1})^{\sigma}+g^{\nu\sigma}(q_{1}-p_{1})^{\mu}+g^{\sigma\mu}(-q_{1}-k_{A})^{\nu}]\nonumber \\
 &  & \times[g^{\mu^{\prime}\nu^{\prime}}(k_{A}^{\prime}+p_{1})^{\sigma^{\prime}}+g^{\nu^{\prime}\sigma^{\prime}}(q_{1}^{\prime}-p_{1})^{\mu^{\prime}}+g^{\sigma^{\prime}\mu^{\prime}}(-q_{1}^{\prime}-k_{A}^{\prime})^{\nu^{\prime}}]
\end{eqnarray}
\begin{eqnarray}
I_{2} & = & \mathrm{Tr}[\gamma_{5}(n\cdot\gamma)(p_{2}\cdot\gamma+m)\gamma_{\sigma}\Lambda_{1/2}(-\mathbf{k}_{B})(\gamma_{0}+1)\Lambda_{1/2}^{-1}(-\mathbf{k}_{B}^{\prime})\nonumber \\
 &  & \times\gamma^{\nu^{\prime}}(q_{2}^{\prime}\cdot\gamma+m)\gamma^{\mu^{\prime}}]g_{\mu\mu^{\prime}}g_{\nu\nu^{\prime}}\nonumber \\
 &  & \times[g^{\mu\nu}(k_{A}+p_{1})^{\sigma}+g^{\nu\sigma}(q_{1}-p_{1})^{\mu}+g^{\sigma\mu}(-q_{1}-k_{A})^{\nu}]
\end{eqnarray}
\begin{eqnarray}
I_{3} & = & \mathrm{Tr}[\gamma_{5}(n\cdot\gamma)(p_{2}\cdot\gamma+m)\gamma_{\sigma}\Lambda_{1/2}(-\mathbf{k}_{B})(\gamma_{0}+1)\Lambda_{1/2}^{-1}(-\mathbf{k}_{B}^{\prime})\nonumber \\
 &  & \times\gamma^{\mu^{\prime}}(q_{3}^{\prime}\cdot\gamma+m)\gamma^{\nu^{\prime}}]g_{\mu\mu^{\prime}}g_{\nu\nu^{\prime}}\nonumber \\
 &  & \times[g^{\mu\nu}(k_{A}+p_{1})^{\sigma}+g^{\nu\sigma}(q_{1}-p_{1})^{\mu}+g^{\sigma\mu}(-q_{1}-k_{A})^{\nu}]
\end{eqnarray}

\begin{eqnarray}
I_{4} & = & \mathrm{Tr}[\gamma_{5}(n\cdot\gamma)(p_{2}\cdot\gamma+m)\gamma^{\mu}(q_{2}\cdot\gamma+m)\gamma^{\nu}\nonumber \\
 &  & \times\Lambda_{1/2}(-\mathbf{k}_{B})(\gamma_{0}+1)\Lambda_{1/2}^{-1}(-\mathbf{k}_{B}^{\prime})\gamma_{\sigma^{\prime}}]g_{\mu\mu^{\prime}}g_{\nu\nu^{\prime}}\nonumber \\
 &  & \times[g^{\mu^{\prime}\nu^{\prime}}(k_{A}^{\prime}+p_{1})^{\sigma^{\prime}}+g^{\nu^{\prime}\sigma^{\prime}}(q_{1}^{\prime}-p_{1})^{\mu^{\prime}}+g^{\sigma^{\prime}\mu^{\prime}}(-q_{1}^{\prime}-k_{A}^{\prime})^{\nu^{\prime}}]
\end{eqnarray}
\begin{eqnarray}
I_{5} & = & \mathrm{Tr}[\gamma_{5}(n\cdot\gamma)(p_{2}\cdot\gamma+m)\gamma^{\mu}(q_{2}\cdot\gamma+m)\gamma^{\nu}\nonumber \\
 &  & \times\Lambda_{1/2}(-\mathbf{k}_{B})(\gamma_{0}+1)\Lambda_{1/2}^{-1}(-\mathbf{k}_{B}^{\prime})\nonumber \\
 &  & \times\gamma^{\nu^{\prime}}(q_{2}^{\prime}+m)\gamma^{\mu^{\prime}}]g_{\mu\mu^{\prime}}g_{\nu\nu^{\prime}}
\end{eqnarray}
\begin{eqnarray}
I_{6} & = & \mathrm{Tr}[\gamma_{5}(n\cdot\gamma)(p_{2}\cdot\gamma+m)\gamma^{\mu}(q_{2}\cdot\gamma+m)\gamma^{\nu}\nonumber \\
 &  & \times\Lambda_{1/2}(-\mathbf{k}_{B})(\gamma_{0}+1)\Lambda_{1/2}^{-1}(-\mathbf{k}_{B}^{\prime})\nonumber \\
 &  & \times\gamma^{\mu^{\prime}}(q_{3}^{\prime}\cdot\gamma+m)\gamma^{\nu^{\prime}}]g_{\mu\mu^{\prime}}g_{\nu\nu^{\prime}}
\end{eqnarray}
\begin{eqnarray}
I_{7} & = & \mathrm{Tr}[\gamma_{5}(n\cdot\gamma)(p_{2}\cdot\gamma+m)\gamma^{\nu}(q_{3}\cdot\gamma+m)\gamma^{\mu}\nonumber \\
 &  & \times\Lambda_{1/2}(-\mathbf{k}_{B})(\gamma_{0}+1)\Lambda_{1/2}^{-1}(-\mathbf{k}_{B}^{\prime})\gamma_{\sigma^{\prime}}]g_{\mu\mu^{\prime}}g_{\nu\nu^{\prime}}\nonumber \\
 &  & \times[g^{\mu^{\prime}\nu^{\prime}}(k_{A}^{\prime}+p_{1})^{\sigma^{\prime}}+g^{\nu^{\prime}\sigma^{\prime}}(q_{1}^{\prime}-p_{1})^{\mu^{\prime}}+g^{\sigma^{\prime}\mu^{\prime}}(-q_{1}^{\prime}-k_{A}^{\prime})^{\nu^{\prime}}]
\end{eqnarray}
\begin{eqnarray}
I_{8} & = & \mathrm{Tr}[\gamma_{5}(n\cdot\gamma)(p_{2}\cdot\gamma+m)\gamma^{\nu}(q_{3}\cdot\gamma+m)\gamma^{\mu}\nonumber \\
 &  & \times\Lambda_{1/2}(-\mathbf{k}_{B})(\gamma_{0}+1)\Lambda_{1/2}^{-1}(-\mathbf{k}_{B}^{\prime})\nonumber \\
 &  & \times\gamma^{\nu^{\prime}}(q_{2}^{\prime}\cdot\gamma+m)\gamma^{\mu^{\prime}}]g_{\mu\mu^{\prime}}g_{\nu\nu^{\prime}}
\end{eqnarray}
\begin{eqnarray}
I_{9} & = & \mathrm{Tr}[\gamma_{5}(n\cdot\gamma)(p_{2}\cdot\gamma+m)\gamma^{\nu}(q_{3}\cdot\gamma+m)\gamma^{\mu}\nonumber \\
 &  & \times\Lambda_{1/2}(-\mathbf{k}_{B})(\gamma_{0}+1)\Lambda_{1/2}^{-1}(-\mathbf{k}_{B}^{\prime})\nonumber \\
 &  & \times\gamma^{\mu^{\prime}}(q_{3}^{\prime}\cdot\gamma+m)\gamma^{\nu^{\prime}}]g_{\mu\mu^{\prime}}g_{\nu\nu^{\prime}}
\end{eqnarray}

\subsection{$\bar{q}_{a}q_{a}\rightarrow\bar{q}_{b}q_{b}$ with $a\protect\neq b$}

For the polarization of $q_{b}$ in the final state, we obtain 
\begin{eqnarray}
\Delta I_{M}^{\bar{q}_{a}q_{a}\rightarrow\bar{q}_{b}q_{b}} & = & C_{\bar{q}_{a}q_{a}\rightarrow\bar{q}_{b}q_{b}}g_{s}^{4}m^{2}\frac{1}{q^{2}q^{\prime2}}\nonumber \\
 &  & \times\mathrm{Tr}\left[\Lambda_{1/2}(-\mathbf{k}_{A}^{\prime})(\gamma_{0}-1)\Lambda_{1/2}^{-1}(-\mathbf{k}_{A})\gamma^{\mu}\right.\nonumber \\
 &  & \times\left.\Lambda_{1/2}(-\mathbf{k}_{B})(\gamma_{0}+1)\Lambda_{1/2}^{-1}(-\mathbf{k}_{B}^{\prime})\gamma^{\nu}\right]\nonumber \\
 &  & \times\mathrm{Tr}\left[\gamma_{5}(n\cdot\gamma)(p_{2}\cdot\gamma-m)\gamma_{\mu}(p_{1}\cdot\gamma-m)\gamma_{\nu}\right],
\end{eqnarray}
where $q=k_{A}+k_{B}$ and $q^{\prime}=k_{A}^{\prime}+k_{B}^{\prime}$
are momenta in propagators. 

\bibliographystyle{plain}
\bibliography{ref-1}

\end{document}